\documentclass[aps,prd,reprint,groupedaddress,superscriptaddress,preprintnumbers]{revtex4-2}

\usepackage{subfigure}
\usepackage{graphicx,natbib}
\usepackage{bm}
\usepackage{color}
\usepackage{amssymb}
\usepackage[margin=0.6in]{geometry}
\usepackage{multirow}
\usepackage{mathrsfs}
\usepackage{hyperref}

\newcommand{\meanv}{\overline{v}}
\newcommand{\AAA}{\bm{A}}
\newcommand{\BB}{\bm{B}}

\newcommand{\UU}{\bm{U}}

\newcommand{\ttau}{\bm{\tau}}
\newcommand{\SSSS}{\mbox{\boldmath ${\sf S}$} {}}

\newcommand{\meanrho}{\overline{\rho}}
\def\cs{c_{\rm s}}
\newcommand{\nab}{\bm{\nabla}}
\def\Rm{{\rm Re}_{_\mathrm{M}}}
\newcommand{\bra}[1]{\langle #1\rangle}
\def\cs{c_{\rm s}}

\newcommand{\meanBB}{\overline{\mbox{\boldmath $B$}}{}}{}
\newcommand{\meanUU}{\overline{\bm{U}}}
\newcommand{\meanEMF}{\overline{\mbox{
\boldmath ${\cal E}$}}{}}{}
\newcommand{\meanmu}{\overline{\mu}}
\newcommand{\mnras}{Mon.\ Not.\ Roy.\ Astron.\ Soc.}

\newcommand{\aap}{Astron.\ and Astrophys.}
\newcommand{\araa}{Ann.\ Rev.\ Astron.\ Astrophys.}
\newcommand{\aapr}{Astron.\ Astrophys.\ Rev.}
\newcommand{\ssr}{Space Sci.\ Rev.}

\begin{document}

\title{Dynamo instabilities in plasmas with inhomogeneous chiral chemical potential}
\preprint{NORDITA-2021-067}

\author{Jennifer~Schober}
\email{jennifer.schober@epfl.ch}
\affiliation{Laboratoire d'Astrophysique, EPFL, CH-1290 Sauverny, Switzerland}

\author{Igor~Rogachevskii}
\affiliation{Department of Mechanical Engineering, Ben-Gurion University of the Negev, P.O. Box 653, Beer-Sheva 84105, Israel}
\affiliation{Nordita, KTH Royal Institute of Technology and Stockholm University, 10691 Stockholm, Sweden}

\author{Axel~Brandenburg}
\affiliation{Nordita, KTH Royal Institute of Technology and Stockholm University, 10691 Stockholm, Sweden}
\affiliation{The Oskar Klein Centre, Department of Astronomy, Stockholm University, AlbaNova, SE-10691 Stockholm, Sweden}
\affiliation{School of Natural Sciences and Medicine, Ilia State University, 0194 Tbilisi, Georgia}
\affiliation{McWilliams Center for Cosmology and Department of Physics, Carnegie Mellon University, Pittsburgh, Pennsylvania 15213, USA}

\date{\today}

\begin{abstract}
We study the dynamics of magnetic fields in chiral magnetohydrodynamics, 
which takes into account the effects of an additional electric current 
related to the chiral magnetic effect in high-energy plasmas.
We perform direct numerical simulations, 
considering weak seed magnetic fields and 
inhomogeneities of the chiral chemical potential $\mu_5$ with a zero mean.
We demonstrate that a small-scale chiral dynamo can occur
in such plasmas if fluctuations of $\mu_5$ are correlated 
on length scales that are much larger than 
the scale on which the dynamo growth rate reaches its maximum.
Magnetic fluctuations grow by many orders of magnitude due to
the small-scale chiral dynamo instability.
Once the nonlinear backreaction of the generated magnetic 
field on fluctuations of $\mu_5$ sets in, 
the ratio of these scales decreases
and the dynamo saturates. 
When magnetic fluctuations grow sufficiently to
drive turbulence via the Lorentz force before reaching maximum field
strength, an additional mean-field dynamo phase is identified. 
The mean magnetic field grows on a scale that is larger than
the integral scale of turbulence after the amplification of
the fluctuating component saturates. 
The growth rate of the mean magnetic field 
is caused by a magnetic $\alpha$ effect
that is proportional 
to the current helicity. 
With the onset of turbulence,
the power spectrum of $\mu_5$ develops 
a universal $k^{-1}$ scaling independently of its initial shape, 
while the magnetic energy spectrum approaches a $k^{-3}$ scaling.
\end{abstract}

\maketitle

\section{Introduction}

The macroscopic dynamics of magnetized plasmas can be described by an 
effective one-fluid model, namely magnetohydrodynamics (MHD).
The set of variables in MHD includes 
the fluid density $\rho$, the velocity $\UU$,
and the magnetic field $\BB$, which are evolved by the continuity equation, the 
Navier-Stokes equation, and the induction equation,
respectively.
Together with an equation of state, this constitutes 
the basic MHD equations in a dynamical theory.
One field of research within MHD is dynamo theory
\citep{M78,KR80,Zeldovich(1983),BS05,Ruediger(2013),Brandenburg2018,RI21} 
which describes the amplification of an initially weak seed 
magnetic field by conversion
of kinetic (mostly turbulent) energy into magnetic energy.
Dynamo theory is used primarily in planetary physics and 
astrophysics, to understand the observed strength and structure 
of magnetic fields in planets 
\citep{Stevenson2003,Christensen2010,MD2019}, 
stars \citep{Parker(1979),Ossendrijver2003,KaepylaeEtAl2008}, and galaxies 
\citep{Ruzmaikin(1988),BeckEtAl1996,Kulsrud1999,SchoberEtAl2013,ChamandySingh2018}.

At very high energies, however, MHD necessarily needs to be extended to 
include the electric current caused by the chiral magnetic 
effect (CME) \citep{Vilenkin:80a}.
This macroscopic quantum effect describes the coupling between
the chiral chemical potential $\mu_5$, i.e., the difference 
between the number density
of left- and right-handed fermions, and magnetic helicity $\mathcal{H}$.
To account for the CME in the modeling of high-energy plasma, $\mu_5$ 
has to be included as an additional 
dynamical variable in the evolution equation describing the physics of the CME.
The effective theory of a plasma with nonzero $\mu_5$
is referred to as chiral MHD 
\citep{GI13,REtAl17,ZB18,HHY19},
which is an extension of classical MHD.
Due to the CME, magnetic field and magnetic helicity can be amplified by many
orders of magnitude by a chiral dynamo instability
\citep{JS97,REtAl17,BSRKBFRK17,Schober2017}.

A central property of chiral MHD is the conservation of total
chirality (the sum of mean magnetic helicity $\langle\mathcal{H}\rangle$ and 
$\langle\mu_5\rangle$ multiplied by the inverse chiral nonlinearity 
parameter), whereas in MHD, $\langle\mathcal{H}\rangle$ is conserved in the limit of vanishing magnetic resistivity.
The conservation law in chiral MHD has important consequences: the conversion of
$\langle\mu_5\rangle$ to $\langle\mathcal{H}\rangle$
leads to a transfer of magnetic energy 
from small to large spatial scales, i.e., a chirally induced inverse cascade 
\citep{BFR12,HKY15,GRS16}.
Depending on the initial condition, $\mu_5$ can also be 
generated at the expense of magnetic helicity \citep{SchoberFujitaDurrer2020}.

The extension from MHD to chiral MHD is required for all systems where
fermions can be considered as being effectively massless, i.e., where the
kinetic energy of the fermions exceed their rest energy significantly.
In this case, chirality flipping reactions are suppressed \citep{CampbellEtAl1992}. 
The critical energy scale where this transition occurs depends on the exact 
value of the chirality flipping rate which is still under debate \citep{BoyarskyEtAl2021}.
Nevertheless, typical examples of such systems are the high-energy plasma generated in 
heavy ion collisions \citep{KH14,KH16,Kharzeev:07,HironoEtAl2014},
and, indeed, signatures of the CME have been 
observed at the Relativistic Heavy Ion Collider \citep{STAR2009}
and the Large Hadron Collider \citep{ALICE2013}.
However, in view of a number of background effects, there is ambiguity 
in the interpretation of the experimental results \citep{KH16}.
Furthermore, there is also the chiral vortical effect (CVE) \citep{KH16}.
It may be important in the nonlinear stage of the magnetic field evolution, 
especially when strong chiral turbulence is produced.
However, since chiral turbulence is usually magnetically dominated, the 
CVE is likely to be subdominant.

Chiral MHD has also been applied to high-energy astrophysical plasmas like the early Universe \citep{JS97,BSRKBFRK17,DvornikovSemikoz2017,Schober2017}
and proto-neutron stars 
\citep{Charbonneau:2009ax,Ohnishi:2014uea,Yamamoto:2015gzz,SiglLeite2016,DvornikovEtAl2020} 
where, in particular, 
the evolution of the magnetic field has been studied. 
Beyond that, chiral MHD can be used to describe the
dynamics of new materials, like Weyl and Dirac semimetals \citep{GalitskiEtAl2018}.
Here, the chirality of the massless quasiparticles allows for the occurrence of the CME and an effective description of the system by chiral MHD.

In the framework of chiral MHD, the effects of coupling 
between magnetic and velocity fields were analyzed by means 
of a mean-field theory \citep{REtAl17}.
This way, a new turbulent $\alpha_\mu$ effect was identified that is 
based on fluctuations of $\mu_5$ and, contrary to the
classical kinetic $\alpha_\mathrm{K}$ effect, is not sourced by kinetic helicity. 
The $\alpha_\mu$ effect causes a mean-field dynamo instability resulting in the
generation of a mean magnetic field at a length scale that is larger
than the integral scale of turbulence.
The mean-field dynamo was observed in direct numerical simulations (DNS) 
\citep{Schober2017,SBR19,SBR20}
and the existence of the $\alpha_\mu$ effect was confirmed.
The mean-field dynamo leads to an even more efficient transfer of magnetic energy to larger spatial scales.

The initial conditions of the previously mentioned studies of
chiral dynamos included a mean chiral chemical potential 
which was extended over the entire simulation domain,
e.g., there was a nonzero volume average $\langle\mu_5\rangle$ or a constant difference between right- and left-handed fermions. 
In our accompanying Letter \citep{SRB21a}, it was first shown in DNS that locally nonzero fluctuations of $\mu_5$ can  also induce a small-scale chiral dynamo, even if the mean chiral chemical potential $\langle\mu_5\rangle$ is vanishing.
As a consequence of this small-scale chiral dynamo instability, magnetically dominated turbulence is driven, 
which leads to the production of a $\langle\mu_5\rangle$ 
and ultimately generates a mean magnetic field via a large-scale turbulent dynamo. 
As argued above, the CVE \citep{KH16} is likely to be subdominant in magnetically 
dominated turbulence and its detailed investigation will therefore be postponed 
to a subsequent study focusing specifically on
this effect.

The present paper serves as a companion to Ref.~\citep{SRB21a} and focuses on
a technical analysis of the properties of chiral dynamos that
are sourced by an inhomogeneous initial $\mu_5$. 
To this end, we are extending the study of Ref.~\citep{SRB21a} by 
systematically exploring simulations with different initial inhomogeneities
of $\mu_5$, starting with a two-dimensional toy model in Sec.~\ref{sec_2Dsine}.
With this model we explore the conditions under 
which a small-scale chiral dynamo can operate. 
In particular we test how the growth rate of 
the chiral dynamo
depends on the separation of scales in the system. 
The value of the maximum possible growth rate of 
the small-scale chiral dynamo is determined. 
In Sec.~\ref{sec_turb_sine}, we present high-resolution simulations 
in which turbulence is generated in a self-consistent way, 
i.e., via the Lorentz force of the magnetic field 
produced by the small-scale chiral dynamo.
We analyze the different contributions to the large-scale 
dynamo growth rate and the evolution of the power spectra 
in chiral MHD with initially vanishing $\langle\mu_5\rangle$.
Conclusions are drawn in Sec.~\ref{sec_conclusion}.

\section{Physical background and methods}
\label{sec_background}

\subsection{Evolution equations of a chiral plasma}

In spatial regions, where the chemical potentials of left-handed $(\mu_L)$ 
and right-handed $(\mu_R)$ fermions differ from one
another, i.e., where the chiral chemical potential 
$\mu_5 \equiv \mu_L - \mu_R$ is nonzero, an additional electric 
current arises due to the chiral magnetic effect. 
This current exists in addition to the Ohmic current and leads to 
an extension of the induction equation
to the case of high-energy plasma
and, therefore, the classical MHD equations. 

In chiral MHD, the set of equations is given by
\begin{eqnarray}
  \frac{\partial \BB}{\partial t} &=& \nab   \times   \left[{\UU}  \times   {\BB}
    + \eta \, \left(\mu_5 {\BB} - \nab \times {\BB}\right) \right] ,
\label{ind-DNS}\\
  \rho{D \UU \over D t}&=& (\nab   \times   {\BB})  \times   \BB  -\nab  p + \nab  {\bm \cdot} \ttau ,
\label{UU-DNS}\\
  \frac{D \rho}{D t} &=& - \rho \, \nab  \cdot \UU ,
\label{rho-DNS}
\end{eqnarray}
together with the evolution equation of $\mu_5$:
\begin{eqnarray}
  \frac{D \mu_5}{D t} &=& \mathscr{D}_5(\mu_5)
  + \lambda \eta \left[{\BB} {\bm \cdot} (\nab   \times   {\BB})
  - \mu_5 {\BB}^2 \right]
  - \mu_5 \nab  \cdot \UU.
\label{mu-DNS}
\end{eqnarray}

In this set of equations, the magnetic field $\BB$ is 
normalized such that the magnetic energy 
density is $\BB^2/2$, $\UU$ is the velocity field, and $\rho$ is the mass density.
The advective derivative is written as $D/D t = \partial/\partial t + \UU \cdot \nab$.
Further, $\eta$ is the microscopic magnetic diffusivity,
$p$ is the fluid pressure,
$\ttau=2\nu \rho \SSSS$ is the stress tensor,
$\SSSS$ is the trace-free strain tensor with components
${\sf S}_{ij}=(U_{i,j}+U_{j,i})/2 -
\delta_{ij} ({\bm \nabla}{\bm \cdot} \UU)/3$
(commas denote partial spatial derivatives), and
$\nu$ is the kinematic viscosity. 
In Eq.~(\ref{mu-DNS}) for the chiral chemical potential $\mu_5$, 
a diffusion term characterized by the diffusion operator $\mathscr{D}_5(\mu_5)$
has been introduced for numerical stability;
see Sec.~\ref{sec_setup} for details.
Further, $\lambda$ is the chiral nonlinearity parameter 
which quantifies the
coupling between magnetic helicity and $\mu_5$.
The system of Eqs.~(\ref{ind-DNS})--(\ref{mu-DNS}) implies 
that total chirality $\chi_{\rm tot} \equiv
\langle \mathcal{H}\rangle + 2 \langle \mu_5 \rangle /\lambda$
is conserved \citep{KH14,KH16}, where angle brackets denote 
volume averaging and $\langle\mathcal{H}\rangle=\langle \AAA \cdot\BB\rangle$ 
is the magnetic helicity, $\BB = \nabla \times \AAA$ is the magnetic field 
strength with the vector potential $\AAA$.
The addition of the term $-\mu_5\nab\cdot \UU$ in Eq.~(\ref{mu-DNS})
relative to Ref.~\cite{REtAl17} does not make a noticeable difference;
see the appendix of Ref.~\cite{BHKRS21}.
This conservation law would need to be extended if the CVE were to be included.

In the system of Eqs.~(\ref{ind-DNS})--(\ref{mu-DNS}),
we do not include the evolution equation 
for the chemical potential $\mu \equiv \mu_\mathrm{L} + \mu_\mathrm{R}$.
The inclusion of this equation allows
to describe the chiral magnetic waves
\citep{KY11}.
The existence of the chiral magnetic waves
requires the presence of a significant equilibrium magnetic field.
In particular, the frequency of
the chiral magnetic waves is proportional
to the equilibrium magnetic field.
However, since we consider dynamo
excited from a very small seed magnetic field,
the chiral magnetic waves do not exist
in our system unless the generated 
mean magnetic field reaches a high enough strength.

\subsection{Initial conditions}

We study the generation of the magnetic field
by fluctuations $\mu'_5$ of the chiral chemical
potential with a zero mean value $\langle \mu_5\rangle(t_0)=0$
at the initial time $t_0$.
Initially, velocity fluctuations vanish and there is a weak 
seed magnetic field $\BB$.

The focus of this study lies on cases in which $\mu_5$ is
inhomogeneous, but we will also discuss the
comparison to runs with an initially homogeneous $\mu_5$.
In particular, the following different cases are considered:
\begin{itemize}
\item[(a)]{Systems with an initial $\mu_5$ in form of 
a sine spatial profile along the $x$ axis, 
i.e., $\mu_5(t_0,\boldsymbol{x}) = \mu_5(t_0,x)= \mu_{5,0} ~ \mathrm{sin}(k_x  x)$. 
The wave number of the sine function $k_x$ will be varied.}
\item[(b)]{ 
Random distributions of $\mu_5$
at different wave numbers that are 
initialized in such that the spectrum of the chiral chemical potential,
$E_5$, takes the form 
of a power-law function in $k$ space,~i.e., 
$E_5(t_0) \propto k^{-n}$.
The power-law exponent $n$ will be varied.
Further, we will consider cases where this initial condition 
includes a nonzero initial $\langle\mu_5\rangle$ and cases
where $\langle\mu_5\rangle(t_0) = 0$. }
\item[(c)]{Systems with a uniform distribution
of the chiral chemical potential,
$\mu_5(t_0,\boldsymbol{x})=\mathrm{const}$, 
that serve as comparison with previous results.}
\end{itemize}

\begin{table*}
\small
\centering
\caption{Summary of all runs presented in this paper.
}
\begin{tabular}{ l | l l | l l l | l l l l | l l l }
 \hline
        & Setup:\hspace{-2cm}   &       & Parameters:\hspace{-2cm}   &             &                & Initial conditions:\hspace{-4cm}  &   &  &   &   Output:\hspace{-2cm}  &  &  \\
        & Dim.  & Res.     & $\eta=\nu$       & $\mu_5$ diffusion       & $\lambda_5$    & $\mu_{5,\mathrm{max}}$  &  $\mu_{5,\mathrm{rms}}$ &  $\langle\mu_{5}\rangle$ &  $\mu_{5}$ structure   &  $\mathrm{max}(B_\mathrm{rms})$   &  $\mathrm{max}(\langle\mu_{5}\rangle)$  &  $\mathrm{max}(\Rm)$ \\
 \hline
\bf{Series H}  &   &   &   &   &     &     & &     & &   &      &  \\
H1     &  3D     &  $128^2$  &  $0.001$     &  $\mathcal{D}_5 = 2.4 \times 10^{-7}$  &  $1\times 10^{2}$      &  $10$       &  $10$       &  $10$     &  const           &  $0.44$     &  $10$       &  $65$                  \\
H2     &  3D     &  $672^3$  &  $0.0002$  &  $\mathcal{D}_5 =1.8 \times 10^{-9}$  &  $4 \times 10^{2}$  &  $42$       &  $42$       &  $42$     &  const           &  $0.41$     &  $42$       &  $4.2 \times 10^{2}$  \\
 \hline
\bf{Series S} &   &   &   &   &     &     & &     & &   &      &  \\
S1        &  2D     &  $128^2$  &  $0.001$     &  $\mathcal{D}_5 =2.4 \times 10^{-7}$  &  $1\times 10^{4}$      &  $10$       &  $7.1$      &  $-7.2 \times 10^{-15}$  &  $\mathrm{sin} (1x)$     &  $0.036$                 &  $0.31$                  &  $14$                    \\
S1L        &  2D     &  $128^2$  &  $0.001$     &  $\mathcal{D}_{5,\mathrm{L}} =0.001$  &  $1\times 10^{4}$      &  $9.9$       &  $7.0$      &  $9.1 \times 10^{-15}$  &  $\mathrm{sin} (1x)$     &  $1.9 \times 10^{-6}$                 &  $4.5 \times 10^{-9}$                  &  $9.5 \times 10^{-8}$                    \\
S1H3        &  2D     &  $128^2$  &  {$0.001$}     & {$\mathcal{D}_{5,\mathrm{H}3} =6.0\times10^{-11}$}  &  {$1\times 10^{4}$}      &  {$10$}       &  {$7.1$}      &  {$-8.2 \times 10^{-15}$}  &  {$\mathrm{sin} (1x)$}     &  {$0.037$}                 &  {$0.31$}                  &  {$14$}                    \\
S2        &  2D     &  $128^2$  &  $0.001$     &  $\mathcal{D}_5 =2.4 \times 10^{-7}$  &  $1\times 10^{4}$      &  $10$       &  $7.1$      &  $2.5 \times 10^{-16}$   &  $\mathrm{sin} (2x)$     &  $0.04$                  &  $0.14$                  &  $6.1$                   \\
S2$\lambda$2  &  2D     &  $128^2$  &  $0.001$     &  $\mathcal{D}_5 =2.4 \times 10^{-7}$  &  $1\times 10^{2}$      &  $10$       &  $7.1$      &  $2.5 \times 10^{-16}$   &  $\mathrm{sin} (2x)$     &  $0.11$                  &  $0.05$                  &  $24$                    \\
S2$\lambda$6  &  2D     &  $128^2$  &  $0.001$     &  $\mathcal{D}_5 =2.4 \times 10^{-7}$  &  $1\times 10^{6}$      &  $10$       &  $7.1$      &  $2.5 \times 10^{-16}$   &  $\mathrm{sin} (2x)$     &  $0.0045$                &  $0.17$                  &  $0.35$                  \\
{S2L}        &   {2D}     &   {$128^2$}  &   {$0.001$}     &   {$\mathcal{D}_{5,\mathrm{L}} =0.001$}  &   {$1\times 10^{4}$}      &   {$10$}       &   {$7.1$ }     &   {$2.5 \times 10^{-16}$}   &   {$\mathrm{sin} (2x)$}     &   {$1.8 \times 10^{-7}$}                 &   {$1.9 \times 10^{-14}$}                  &   {$1.6 \times 10^{-12}$}           \\
{S2H3}        &   {2D}     &   {$128^2$}  &   {$0.001$}     &   {$\mathcal{D}_{5,\mathrm{H}3} =6.0\times10^{-11}$}  &   {$1\times 10^{4}$}      &   {$10$}       &   {$7.1$ }     &   {$2.5 \times 10^{-16}$}   &   {$\mathrm{sin} (2x)$}     &   {$0.04$}                 &   {$0.14$}                  &   {$6.4$}           \\
S3        &  2D     &  $128^2$  &  $0.001$     &  $\mathcal{D}_5 =2.4 \times 10^{-7}$  &  $1\times 10^{4}$      &  $10$       &  $7.1$      &  $-1.2 \times 10^{-14}$  &  $\mathrm{sin} (3x)$     &  $0.029$                 &  $0.039$                 &  $3.6$                   \\
S4        &  2D     &  $128^2$  &  $0.001$     &  $\mathcal{D}_5 =2.4 \times 10^{-7}$  &  $1\times 10^{4}$      &  $9.9$      &  $7.1$      &  $-1.3 \times 10^{-14}$  &  $\mathrm{sin} (4x)$     &  $0.015$                 &  $0.091$                 &  $2.3$                   \\
S5        &  2D     &  $128^2$  &  $0.001$     &  $\mathcal{D}_5 =2.4 \times 10^{-7}$  &  $1\times 10^{4}$      &  $10$       &  $7.1$      &  $-2.3 \times 10^{-16}$  &  $\mathrm{sin} (5x)$     &  $4.1 \times 10^{-7}$  &  $5.4 \times 10^{-11}$  &  $3.1 \times 10^{-9}$    \\
S6        &  2D     &  $128^2$  &  $0.001$     &  $\mathcal{D}_5 =2.4 \times 10^{-7}$  &  $1\times 10^{4}$      &  $9.9$      &  $7$        &  $-1.1 \times 10^{-14}$  &  $\mathrm{sin} (6x)$     &  $5.7 \times 10^{-9}$  &  $1.1 \times 10^{-14}$  &  $1.4 \times 10^{-12}$  \\
S8        &  2D     &  $128^2$  &  $0.001$     &  $\mathcal{D}_5 =2.4 \times 10^{-7}$  &  $1\times 10^{4}$      &  $9.8$      &  $7$        &  $-1.2 \times 10^{-14}$  &  $\mathrm{sin} (8x)$     &  $9 \times 10^{-9}$      &  $1.4 \times 10^{-14}$  &  $1.7 \times 10^{-12}$  \\
{S8L}       &  {2D}     &  {$128^2$}  &  {$0.001$}     &  {$\mathcal{D}_{5,\mathrm{L}} =0.001$}  &  {$1\times 10^{4}$}      &  {$5.8$}      &  {$4.2$ }       & { $-1.3 \times 10^{-14}$}  & { $\mathrm{sin} (8x)$}     &  {$5.9 \times 10^{-9}$ }     &  {$1.3 \times 10^{-14}$ } &  {$1.9 \times 10^{-12}$}  \\
{S8H3}       &  {2D}     &  {$128^2$}  &  {$0.001$}     &  {$\mathcal{D}_{5,\mathrm{H}3} =6.0\times10^{-11}$}  &  {$1\times 10^{4}$}      &  {$9.8$}      &  {$7.1$ }       & { $-1.2 \times 10^{-14}$}  & { $\mathrm{sin} (8x)$}     &  {$9 \times 10^{-9}$ }     &  {$1.4 \times 10^{-14}$ } &  {$1.7 \times 10^{-12}$}  \\
S23D      &  3D     &  $672^3$  &  $0.0002$  &  $\mathcal{D}_5 =1.8 \times 10^{-9}$  &  $4 \times 10^{2}$  &  $50$       &  $35$       &  $-7.1 \times 10^{-17}$  &  $\mathrm{sin} (2x)$     &  $0.4$                   &  $0.2$                   &  $3.3 \times 10^{2}$     \\
{S23D$\lambda$4}  &  {3D}     &  {$672^3$}  &  {$0.0002$}  &  {$\mathcal{D}_5=1.8 \times 10^{-9}$}  &  {$4 \times 10^{4}$}  &  {$50$}       &  {$35$}       &  {$-1.6 \times 10^{-15}$}  &  {$\mathrm{sin} (2x)$}    &  {$0.068$  }               &  {$0.027$  }               &  {$45$      }              \\
{S23D$\lambda$8}  &  {3D}     &  {$672^3$}  &  {$0.0002$}  &  {$\mathcal{D}_5=1.8 \times 10^{-9}$}  &  {$4 \times 10^{8}$}  &  {$50$}       &  {$35$}       &  {$-1.6 \times 10^{-15}$}  &  {$\mathrm{sin} (2x)$}    &  {$0.0011$  }              &  {$0.029$   }              & { $0.016$  }               \\
{S23DL}    &  {3D}     &  {$672^3$}  &  {$0.0002$}  &  {$\mathcal{D}_{5,\mathrm{L}} =0.0002$}  &  {$4 \times 10^{2}$}  &  {$50$}       &  {$35$}       & { $-7.1 \times 10^{-17}$}  &  {$\mathrm{sin} (2x)$}     & { $0.32$}                   &  {$0.16$  }                 &  {$4 \times 10^{2}$}     \\
S203D     &  3D     &  $672^3$  &  $0.0002$  &  $\mathcal{D}_5 =1.8 \times 10^{-9}$  &  $4 \times 10^{2}$  &  $50$       &  $35$       &  $-1.6 \times 10^{-15}$  &  $\mathrm{sin} {(20x)}$  &  $0.02$                  &  $2.4 \times 10^{-5}$    &  $2.1$                   \\
 \hline
\bf{Series R}  &   &   &   &   &     &     & &     & &   &      &  \\
R$-$2m  &  3D     &  $672^2$  &  $0.0002$  &  $\mathcal{D}_5 =1.8 \times 10^{-9}$  &  $4 \times 10^{2}$  &  $46$       &  $15$       &  $-4.5$                  &  $E_5(k)\propto k^{-2}$  &  $0.19$     &  $4.5$      &  $2.1 \times 10^{2}$  \\
R$-$2     &  3D     &  $672^3$  &  $0.0002$  &  $\mathcal{D}_5 =1.8 \times 10^{-9}$  &  $4 \times 10^{2}$  &  $50$       &  $14$       &  $0$                     &  $E_5(k)\propto k^{-2}$  &  $0.18$     &  $0.18$     &  $2.7 \times 10^{2}$  \\
R$-$2\_CMW1  &  3D    &  $672^3$  &  $0.0002$  &  $\mathcal{D}_5 = 1.8 \times 10^{-9}$  &  $4 \times 10^{2}$  &  $51$       &  $14$       &  $-9.8 \times 10^{-16}$  &  $E_5(k)\propto k^{-2}$  &  $0.18$    &  $0.28$     &  $2.1 \times 10^{2}$  \\
R$-$2\_CMW2  &  3D     &  $672^3$  &  $0.0002$  &  $\mathcal{D}_5 = 1.8 \times 10^{-9}$  &  $4 \times 10^{2}$  &  $51$       &  $14$       &  $-9.8 \times 10^{-16}$  &  $E_5(k)\propto k^{-2}$  &  $0.19$     &  $0.26$     &  $2.2 \times 10^{2}$                \\
R$-$1     &  3D     &  $672^2$  &  $0.0002$  &  $\mathcal{D}_5 =1.8 \times 10^{-9}$  &  $4 \times 10^{2}$  &  $86$       &  $16$       &  $0$                     &  $E_5(k)\propto k^{-1}$  &  $0.095$  &  $0.052$  &  $1.3 \times 10^{2}$  \\
R$+$1     &  3D     &  $672^3$  &  $0.0002$  &  $\mathcal{D}_5 =1.8 \times 10^{-9}$  &  $4 \times 10^{2}$  &  $54$       &  $13$       &  $-2.1 \times 10^{-15}$  &  $E_5(k)\propto k^{+1}$  &  $0.068$  &  $0.055$  &  $16$                   \\
\hline

\end{tabular}
\label{tab_DNSoverview}
\end{table*}

\begin{table}[h!]
\centering
\caption{Different characteristic wave numbers and averages.
}
\begin{tabular}{p{0.14\linewidth}  p{0.36\linewidth} p{0.47\linewidth}}
 \hline
    Name           & Definition      & Description \\
 \hline
\multicolumn{3}{l}{Wave numbers:}  \\
    $k_1$          & $\frac{2\pi}{L}=1$        & Minimum wave number in the domain with length $L=2\pi$ \\
    $k_5$          & $\frac{\mu_{5,\mathrm{max}}}{2}$  & Wave number on which the small-scale chiral instability has its maximum \\
    $k_\mathrm{p}$ & ...  & Wave number on which $E_\mathrm{M}$ attains its maximum \\
    $k_{\mu_5,\mathrm{eff}}$ & $\left(\frac{\int E_5(k) k^{-1}~\mathrm{d}k}{\int E_5(k)~\mathrm{d}k}\right)^{-1}$ & Effective wave number on which $\mu_5$ is correlated \\
    $k_{\mathrm{int}}$ & $\left(\frac{\int E_\mathrm{M}(k) k^{-1}~\mathrm{d}k}{\int E_\mathrm{M}(k)~\mathrm{d}k}\right)^{-1}$ & Effective wave number on which $\BB$ is correlated = integral scale of turbulence   \\ 
 \hline
\multicolumn{3}{l}{Magnetic~field:}\\
    $B_\mathrm{rms}$ &  $\left(2 \int
E_\mathrm{M}(k)~\mathrm{d}k\right)^{1/2}$ & Rms magnetic field strength  \\ 
    $b$            & $\left(2 \int_{k_5}^{k_\mathrm{max}} E_\mathrm{M}(k)~\mathrm{d}k\right)^{1/2}$  & Field strength of small-scale magnetic fluctuations  \\
    $\langle B \rangle_{\mathrm{int}}$ & $\left(\frac{\int E_\mathrm{M}(k)^{2}~\mathrm{d}k}{\int E_\mathrm{M}(k)~\mathrm{d}k}\right)^{1/2}$ & Magnetic field strength on the integral scale of turbulence   \\
 \hline
\multicolumn{3}{l}{Growth~rate~of~magnetic~field:}  \\
    $\gamma_\mathrm{rms}$     & $\frac{\mathrm{d}\mathrm{ln}{(B_\mathrm{rms})}}{\mathrm{d} t}$  & Measured growth rate of $B_\mathrm{rms}$  \\
    $\gamma_b$     & $\frac{\mathrm{d}\mathrm{ln}{(b)}}{\mathrm{d} t}$  & Measured growth rate of $b$  \\
    $\gamma_{\mathrm{int}}$     & $\frac{\mathrm{d}\mathrm{ln}{(\langle B \rangle_{\mathrm{int}})}}{\mathrm{d} t}$  & Measured growth rate of $\langle B \rangle_{\mathrm{int}}$ \\
    $\gamma_5$     & $\frac{\eta\mu_{5,\mathrm{max}}^2}{4}$  & Theoretically predicted growth rate of the small-scale chiral dynamo  \\
    $\gamma_\alpha$     & $\frac{(\mathrm{max}(\overline{v_{5}}, \alpha_\mu, \alpha_\mathrm{M}))^2}{4(\eta + \eta_\mathrm{M})}$  & Theoretically predicted growth rate of the mean-field dynamo  \\
 \hline
\multicolumn{3}{l}{Magnetic~helicity:}\\
    $\langle \mathcal{H} \rangle$ & $ \frac{\int \mathcal{H}(\mathbf{x})~\mathrm{d}V}{V}$ & Volume average of the magnetic helicity  \\ 
    $\langle \mathcal{H}\rangle_{\mathrm{int}}$ & $\frac{\int H_\mathrm{M}(k)E_\mathrm{M}(k)~\mathrm{d}k}{\int E_\mathrm{M}(k)~\mathrm{d}k}$ & magnetic helicity on the integral scale of turbulence   \\ 
 \hline
\multicolumn{3}{l}{Chiral~chemical~potential:} \\
    $\mu_{5,\mathrm{rms}}$ & $\left(\int E_5(k)~\mathrm{d}k\right)^{1/2}$  & Rms value of the chiral chemical potential  \\ 
    $\mu_{5,\mathrm{max}}$ & $\mathrm{max}(\mu_{5}(\mathbf{x}))$ & Maximum of the chiral chemical potential \\ 
    $\langle \mu_5 \rangle$ & $ \frac{\int \mu_5(\mathbf{x})~\mathrm{d}V}{V}$ & Volume average of the chiral chemical potential  \\ 
    $\langle \mu_5 \rangle_{\mathrm{int}}$ & $\left(\frac{\int E_\mathrm{M}(k) E_5(k)~\mathrm{d}k}{\int E_\mathrm{M}(k)~\mathrm{d}k}\right)^{1/2}$ & Chiral chemical potential on the integral scale of turbulence   \\ 
 \hline
\multicolumn{3}{l}{Velocity~field:}\\
    $u$            & $\left(2 \int_{k_5}^{k_\mathrm{max}} E_\mathrm{K}(k)~\mathrm{d}k\right)^{1/2}$  & Field strength of small-scale velocity fluctuations  \\
 \hline
\end{tabular}
\label{tab_wave numbers}
\end{table}

\subsection{Small-scale chiral dynamo}

The initial condition (c) with the homogeneous distribution 
of $\mu_5$ has been used in previous studies and is well understood.
For a spatially constant $\mu_5$, a plane-wave ansatz for
the linearized induction equation~(\ref{ind-DNS}) with the 
CME term and a vanishing velocity field yields 
a dynamo instability that is characterized by the growth rate
\citep{JS97}
\begin{equation}
   \gamma(k) = |v_5 k| - \eta k^2,
\label{eq_gammalam}
\end{equation}
with $k$ being the wave number and $v_5 \equiv \eta \mu_5$.
The maximum growth rate of this instability is
\begin{eqnarray}
   \gamma_5 = \frac{v_5^2}{4 \eta},
\label{eq_gammalam_max}
\end{eqnarray}
and it is attained at the wave number
\begin{equation}
   k_5 = \frac{|\mu_5|}{2}.
\label{eq_klam_max}
\end{equation}

The chiral dynamo instability
is associated with the $\nab\times(v_5 \BB)$ term
in the induction equation~(\ref{ind-DNS}) of chiral MHD.
We note that, while this term is formally similar to the kinetic $\alpha$ effect, $\alpha_\mathrm{K}$ 
(that is related to the kinetic helicity), i.e., it is similar
to the $\nab\times(\alpha_\mathrm{K} \overline{\BB})$ term in the induction 
equation in the classical mean-field MHD, 
the effect described by the $v_5$ term
is not caused by turbulence, but rather 
by a quantum effect related to the handedness of fermions.
By analogy with the classical dynamo caused by the kinetic $\alpha$ 
effect, the small-scale chiral dynamo is referred to as
the $v_5$ dynamo. 
In the presence of shear, its growth rate is modified in ways that 
are similar to those of the classical $\alpha\Omega$ dynamo \citep{REtAl17},
except that, again, this chiral dynamo is not related to a turbulent flow.

\subsection{Production of the mean chiral chemical potential}

Fluctuations of the chiral chemical
potential $\mu'_5$ cause an exponential growth 
of the magnetic field by the $v_5$ dynamo.
During the $v_5$ dynamo phase, magnetic fluctuations produce
velocity fluctuations by the Lorentz force, i.e., the term 
$(\nab \times {\BB}) \times \BB $ on the right-hand side 
of Eq.~(\ref{UU-DNS}).

Since the initial mean chiral chemical potential is zero, 
and the magnetic helicity 
$\langle {\bm a} {\bm \cdot} {\bm b} \rangle$
of the seed magnetic field vanishes, 
the total initial chirality vanishes as well, $\chi_{\rm tot}(t_0)=0$.
Here ${\bm a}$ and ${\bm b}$ are fluctuations of
the vector potential and the magnetic field.
Due to the conservation of total chirality, it is zero
at all times: $\chi_{\rm tot}(t)=0$.
Initial fluctuations of a chiral chemical
potential $\mu'_5$ with a wide range of scales, however,
produce magnetic fluctuations
${\bm b}$ by the $v_5$ dynamo.
In particular, for a wide spectrum
in ${\bm k}$ space, fluctuations of the chiral chemical
potential at larger scales then serve as a 
\textit{mean field} for fluctuations on smaller scales, so that
the chiral dynamo instability excites magnetic
fluctuations at small scales, and produces
small-scale magnetic helicity
$\langle {\bm a} {\bm \cdot} {\bm b} \rangle$
during the dynamo action.
Due to the conservation of total chirality,
the production of the small-scale magnetic helicity
$\langle {\bm a} {\bm \cdot} {\bm b} \rangle$
causes the buildup of the mean chiral chemical potential:
\begin{eqnarray}
   \langle \mu_5 \rangle = - \lambda \langle {\bm a} {\bm \cdot} {\bm b} \rangle /2 .
\end{eqnarray}
The small-scale chiral dynamo produces magnetically driven 
turbulence and enhances turbulent kinetic energy. 
The latter increases the fluid and magnetic Reynolds numbers,
$\mathrm{Re} \equiv U_\mathrm{rms}/(\nu k_\mathrm{int})$ and
$\mathrm{Re}_\mathrm{M}\equiv U_\mathrm{rms}/(\eta k_\mathrm{int})$, 
where
\begin{eqnarray}
   k_\mathrm{int}^{-1} \equiv \frac{\int_1^{k_\mathrm{max}} E_\mathrm{M}(k)\,k^{-1}\,\mathrm{d}k}{\int_1^{k_\mathrm{max}} E_\mathrm{M}(k)~\mathrm{d}k},
\label{eq_kint}
\end{eqnarray}
is the integral scale of magnetically driven turbulence 
\footnote{We note that the expression (\ref{eq_kint}) would 
be ill-defined at $k=0$.
Therefore integration starts at $k=1$ which is the minimum 
possible value of $k_\mathrm{int}$.}.
When $\mathrm{Re}_\mathrm{M}$ is large enough,
the mean chiral dynamo instability is excited,
which can result in the generation of the mean magnetic field.

\subsection{Contributions to the mean-field dynamo}

The mean induction equation is given by
\begin{eqnarray}
  \frac{\partial \meanBB}{\partial t} &=&
     \nab   \times   \left[\meanUU  \times   \meanBB
 + (\meanv_5 + \alpha) \meanBB
     - (\eta+ \, \eta_{_{T}})\nab   \times   \meanBB\right],
\nonumber\\
\label{ind4-eq}
\end{eqnarray}
where $\meanv_5=\eta\overline{\mu_5}$ and the overbars indicate averages. 
In comparison to Eq.~(\ref{ind-DNS}), there are two 
additional contributions in Eq.~(\ref{ind4-eq}): 
$\alpha$ that increases the growth rate if it has the 
same sign as $\meanv_5$ or if $\alpha \gg \meanv_5$ 
and the turbulent diffusion 
$\eta_{_{T}} \approx U_\mathrm{rms}/(3 k_\mathrm{int})$. 

The $\alpha$ effect itself also has different contributions. 
In particular, for chiral MHD with a homogeneous $\mu_5$, 
the $\alpha_\mu$ effect has been derived 
in Ref.~\citep{REtAl17} and confirmed by DNS in Ref.~\citep{Schober2017}.
It is related to an interaction between 
fluctuations of the magnetic field $\mathbf{b}$ and the 
chiral chemical potential $\mu_5'$.
For very small mean magnetic energy (in comparison to the 
turbulent kinetic energy),
$\alpha_\mu$ has the form \citep{REtAl17}
\begin{eqnarray}
  \alpha_\mu = -{2 \over 3} \meanv_5~\mathrm{ln}(\Rm).
\label{eq_alpha5}
\end{eqnarray}

When the turbulent magnetic energy 
$\overline{{\bm b}^2}$ is much larger than the 
turbulent kinetic energy $\overline{{\bm u}^2}$
(so called magnetically driven turbulence), 
the magnetic $\alpha$ effect,
\begin{eqnarray}
\alpha_{\rm M} = C_{\rm M} \, \tau_{\rm c} \, 
\chi_{\rm c} ,
\label{alpha-mag}
\end{eqnarray}
plays a key role in the mean-field dynamo,
where $\chi_{\rm c} =\overline{ {\bm b} {\bm \cdot} ({\bm \nabla} \times{\bm b}) }$ is the current helicity.
For weakly inhomogeneous turbulence,
the current helicity is estimated as
$\chi_{\rm c} \approx \langle{{\bm a} {\bm \cdot} {\bm b}}\rangle_\mathrm{int} k_\mathrm{int}^2$,
i.e., it is proportional to the small-scale magnetic helicity
$\overline{{\bm a} {\bm \cdot} {\bm b}}$ (see Ref.~\citep{KR99}).
The correlation time of the magnetically driven turbulence is
the Alfv\'{e}n time
$\tau_{\rm c} = (u_{\rm A}k_\mathrm{int})^{-1}$, based on 
the integral scale given $k_\mathrm{int}^{-1}$ and the
Alfv\'{e}n speed $u_{\rm A}=\overline{{\bm b}^2}^{1/2}\approx B_\mathrm{rms}$.
The mean fluid density entering in the Alfv\'{e}n speed 
$u_{\rm A}$ and $\alpha_{\rm M}$ is unity and
for large magnetic Reynolds numbers the coefficient $C_{\rm M}= 2 (q-1)/(q+1)$
depends on the exponent $q$ of the magnetic energy spectrum $k^{-q}$.
Finally, there can be a contribution of the kinetic $\alpha$ effect that is caused by kinetic helicity:
\begin{eqnarray}
\alpha_{\rm K} = - {1 \over 3} \tau_{\rm c} \, \chi_{\rm K} .
\label{alpha-kin}
\end{eqnarray}
However, kinetic helicity $\chi_{\rm K} \approx \langle{{\bm u} {\bm \cdot} {\bm \omega}}\rangle_\mathrm{int}$ is not produced efficiently in magnetically driven turbulence and therefore $\alpha_{\rm K}$ is a subdominant effect in the system considered in this work. 
Here ${\bm \omega}$ are vorticity fluctuations.
This will be demonstrated later. 

During the dynamo action, the small-scale magnetic helicity 
$\overline{{\bm a} {\bm \cdot} {\bm b}}$
and the current helicity $\chi_{\rm c}$ are evolving.
The budget equation for $\chi_{\rm c}$ follows from the 
dynamic equation for the magnetic helicity
$\overline{{\bm a} {\bm \cdot} {\bm b}}$.
In the presence of a nonzero mean magnetic field,
this equation reads \citep{REtAl17}
\begin{eqnarray}
{\partial \over \partial t}  \overline{{\bm a} {\bm \cdot} {\bm b}} + {\rm div} \, {\bm F}
= 2 \meanv_5 \overline{{\bm b}^2} - 2 \meanEMF \cdot \meanBB
- 2 \eta \, \overline{{\bm b} \, ({\bm \nabla} \times {\bm b})}  ,
\label{MH1}
\end{eqnarray}
where $\meanEMF \equiv \overline{{\bm u} {\bm \times} {\bm b}}=\alpha_{\rm M} \meanBB - \eta_T \, ({\bm \nabla} \times \meanBB)$
is the turbulent electromotive force and $ {\bm F}$ is the flux of $\overline{{\bm a} {\bm \cdot} {\bm b}}$.
Here we consider the case when the kinetic $\alpha$ effect
caused by the kinetic helicity 
and the $\alpha_\mu$ effect \citep{REtAl17,Schober2017} 
are much smaller than the magnetic $\alpha$ effect. 
This is a typical situation for the chiral mean-field dynamo 
in a nonuniform $\mu_5$ (see below).
Near maximum field strength, two leading terms,
$2 \meanv_5  \overline{{\bm b}^2} - 2 \alpha_{\rm M}  \meanBB^2$, 
in Eq.~(\ref{MH1}) compensate each other, so that
the magnetic $\alpha$ effect reaches the value
\begin{equation}
  \alpha_{\rm M}^{\rm sat} = \eta \, \meanmu_{5} \, {\overline{{\bm b}^2}
  \over \meanBB^2} ,
\label{eq_alpha_mag_sat}
\end{equation}
where we took into account that for large magnetic Reynolds 
numbers the last term on the right-hand side of Eq.~(\ref{MH1}) 
vanishes. 
This term describes the dissipation rate of the magnetic 
helicity with the dissipation time scale
which is $\Rm$ times larger than the correlation time in 
the integral scale of turbulence, where $\Rm$ is the 
magnetic Reynolds number.
We also take into account that the term 
$- \eta_T \, ({\bm \nabla} \times \meanBB)$ in the turbulent electromotive force
is responsible for the magnetic diffusion of the mean magnetic field.

Overall, the growth rate of the mean magnetic field in the 
mean-field dynamo phase is given by
\begin{eqnarray}
\gamma(k) = (\meanv_5 + \alpha) k -
(\eta+\eta_\mathrm{T})k^2,
\label{eq_gammaalpha}
\end{eqnarray}
where $\alpha$ represents the maximum of the different contributions.
In the limit of large $\Rm$, $|\alpha| \gg |\meanv_5|$ 
and $\eta_\mathrm{T} \gg \eta$, so that the maximum growth rate is
\begin{eqnarray}
   \gamma_\alpha = \frac{\alpha^2}{4 \eta_\mathrm{T}},
\label{eq_gammaalpha_max}
\end{eqnarray}
and it is attained at the characteristic wave number
\begin{eqnarray}
   k_\alpha = \frac{\alpha}{2 \eta_\mathrm{T}},
\label{eq_kalpha}
\end{eqnarray}
which is less than the minimum wave number in the system. 
In this study, we show that the mean $\overline{v}_5$ effect,
where $\overline{v}_5=\eta \overline{\mu}_5$, another mechanism
of mean-field dynamo generation, is inefficient and that
$\alpha=\alpha_\mathrm{M}$

\subsection{Numerical setup}
\label{sec_setup}

We use the \textsc{Pencil Code} \cite{JOSS} to
solve equations~(\ref{ind-DNS})--(\ref{mu-DNS})
in a three-dimensional periodic domain of size $L^3 = (2\pi)^3$ with a resolution of
up to $672^3$.
This code employs a third-order accurate time-stepping method \cite{Wil80}
and sixth-order explicit finite differences in space \citep{BD02,Bra03}.
An overview of all runs presented in this paper is given
in Table~\ref{tab_DNSoverview}.
We note that runs R$-$2, R$-$1, and R$+$1 have also been discussed
in the companion Letter\ \citep{SRB21a}.
A list of notations is given in Table~\ref{tab_wave numbers}.

The smallest wave number covered in the numerical domain is 
$k_1 = 2\pi/L = 1$ which we use for the
normalization of length scales.
All velocities are normalized to the sound speed $\cs = 1$ and further the
mean fluid density is unity, $\meanrho = 1$.
Further, the magnetic Prandtl number is $1$,
i.e., the magnetic diffusivity equals the viscosity.
Time is normalized by the diffusion time $t_\eta = (\eta k_1^2)^{-1}$.

For numerical stability, the diffusion of $\mu_5$ is required 
and has been introduced by
the diffusion operator in Eq.~(\ref{mu-DNS}).
In our previous work with a uniform initial $\mu_5$, we have 
always used Laplacian diffusion, i.e.,
$\mathscr{D}_5(\mu_5)= \mathcal{D}_{5,\mathrm{L}}\Delta \mu_5$ where
$\mathcal{D}_{5,\mathrm{L}}$ is a constant and was usually
set to the same value as $\eta$.
In the present case, this would lead to an excessive
loss of $\mu_5$ fluctuations.
For this work, we focus the diffusion to the very 
smallest length scales such that $\mu_5$ on intermediate 
scales is not affected significantly. 
Therefore we use second-order hyperdiffusion which is given by 
$\mathscr{D}_5(\mu_5)= - \mathcal{D}_5 \, \nabla^4 \mu_5$.
The hyperdiffusion coefficient $\mathcal{D}_5$
is set to a value that produces
the same diffusion rate on the Nyquist wave number
as the one of the magnetic field 
for the corresponding value of $\eta$.
In the Appendix, we present
the results of simulations that have been repeated with Laplacian diffusion and
third-order hyperdiffusion (with the diffusion constant $\mathcal{D}_{5,\mathrm{H}3}$) for comparison.

\section{2D DNS with a spatially inhomogeneous initial $\mu_5$}
\label{sec_2Dsine}

In this section we analyze series S which includes 2D simulations with a $\mu_5$ that is set up 
as a sine spatial profile of $\mu_5$ with different wave numbers.
This serves as a simple toy model for an inhomogeneous initial $\mu_5$ and allows us to understand 
the main differences from previously studied simulations which were set up with a constant initial value 
of $\mu_5$ throughout the numerical domain.

\subsection{Onset of the small-scale chiral dynamo}

\begin{figure}
  \includegraphics[width=0.45\textwidth]{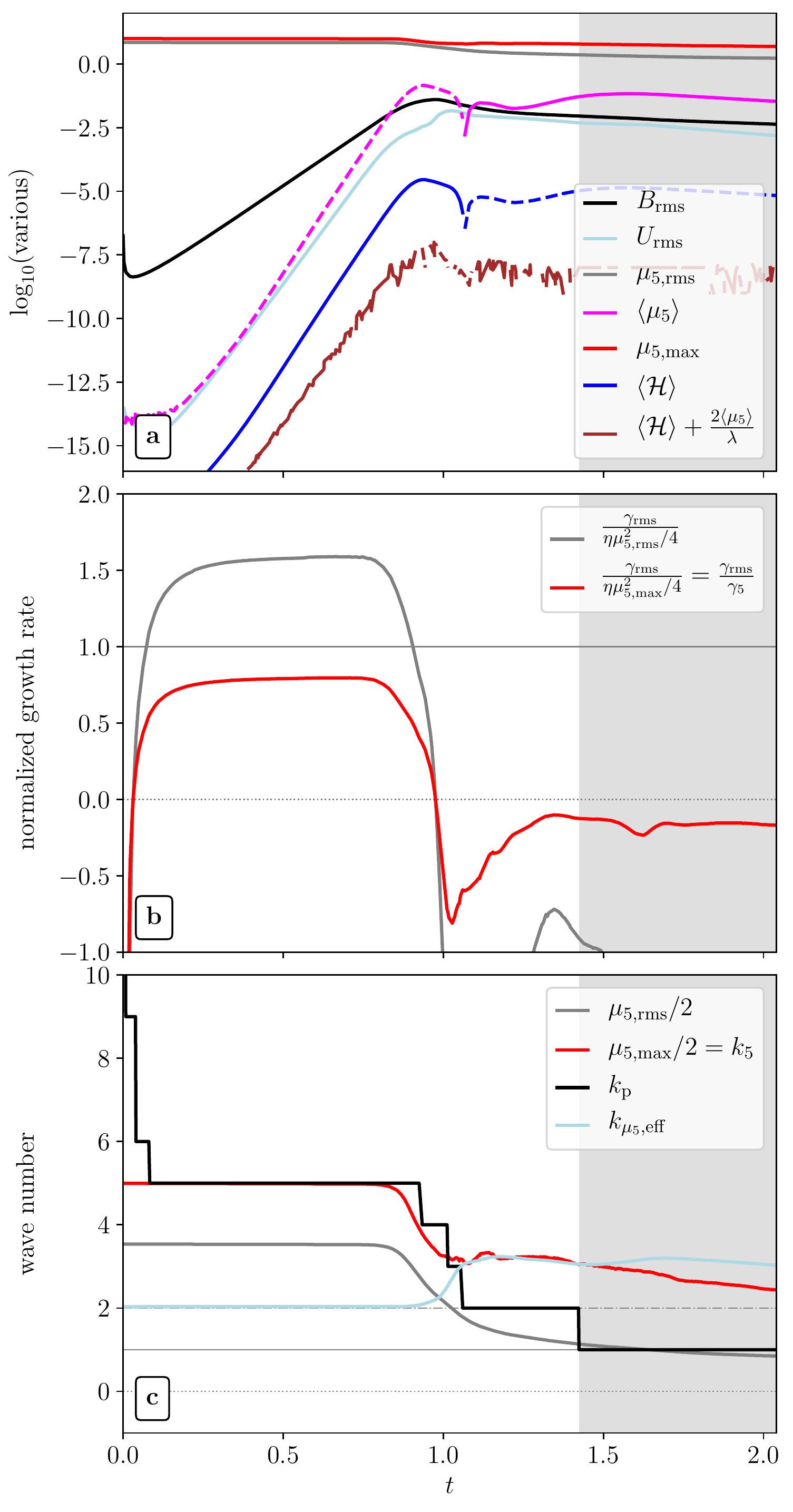}
  \caption{Analysis for run S2: 
  \textit{(a)} 
  Time series of the rms values of $B$, $U$, and $\mu_5$. 
  For $\mu_5$, the evolution of the volume average ($\langle\mu_5\rangle$) and the maximum value are also shown ($\mu_{5,\mathrm{max}}$). 
  Finally, the volume average of magnetic helicity ($\langle\mathcal{H}\rangle$) and the conserved quantity ($\langle\mathcal{H}\rangle + 2 \langle\mu_5\rangle/\lambda$) are plotted. 
  Solid line styles indicate positive sign and
  negative values are shown as dashed lines.
  \textit{(b)} The measured growth rate of $B_\mathrm{rms}$
  normalized by $\eta \mu_{5,\mathrm{rms}}^2/4$ (gray line)
  and $\gamma_5 = \eta \mu_{5,\mathrm{max}}^2/4$ (red line).
  \textit{(c)} Different characteristic wave numbers in the simulation:
  $\mu_{5,\mathrm{rms}}/2$, $k_5 = \mu_{5,\mathrm{max}}/2$, the wave number
  $k_\mathrm{p}$ on which $E_\mathrm{M}$ reaches its maximum, and the
  effective correlation wave number of $\mu_5$, $k_{\mu_5,\mathrm{eff}}$.
  For times larger than $t_{k_1}$, i.e., when the peak of the
magnetic energy spectrum has reached the wave number $k_1$, the plots are
shaded in gray.}
\label{fig_S2_ts}
\end{figure}

\begin{figure}
  \includegraphics[width=0.45\textwidth]{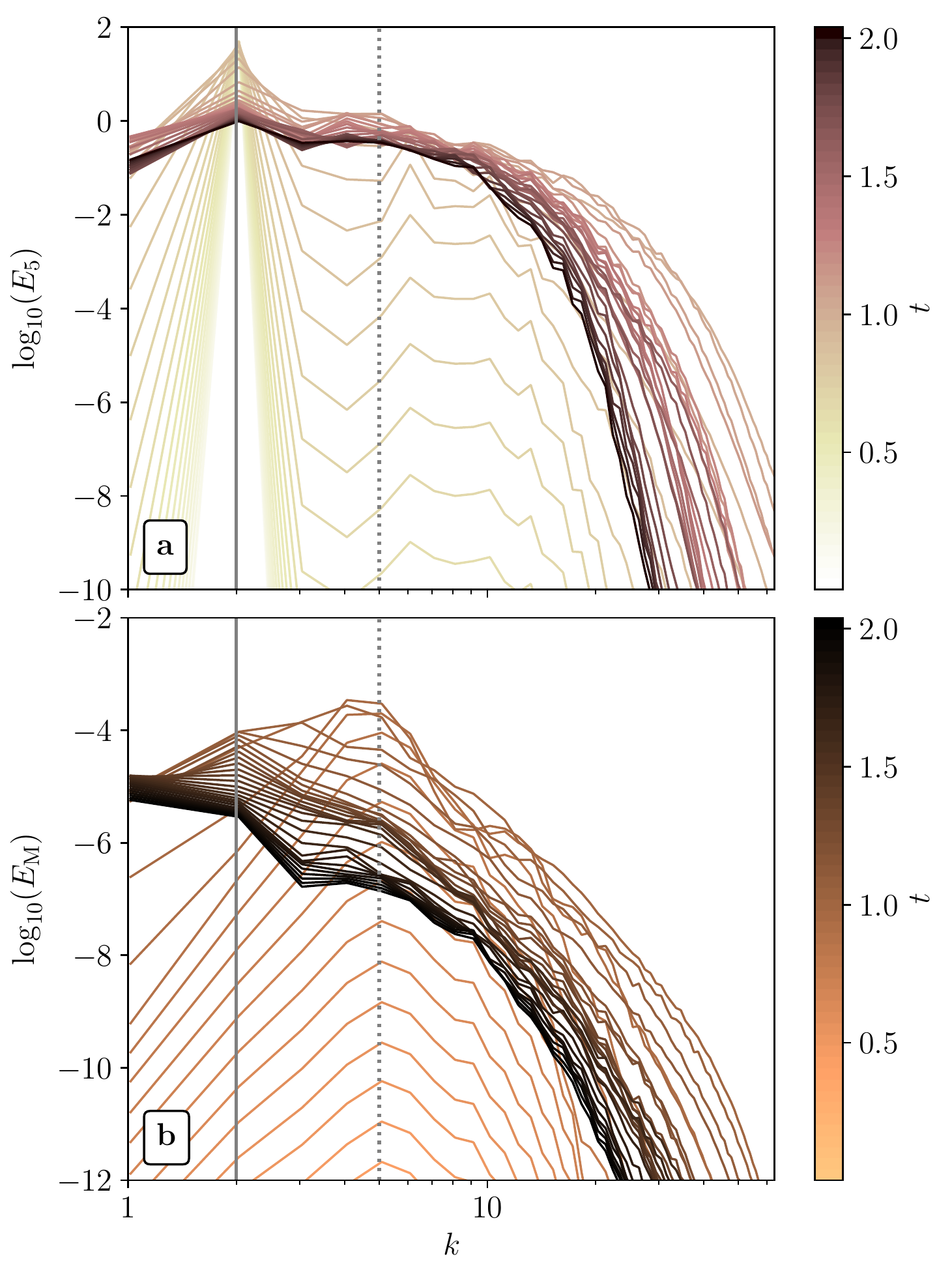}
  \caption{Evolution of the power spectra in run S2.  
  Lines with different colors correspond to different times as indicated by the color bars. 
  The solid vertical line shows the position of the initial sine
  wave number of $\mu_5$, $k=2$, and the dotted vertical line the
  shows the initial instability scale related to the $v_5$ dynamo,
  $\mu_{5,\mathrm{max}}(t_0)/2$.
  \textit{(a)} Power spectrum of $\mu_5$, $E_5(k,t)$.
  \textit{(b)} Magnetic energy spectrum $E_\mathrm{M}(k,t)$. 
  }
\label{fig_S2_spec}
\end{figure}

We first consider run S2 as a representative example. 
Its initial rms value is $\mu_{5,\mathrm{rms}}\approx 7.07$
and the maximum value of the sine function in the domain is
$\mu_{5,\mathrm{max}}=10$.
In Fig.~\ref{fig_S2_ts}a the time series of several characteristic quantities is presented. 
The existence of fluctuations of $\mu_5$ causes an instability in the
magnetic field $B_\mathrm{rms}$, which increases by $7$ orders of
magnitude.
Along with the exponential increase of magnetic energy,
a mean value of $\mu_5$ is generated, reaching a maximum of
$|\langle \mu_5 \rangle| \approx 0.1$ at the time $t\approx 1.0$.
We have repeated run S2 repeated with $32^2$, $64^2$, and $256^2$ grid cells,
respectively, and found that this maximum value of  
$|\langle \mu_5 \rangle|$ is independent on the resolution.
Initially, $\langle \mu_5 \rangle$ is generated
with a negative sign and is roughly compensated by the positive
$\lambda \langle \mathcal{H} \rangle/2$ during the $v_5$ dynamo phase.
Both, $\langle \mathcal{H} \rangle$ and $\langle \mu_5 \rangle$ flip
signs at $t\approx 1.1$, shortly after the end of the kinematic dynamo
amplification.

The measured growth rate $\gamma$ of $B_\mathrm{rms}$,
$\gamma_\mathrm{rms}$ 
is compared to the theoretically predicted maximum rate 
of the $v_5$ dynamo, Eq.~(\ref{eq_gammalam_max}), in Fig.~\ref{fig_S2_ts}b. 
Therefore we test two different values of $\mu_5$ in 
Eq.~(\ref{eq_gammalam_max}), the rms and the maximum value. 
Using $\mu_{5,\mathrm{rms}}$ tends to underestimate the 
observed $\gamma$ by approximately $50$\% while 
$\mu_{5,\mathrm{max}}$ predicts a slightly larger growth 
rate than that observed (the ratio $\gamma/(\eta \mu_{5,\mathrm{max}}^2/4)$ 
reaches up to $\approx 0.75$).
Theoretically it can be expected that the growth rate of the magnetic field
is highest in the region of the numerical domain where $\mu_5$ reaches
it maximum, i.e., where the amplitude of the sine wave is highest.
The evolution of the observed $B_\mathrm{rms}$ should then be
dominated by these local instabilities. 
Therefore, we would expect that $\mu_{5,\mathrm{max}}$ 
should determine $\gamma$.

However, it could be the case that the instability cannot develop 
sufficiently, especially if the spatial maximum of $\mu_5$ is
localized in a small region. 
This is, in particular, critical if the characteristic instability
length scale of the $v_5$ dynamo, given by Eq.~(\ref{eq_klam_max}),
is larger than the region in which $\mu_5$ is correlated.
For a direct comparison between the two different scales, 
we introduce the correlation length of $\mu_5$ as
\begin{equation}
   k_{\mu_5,\mathrm{eff}}^{-1} \equiv \frac{\int E_5(k) \, k^{-1}\,\mathrm{d}k}{\int E_5(k)~\mathrm{d}k},
\label{eq_kmu5eff}
\end{equation}
where $E_5(k)$ is the power spectrum of $\mu_5$.  
In the case of a sine function spatial profile, 
$k_{\mu_5,\mathrm{eff}}$ corresponds, initially, 
roughly to the wave number of the sine function.
A chiral dynamo instability
can only develop properly if $k_{\mu_5,\mathrm{eff}}\ll k_5$.
The evolution of different characteristic wave numbers 
in the simulation S2 is presented in Fig.~\ref{fig_S2_ts}c. 
In the beginning, the $k_5$ is larger than $k_{\mu_5,\mathrm{eff}}$ 
by a factor of $2.5$ and the peak of the magnetic energy spectrum, 
$k_\mathrm{p}$ occurs in $\mu_{5,\mathrm{max}}/2$.
At later times, $k_{\mu_5,\mathrm{eff}}$ changes through the 
backreaction of the magnetic field on the $E_5$ spectrum, 
ultimately becoming larger than $\mu_{5,\mathrm{max}}/2$ for 
times larger than $\approx 1$.
This coincides roughly with the magnetic energy maximum of the chiral dynamo.

The change of $k_{\mu_5,\mathrm{eff}}$ from $2$ to larger 
values can be  directly seen in the evolution of the power 
spectra $E_5$ in Fig.~\ref{fig_S2_spec}a. 
With the amplification of magnetic energy, shown in 
Fig.~\ref{fig_S2_spec}b, the $E_5$ spectrum also grows
for wave numbers both larger and smaller than $k=2$.
In fact, the evolution of $E_5$ seems to follow the one of 
$E_\mathrm{M}$.
Here, the wave number on which the instability develops most 
quickly is clearly $k = 5$ which corresponds to $k_5 = \mu_{5,\mathrm{max}}/2$. 
This is another indication that the growth rate of the 
$v_5$ dynamo is indeed given by the maximum of $\mu_5$ in the spatial domain. 
However, the instability scale, $k_5 = \mu_{5,\mathrm{max}}/2 = 5$ 
(indicated by the dotted vertical line), is close to the effective 
scale of $\mu_5$, $k_{\mu_5,\mathrm{eff}} = 2$ which could compromise 
the actual growth rate of $B_\mathrm{rms}$. 
We will test that statement by varying the wave number 
of the sine function in series S.

\subsection{The role of the effective correlation length of $\mu_5$}
\label{subsec_S2D_correlationlength}

\begin{figure}
  \includegraphics[width=0.45\textwidth]{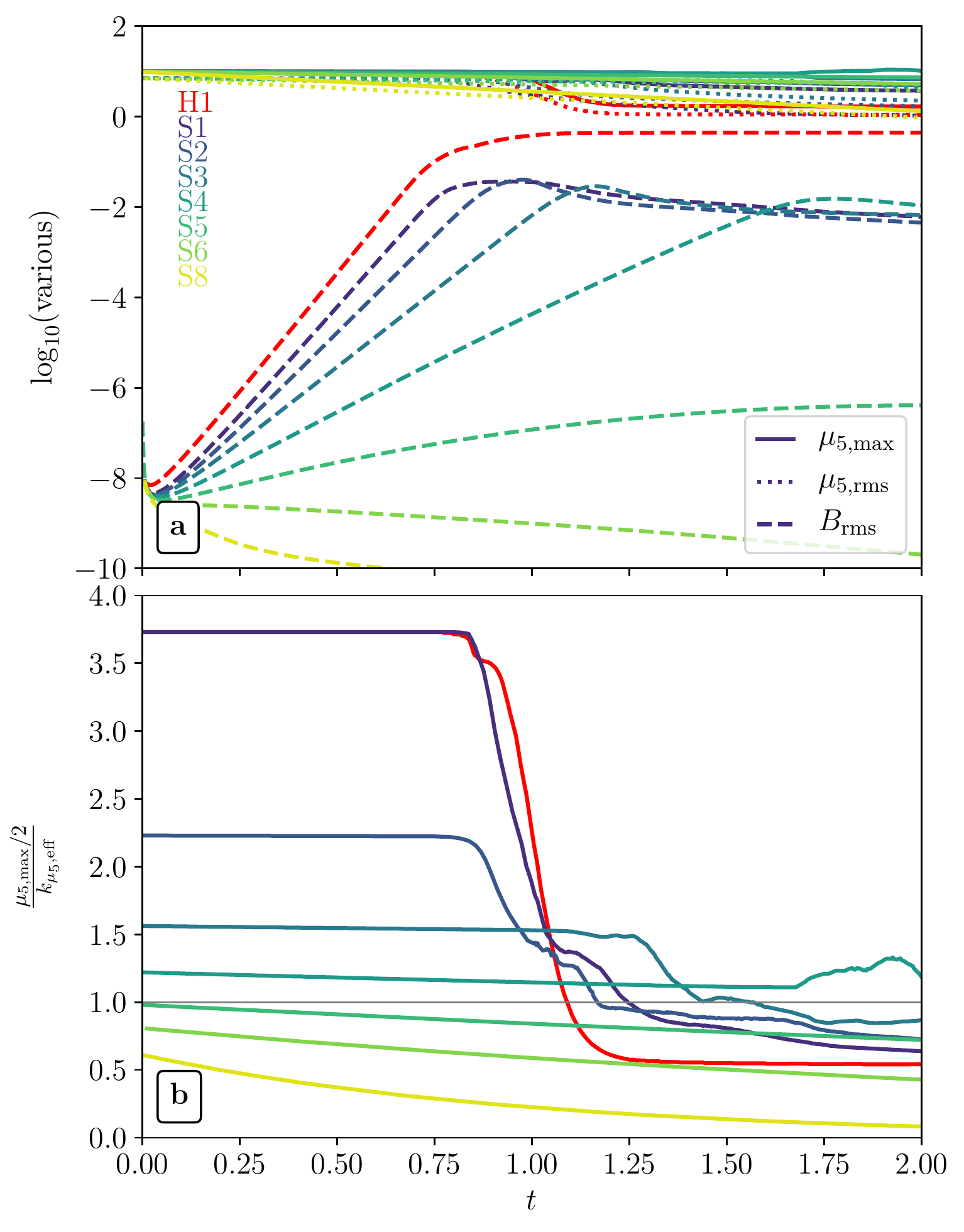}\\
  \caption{Analysis for 2D runs from series S (and comparison run H1): 
  \textit{(a)} Time series of $B_\mathrm{rms}$, $\mu_\mathrm{rms}$, and $\mu_\mathrm{max}$. 
  \textit{(b)} Ratio of $\mu_{5,\mathrm{max}}/2$ over $k_{\mu_5, \mathrm{eff}}$. 
  }
\label{fig_singlesin_ts}
\end{figure}

\begin{figure}
  \includegraphics[width=0.45\textwidth]{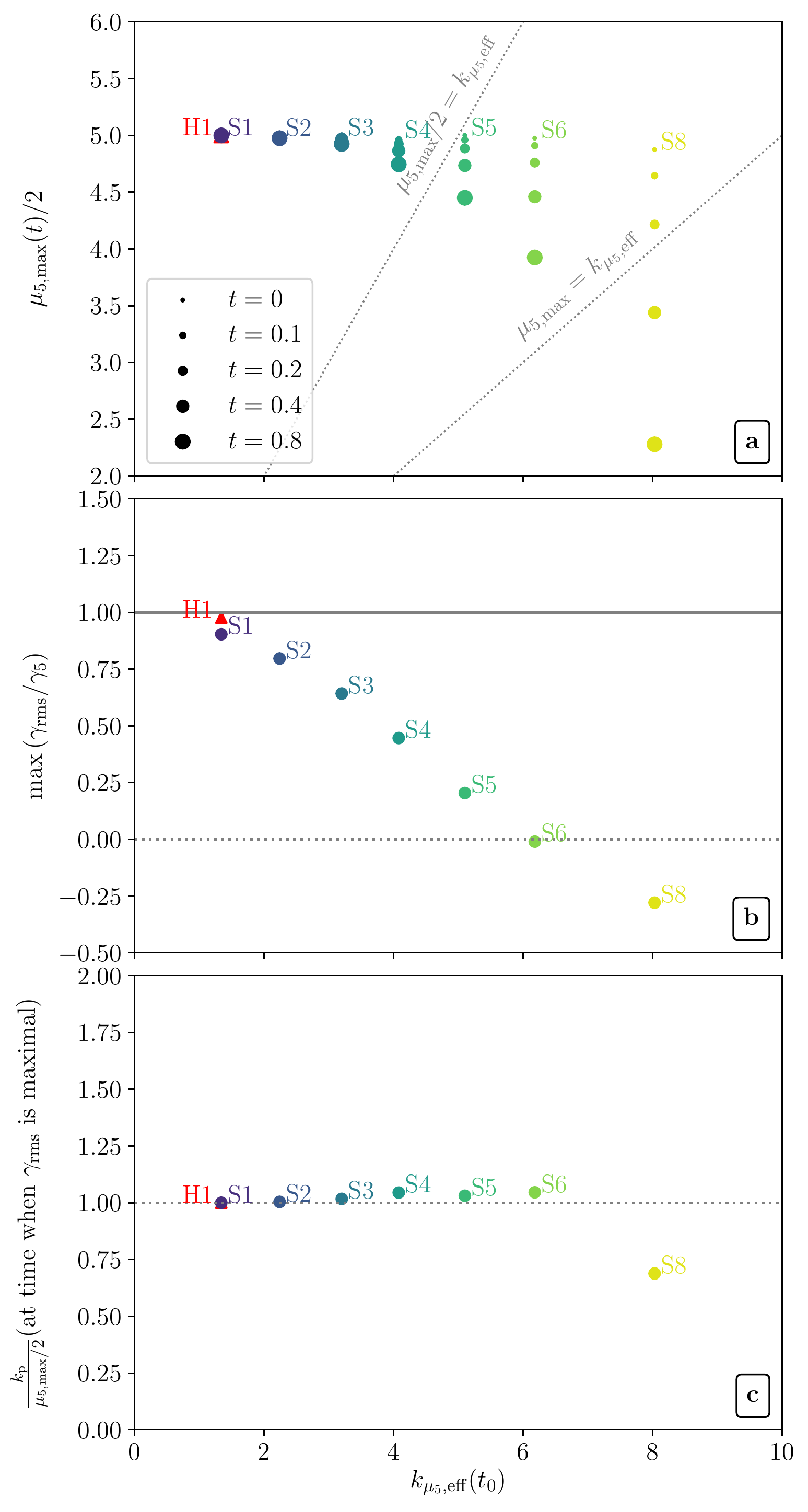}
  \caption{Analysis for 2D runs from series S (and comparison run H1): 
  Measured quantities vs. the initial effective correlation length of 
  $\mu_5$, $k_{\mu_5, \mathrm{eff}}(t_0)$. 
  \textit{(a)} The value of $\mu_{5,\mathrm{max}}/2$ measured in DNS. 
  The size of the symbols indicates time. 
  For comparison the critical lines 
  $\mu_{5,\mathrm{max}}/2 = k_{\mu_5, \mathrm{eff}}$ and $\mu_{5,\mathrm{max}} = k_{\mu_5, \mathrm{eff}}$ 
  are presented. 
  \textit{(b)} The maximum (over the entire simulation time) 
  of the measured growth rate $\gamma_\mathrm{rms}$, normalized by 
  $\gamma_5$.
  \textit{(c)} The ratio of the peak of the magnetic energy spectrum, 
  $k_\mathrm{p}$, and $\mu_{5,\mathrm{max}}/2$ at the time when the maximum growth rate is reached. }
\label{fig_singlesin_comp}
\end{figure}

The run S2 is now compared to run H1 with an initially constant $\mu_5$
that has the same value as the amplitude of the sine wave in S2,
i.e., $\mu_{5,\mathrm{max}}$ in S2, and with the remaining runs of
series S.
In the latter, all runs are initialized with the same amplitude of $\mu_5$
but different wave numbers; see Table~\ref{tab_DNSoverview} for details.

The time evolution of $B_\mathrm{rms}$, $\mu_{5,\mathrm{rms}}$, and 
$\mu_{5,\mathrm{max}}$ for all runs from series S and run 
H1 is presented in Fig.~\ref{fig_singlesin_ts}a.
The largest growth rate is observed 
for run H1, but run S1 has only a slightly smaller growth rate.
S1 has the lowest effective correlation length of 
$\mu_5$ ($k_{\mu_5, \mathrm{eff}}\approx1$). 
With increasing values of $k_{\mu_5, \mathrm{eff}}$, the 
amplification of $B_\mathrm{rms}$ becomes slower. 
In runs S6 and S8, 
$B_\mathrm{rms}$ decays.
In  Fig.~\ref{fig_singlesin_ts}b, the ratio of $\mu_{5,\mathrm{max}}/2$ 
over $k_{\mu_5, \mathrm{eff}}$ is presented for all runs. 
Interestingly, an increase of $B_\mathrm{rms}$ by a factor of 
$10$ is observed for S5, despite 
$\mu_{5,\mathrm{max}}/(2 k_{\mu_5, \mathrm{eff}})$ 
being less than $1$ from the initial time. 

In particular, for runs with spatial sine profiles with high wave numbers, 
the growth rate of the $v_5$ dynamo instability decreases due to a dissipation of $\mu_5$. 
Hyperdiffusion is applied in most runs of series S. 
Nevertheless, for runs 
that are set with an inhomogeneity in $\mu_5$ at high wave numbers,  
in particular for runs S4-S8, significant 
dissipation of $\mu_5$ leads to a constantly 
decreasing $\gamma_\mathrm{rms}$. 
The dissipation of $\mu_5$ can also be seen in Fig.~\ref{fig_singlesin_comp}a, 
where the value of $\mu_{5,\mathrm{max}}/2$ is shown at different times as a 
function of $k_{\mu_5, \mathrm{eff}}$ (at $t_0$) 
for all runs of series S and run H1. 
As long as $\mu_{5,\mathrm{max}}/2 \ll k_{\mu_5, \mathrm{eff}}$, the observed 
dynamo growth rate should be close to the maximum theoretical value, 
$\eta \mu_{5,\mathrm{max}}^2/4$. 
Indeed, it can be seen in Fig.~\ref{fig_singlesin_comp}b that the
observed growth rate (maximum value across the entire simulation time),
becomes smaller than $\eta \mu_{5,\mathrm{max}}^2/4$ with increasing
$k_{\mu_5, \mathrm{eff}}$.
Once $\mu_{5,\mathrm{max}}$ drops below $k_{\mu_5, \mathrm{eff}}$, 
no dynamo instability can occur, which is the case for runs S6 and S8. 
In all cases where a dynamo instability occurs, analysis of the magnetic
energy spectra shows that the maximum growth rate is attained for the
scale $\mu_{5,\mathrm{max}}/2$; see Fig.~\ref{fig_singlesin_comp}c.
Even run S6, where the rms magnetic field never increases, shows a peak
at $\mu_{5,\mathrm{max}}/2$ at the time when 
$\gamma_\mathrm{rms}$ is maximum.

\subsection{Termination of growth caused by alternation of the spatial distribution of $\mu_5$}

\begin{figure}[t]
  \includegraphics[width=0.45\textwidth]{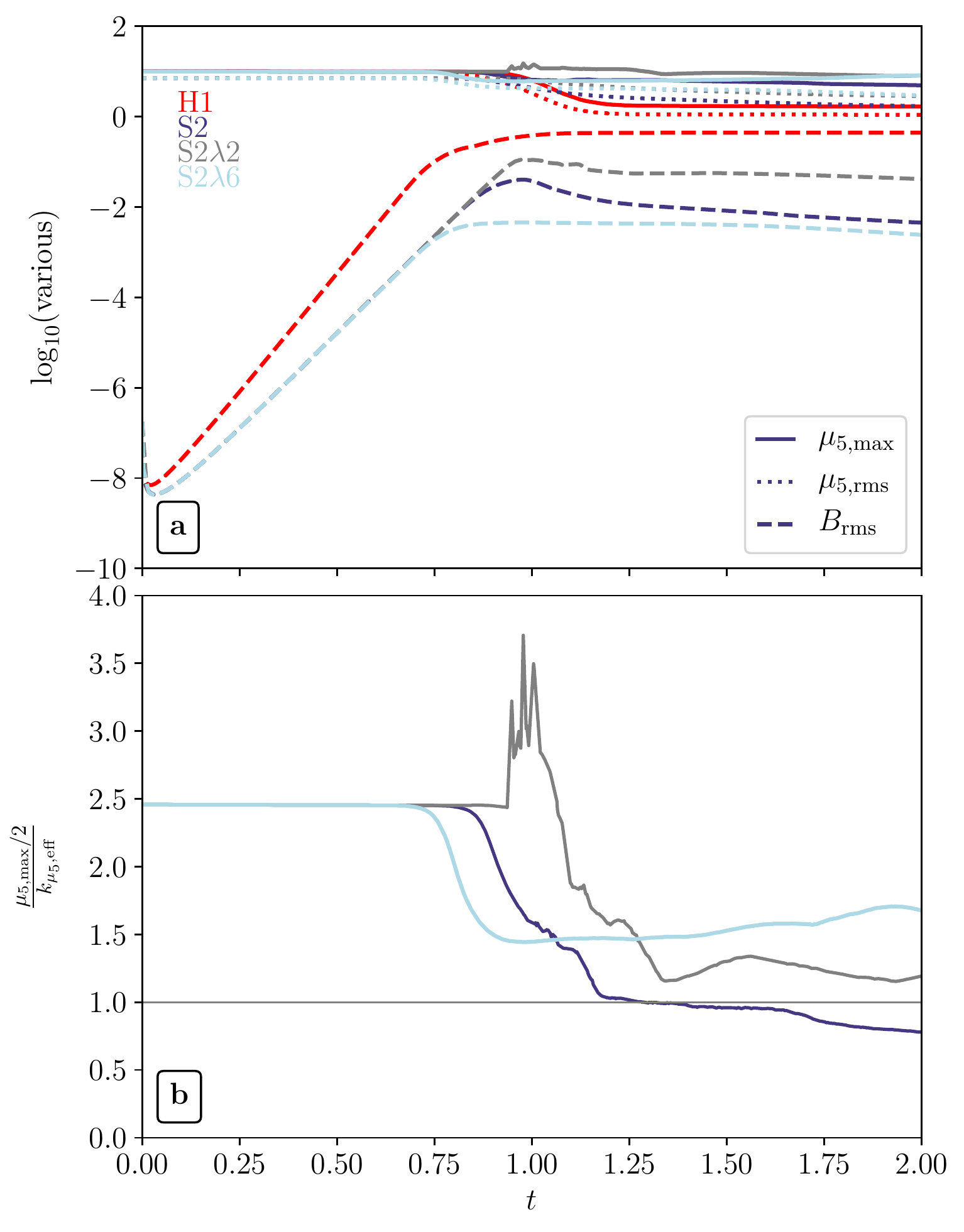}
  \caption{Analysis for 2D runs from series S with an initial sine 
  wave with $k=2$ and different values of $\lambda$ (and the comparison run H1).
  \textit{(a)} Time series of $B_\mathrm{rms}$, $\mu_\mathrm{rms}$, and $\mu_\mathrm{max}$. 
  \textit{(b)} Ratio of $\mu_{5,\mathrm{max}}/2$ to $k_{\mu_5, \mathrm{eff}}$. }
\label{fig_singlesin_ts_lambda}
\end{figure}

In this section we analyze the mechanism that limits the growth of the $v_5$ dynamo. It differs from that where there is an initially nonvanishing $\langle\mu_5\rangle$.
For an initial $\mu_5$ with zero mean value and simultaneously
vanishing $\langle\mathcal{H}\rangle$, 
the total chirality, $\langle\mathcal{H}\rangle + 2 \langle\mu_5\rangle/\lambda$, which is 
the conserved quantity in the system, is zero and stays zero, 
except for noise related to numerical precision; see Fig.~\ref{fig_S2_ts}a
where $\langle\mathcal{H}\rangle + 2 \langle\mu_5\rangle/\lambda$ grows
initially but never reaches values above $10^{-8}$.
Hence, the termination of further growth cannot be caused by $\langle\mathcal{H}\rangle$ reaching
a value comparable to $2 \langle\mu_5\rangle/\lambda$ which is the case
for simulations with constant initial $\mu_5$, like for example H1.

Direct comparison of run H1 with series S in 
Fig.~\ref{fig_singlesin_ts}a, shows that the $v_5$ 
dynamo is much less efficient for runs with 
$\langle\mu_5\rangle(t_0) = 0$.
The maximum of $B_\mathrm{rms}$ is smaller in all runs S 
than in run H1 by at least a factor of $10$.
Looking at Fig.~\ref{fig_singlesin_ts}b, it appears that, 
in series S, the $v_5$ dynamo reaches its maximum when 
the ratio $\mu_{5,\mathrm{max}}/2$ to $k_{\mu_5, \mathrm{eff}}$ 
drops to the order of $1$, quenching the dynamo instability. 
The decrease of the ratio of $\mu_{5,\mathrm{max}}/2$ to 
$k_{\mu_5, \mathrm{eff}}$ is largely caused by a change of the 
spatial structure of $\mu_{5}$ which happens when the 
magnetic field grows and the $\lambda$ term in 
Eq.~(\ref{mu-DNS}) becomes important. 
By that time the spectrum $E_5$ has changed significantly, 
leading to an increase of $k_{\mu_5, \mathrm{eff}}$; see 
Fig.~\ref{fig_S2_spec}a for an example.

The restructuring of the spectrum $E_5$ and the  
accompanying increase of $k_{\mu_5, \mathrm{eff}}$
depends on the strength of the coupling between the 
magnetic field and the chiral chemical potential $\mu_5$.
This strength of the coupling is
controlled by the parameter $\lambda$.
To test this hypothesis, we run two more simulations with 
the same initial conditions 
and the same parameters as in run S2, expect for 
the parameter $\lambda$ which is decreased by a factor of $10^2$ 
in run S2$\lambda$2 and increased by the same factor in run S2$\lambda$6.
The results for these runs are presented in Fig.~\ref{fig_singlesin_ts_lambda}. 
Indeed, the run with the smallest $\lambda$ (S2$\lambda$2) reaches the highest
$B_\mathrm{rms}$ while run S2$\lambda$6 saturates at a value that is $10$
times less.
Figure~\ref{fig_singlesin_ts_lambda}b shows that the ratio
$\mu_{5,\mathrm{max}}/2$ over $k_{\mu_5, \mathrm{eff}}$ drops
earlier for runs with larger $\lambda$.
We note that, even though run S2$\lambda$2 has a parameter $\lambda$ that
is $10^2$ larger than the one for run H1, the dynamo with initially
vanishing $\langle\mu_5\rangle$ is still not as efficient as for a
uniform distribution of $\mu_5$.

\begin{figure}
  \includegraphics[width=0.45\textwidth]{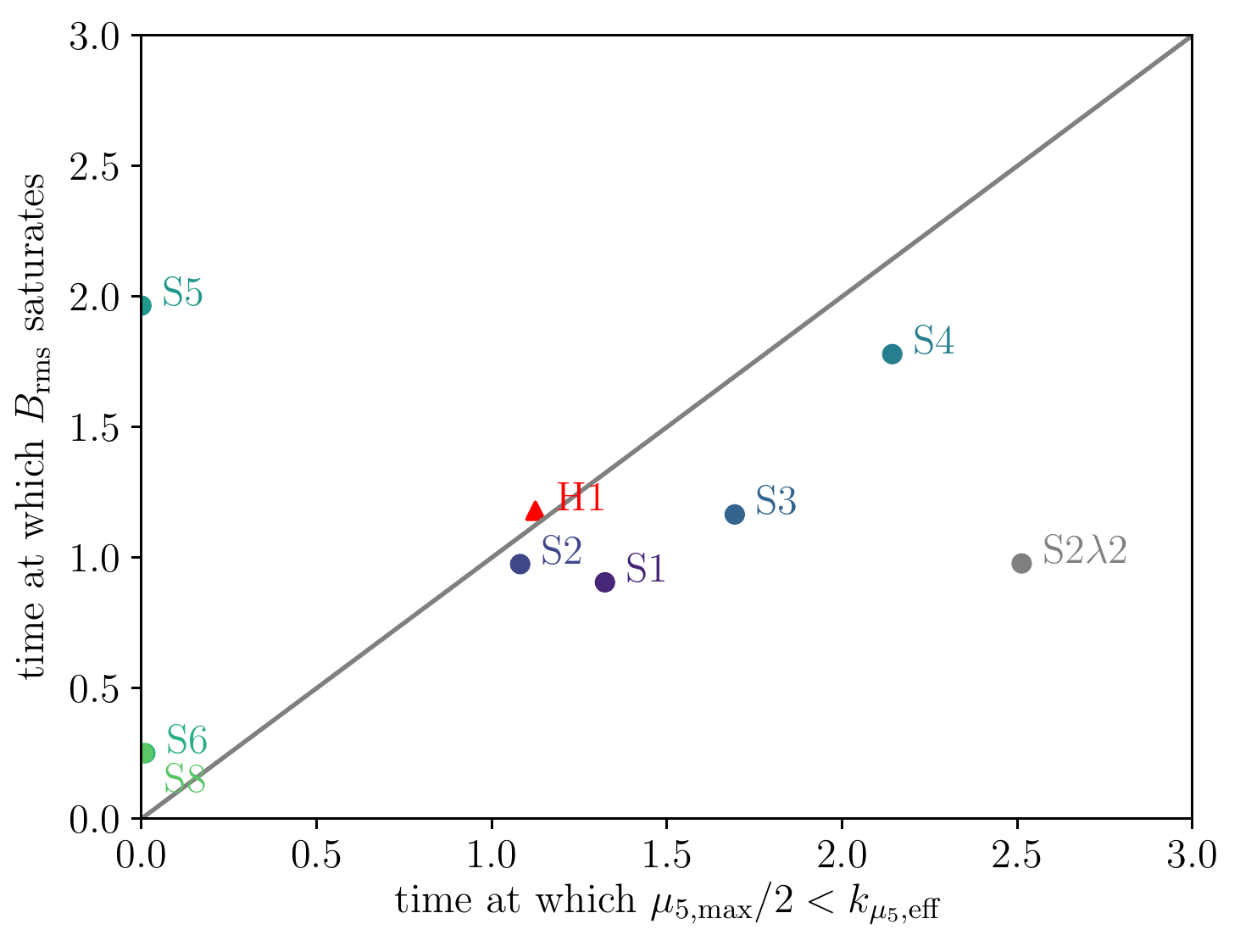}\vspace{-0.5cm}
  \caption{Comparison between the time at which $B_\mathrm{rms}$ 
  saturates and the time at which the laminar dynamo instability 
  scale, $\mu_{5,\mathrm{max}}/2$, becomes smaller than 
  $k_{\mu_5,\mathrm{eff}}$ for all 2D runs from series S and H1.
  }
\label{fig_singlesin_saturation_time}
\end{figure}

In Fig.~\ref{fig_singlesin_saturation_time}, the time at which the dynamo reaches maximum energy is compared to the time at which 
$\mu_{5,\mathrm{max}}/2$ becomes smaller than $k_{\mu_5,\mathrm{eff}}$
for all 2D runs from series S and H1. 
Here, dynamo limitation is defined as the time when 
$\gamma_\mathrm{rms}$ drops below $10^{-4}$ which is more than $2$ orders of magnitude below the maximum possible growth rate
in the system, $\eta \mu_5^2/4 = 2.5\times10^{-2}$.
All runs except for the ones which have initial small-scale 
$\mu_5$ fields, S5, S6, and S7, and S2$\lambda$2, lie on the 
linear correlation in Fig.~\ref{fig_singlesin_saturation_time}.
Regarding S2$\lambda$2, $k_{\mu_5,\mathrm{eff}}$ never drops below 
$\mu_{5,\mathrm{max}}/2$; however at the time when the magnetic 
field reaches its maximum, the ratio 
$(\mu_{5,\mathrm{max}}/2)/k_{\mu_5,\mathrm{eff}}$ drops to 
$\approx1.1$, possibly reducing the scale separation 
sufficiently to quench the $v_5$ dynamo. 
 
\section{Mean-field dynamos driven by an inhomogeneous $\mu_5$}
\label{sec_turb_sine}

For a homogeneous initial $\mu_5$, the magnetic field generated by
the $v_5$ dynamo drives turbulence, which eventually causes
mean-field dynamo action \citep{REtAl17, Schober2017}
if the plasma parameters are supercritical.
Specifically, the criteria for the occurrence of a mean-field dynamo are as follows:\\
\textit{(i)} The Reynolds numbers have to be much larger than 1.\\
\textit{(ii)} The $\lambda$ parameter should be small enough, so
that $\langle\AAA\cdot \BB\rangle$ is still less than 
$2 \langle \mu_5 \rangle/\lambda$ at the time when turbulence sets in. 

The objective of this section is to determine whether conditions
\textit{(i)} and \textit{(ii)} can be satisfied in a plasma with an
initially inhomogeneous $\mu_5$ with zero mean and a mean-field dynamo can be excited.
In particular the role of condition \textit{(ii)}, which regulates the dynamo limitation, 
is unclear for systems 
with an initial spatial profile of $\mu_5$ as a sine function 
(3D runs from series S) or with fluctuations of $\mu_5$ over an
extended range of spatial scales (run series R).
As before, the simulations with nonuniform initial $\mu_5$ 
will be compared to a run in which the initial $\mu_5$
is constant in space (run H2).

\subsection{DNS of an initial $\mu_5$ with sine spatial profile}
\label{sec_sin}

Run S23D is set up in a way that should allow the development of
turbulence for an initial spatial profile of $\mu_5$ in the form
of a sine function with wave number $2$.
In comparison to the runs S1--S8, this run
is performed in 3D space instead of 2D, it has higher resolution, and
the magnetic resistivity and viscosity values are $5$ times lower
than in the 2D runs.
To reach higher magnetic field strengths, and therefore higher fluid
velocities in the simulation, the initial amplitude of the sine function
is set to a value which is $5$ times higher than in the 2D runs.
In S23D, we have $\mu_{5,\mathrm{max}} = 50$, implying a characteristic
wave number of the $v_5$ dynamo instability of $k_5=25$.
Furthermore, in comparison to the 2D runs in series S, 
the chiral feedback parameter $\lambda$ is reduced to delay 
the backreaction of $\BB$ on $\mu_5$.
Both, the higher initial amplitude of $\mu_5$ and the lower value of
$\lambda$ lead to an extended period of dynamo action and thereby 
higher magnetic field strengths.
To test the importance of scale separation in the development of turbulence 
from an inhomogeneous chiral chemical potential, we perform
a second high-resolution run with an initial $\mu_5$ spatial profile in
the form of a sine function with wave number $k=20$ (run S203D).

\begin{figure}
  \includegraphics[width=0.45\textwidth]{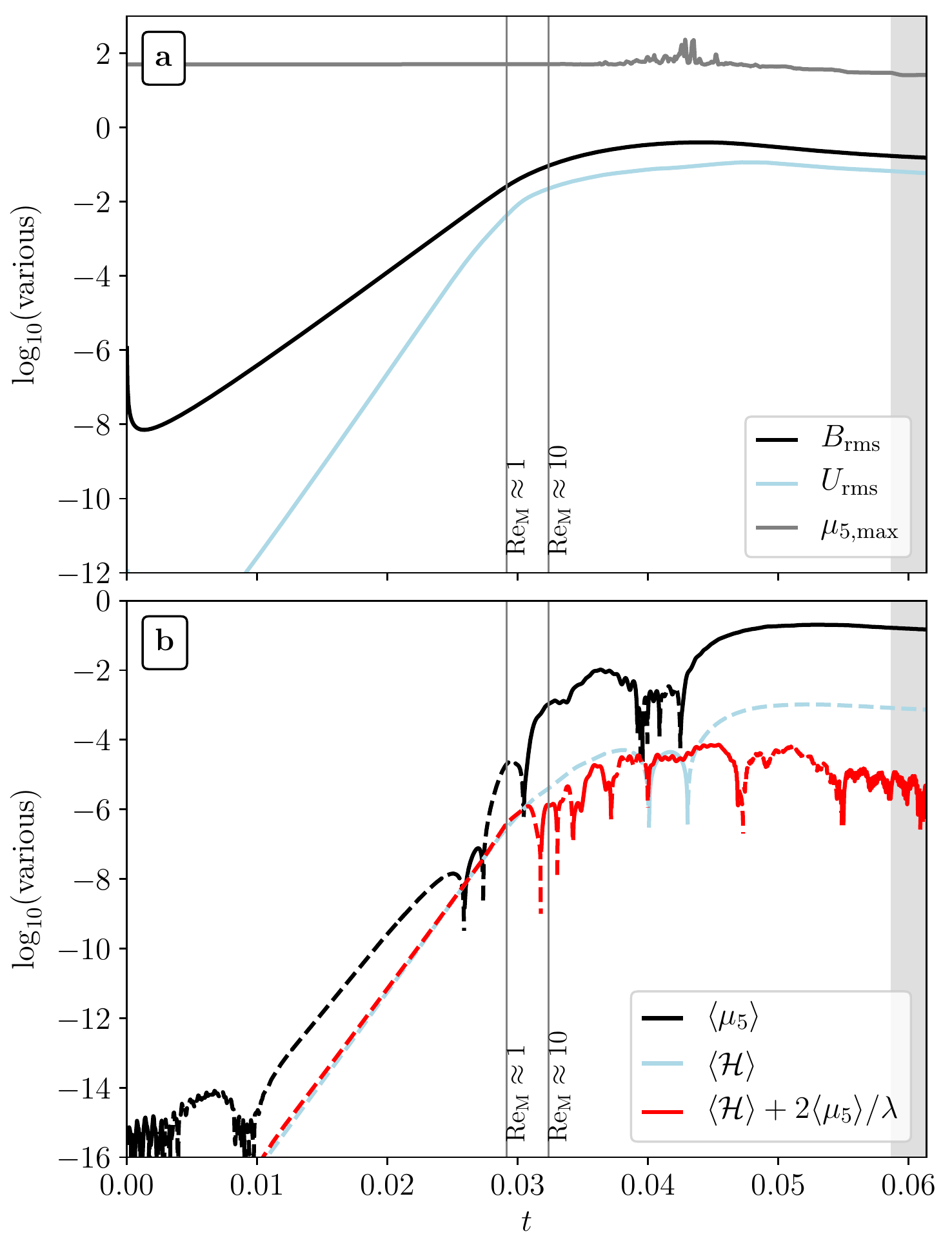}
\caption{Time evolution of run S23D.
\textit{(a)} The rms value of $\BB$, $\UU$, and $\mu_5$ as well as 
the maximum value of $\mu_5$.
\textit{(b)} The volume averages of $\mu_5$ and $\AAA\cdot\BB$
and the conserved quantity in chiral MHD, 
$\langle\AAA\cdot\BB\rangle + 2\langle\mu_5\rangle/\lambda$.
Here, a positive sign is indicated by solid line style and a 
negative sign by dashed line style.
 }
\label{fig_ts_S3D}
\end{figure}

\begin{figure*}
  \includegraphics[width=0.3\textwidth]{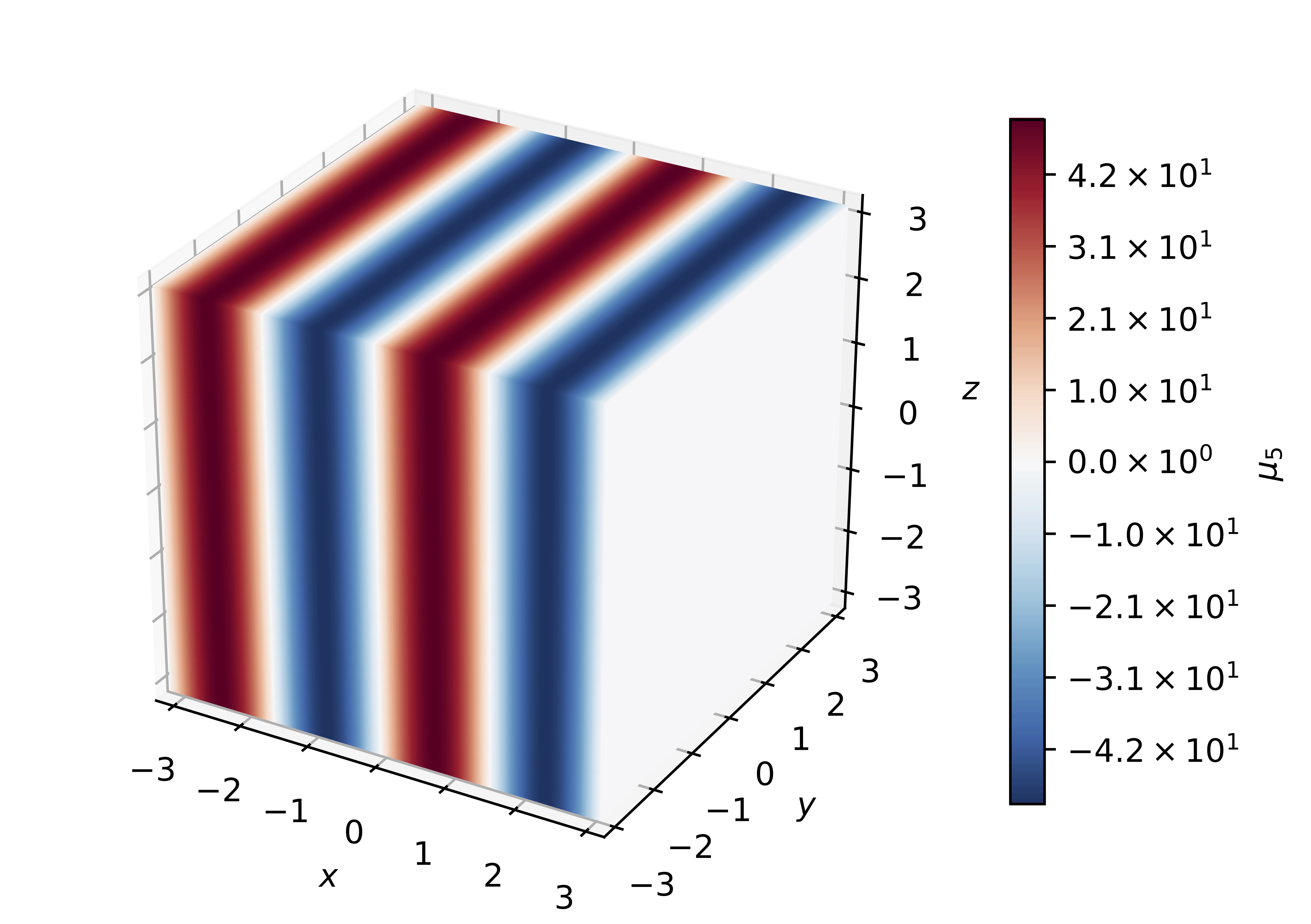}
  \includegraphics[width=0.3\textwidth]{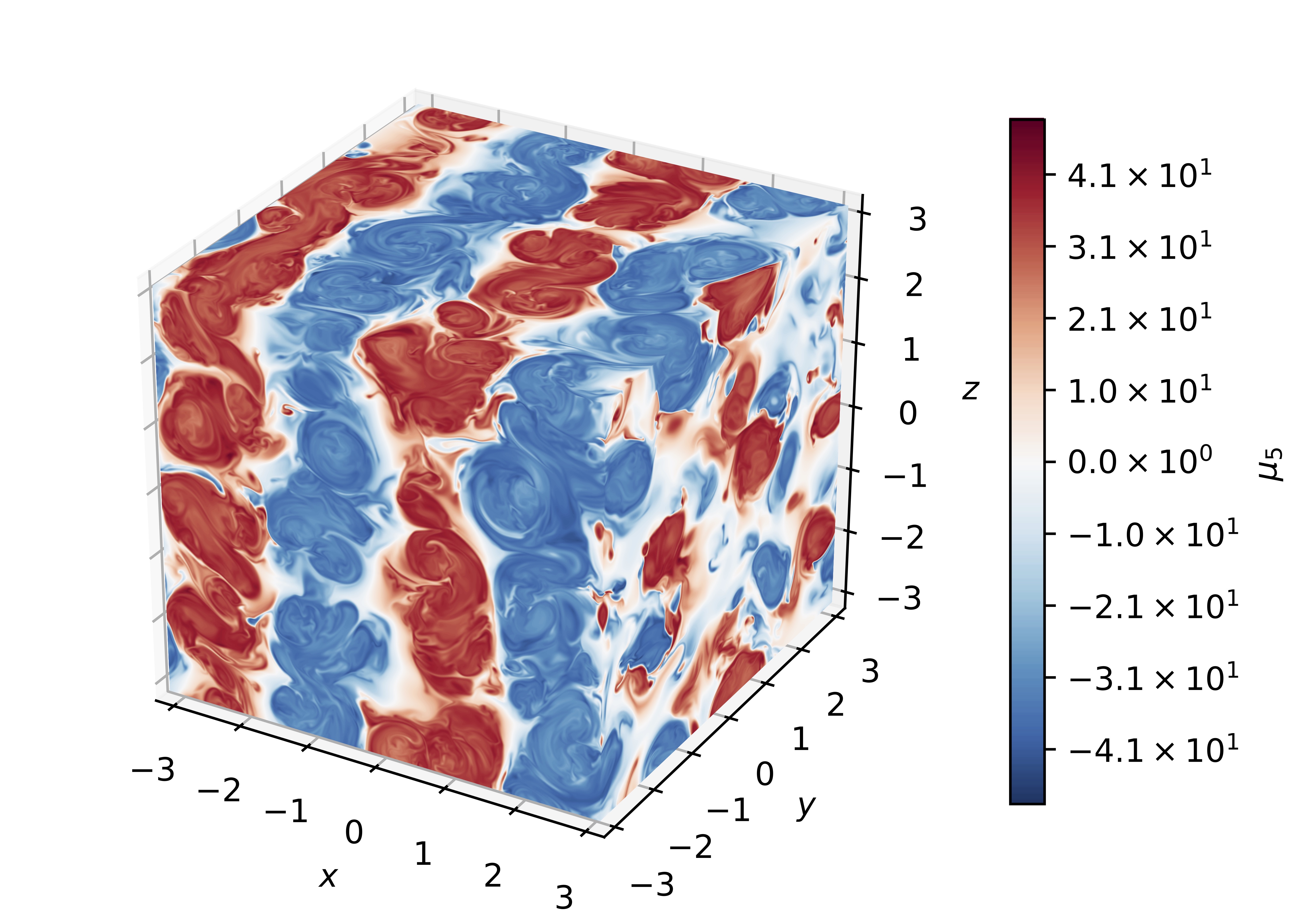}
  \includegraphics[width=0.3\textwidth]{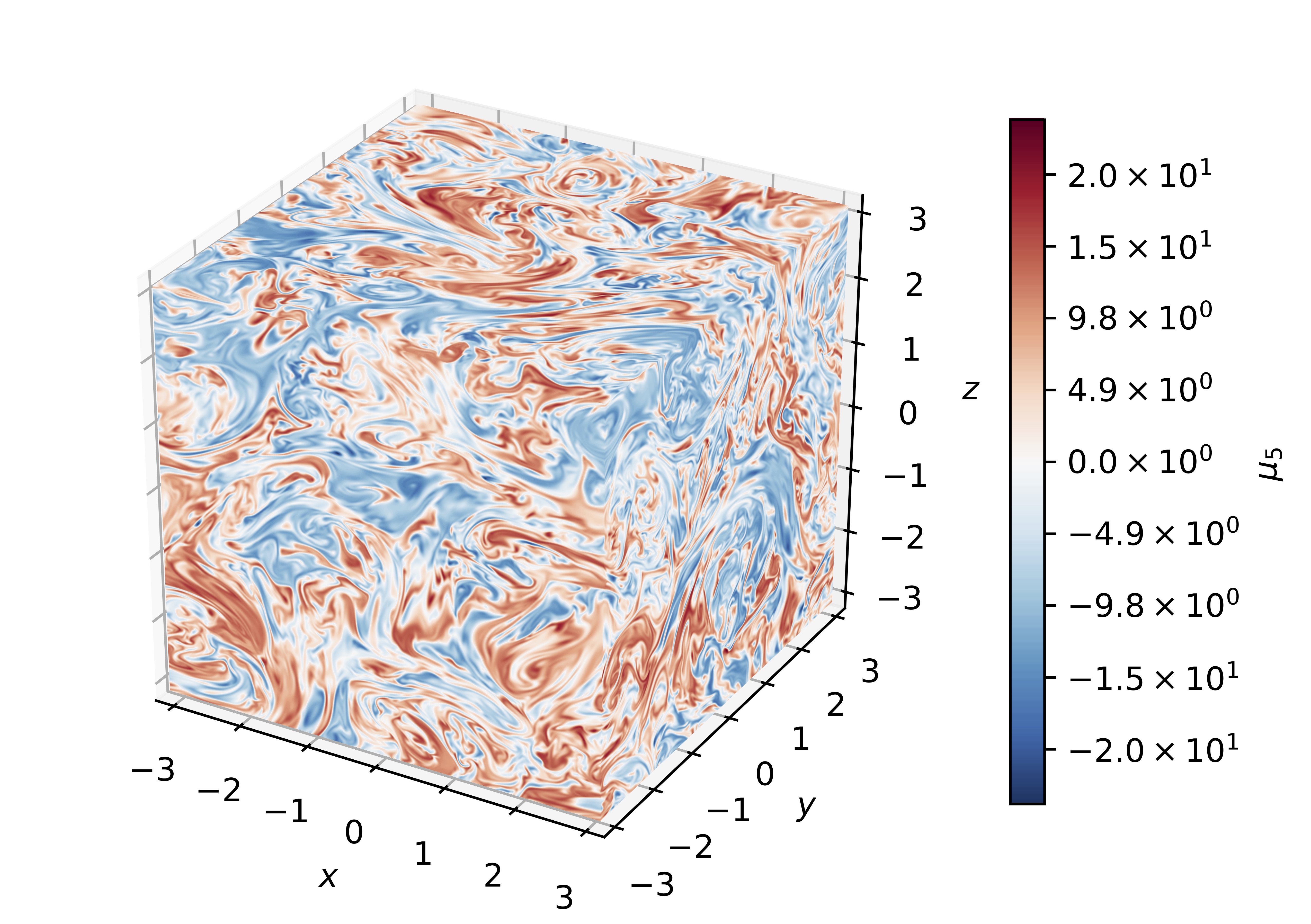} \\
  \includegraphics[width=0.3\textwidth]{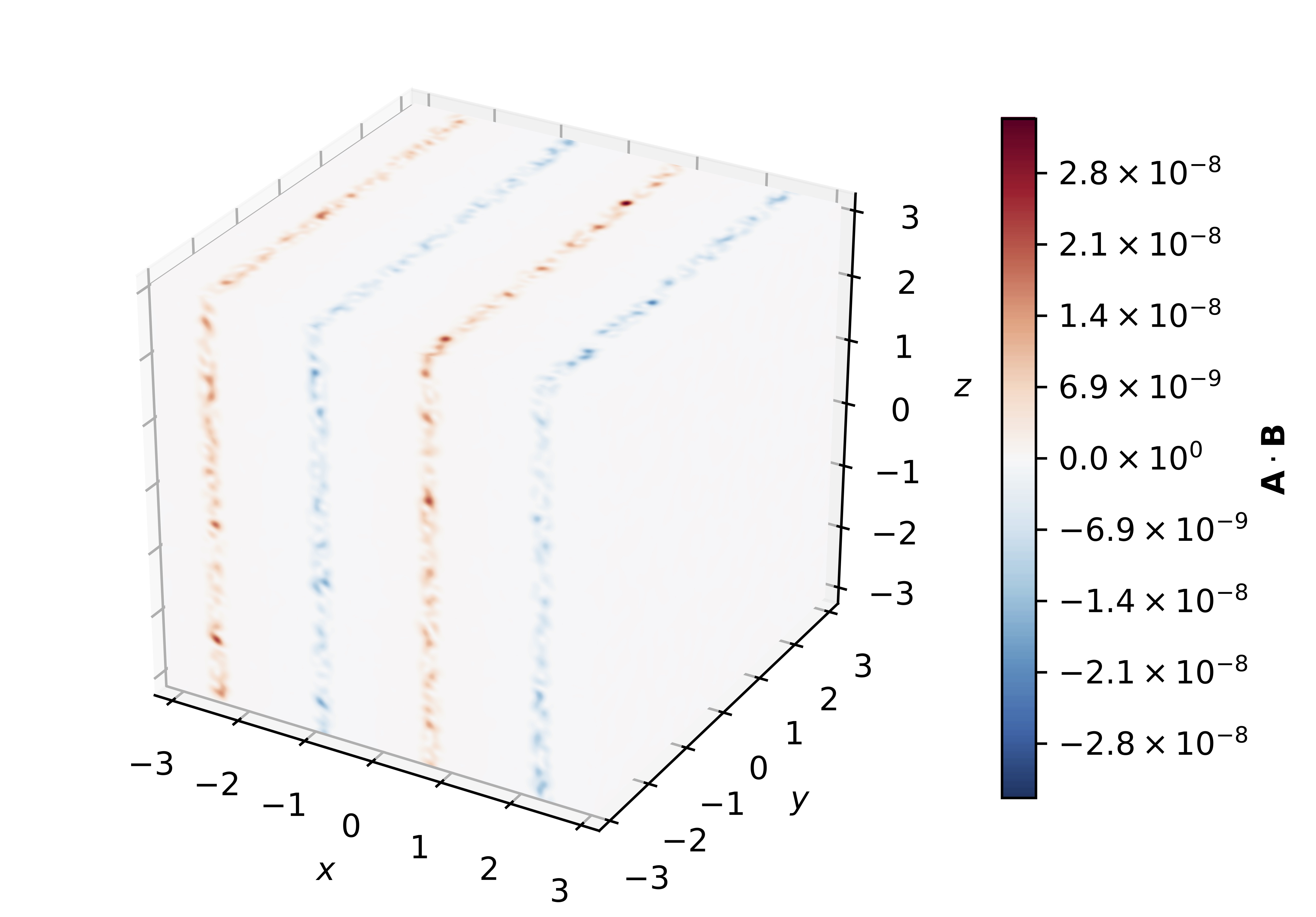}
  \includegraphics[width=0.3\textwidth]{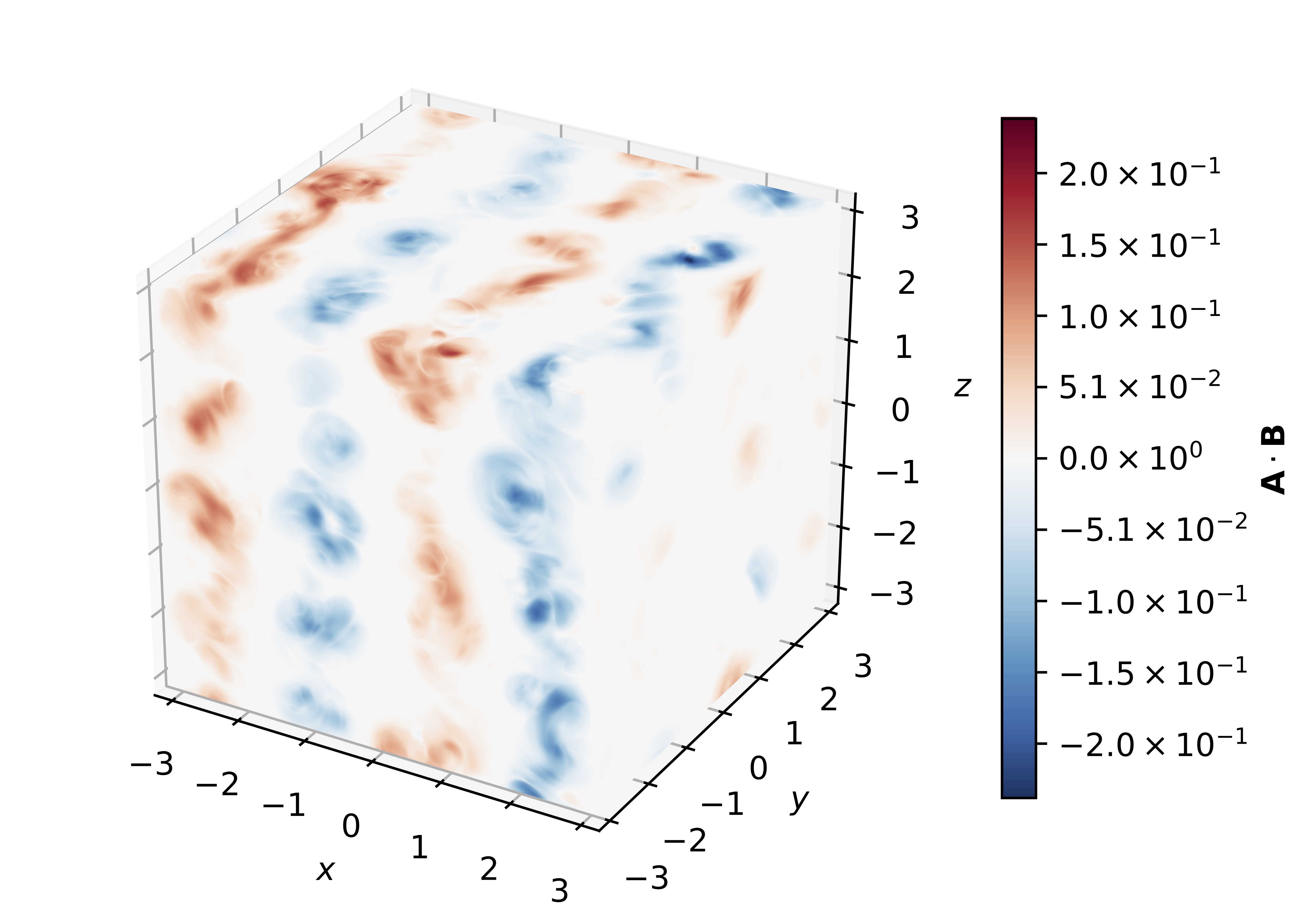}
  \includegraphics[width=0.3\textwidth]{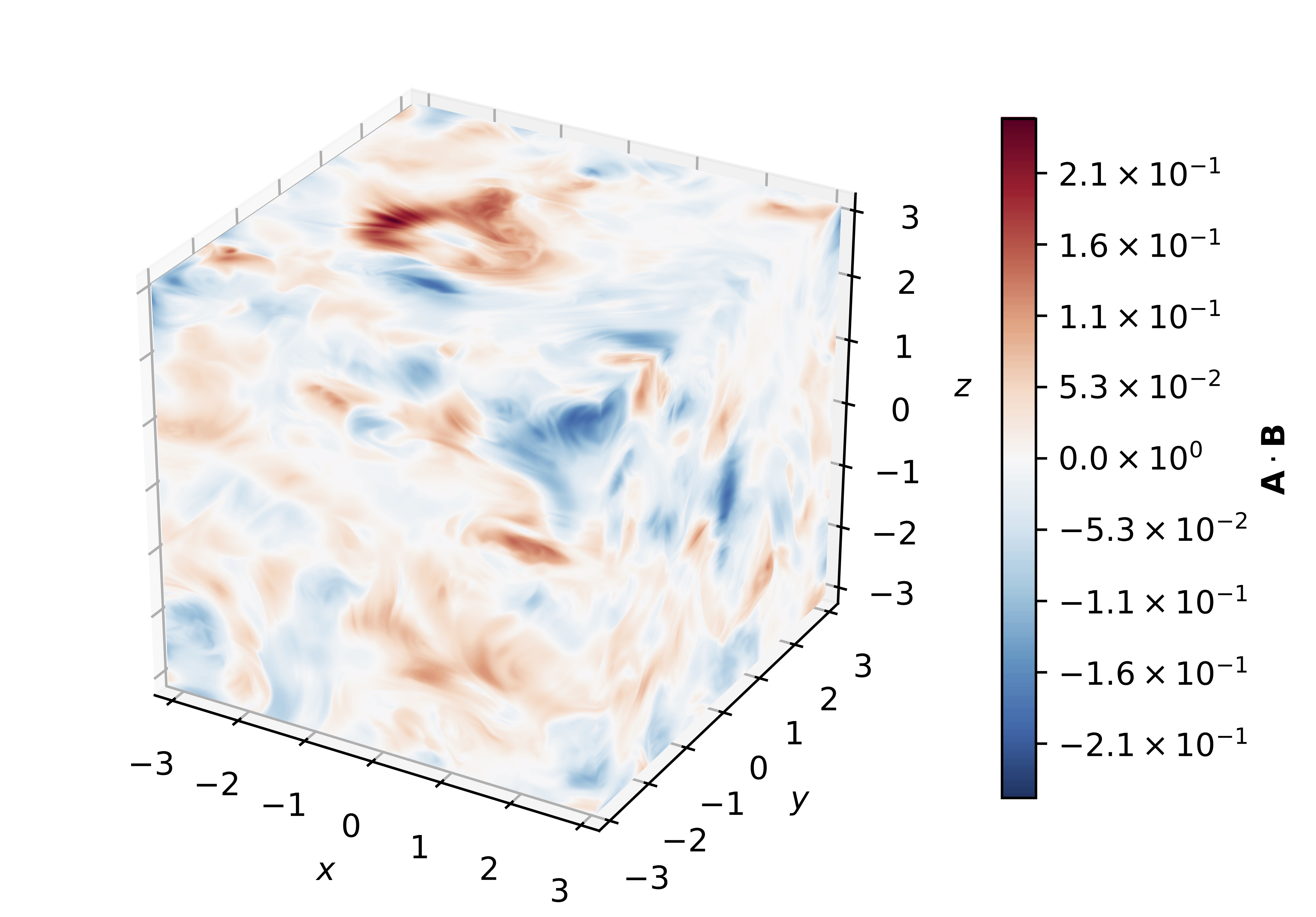} \\
  \includegraphics[width=0.3\textwidth]{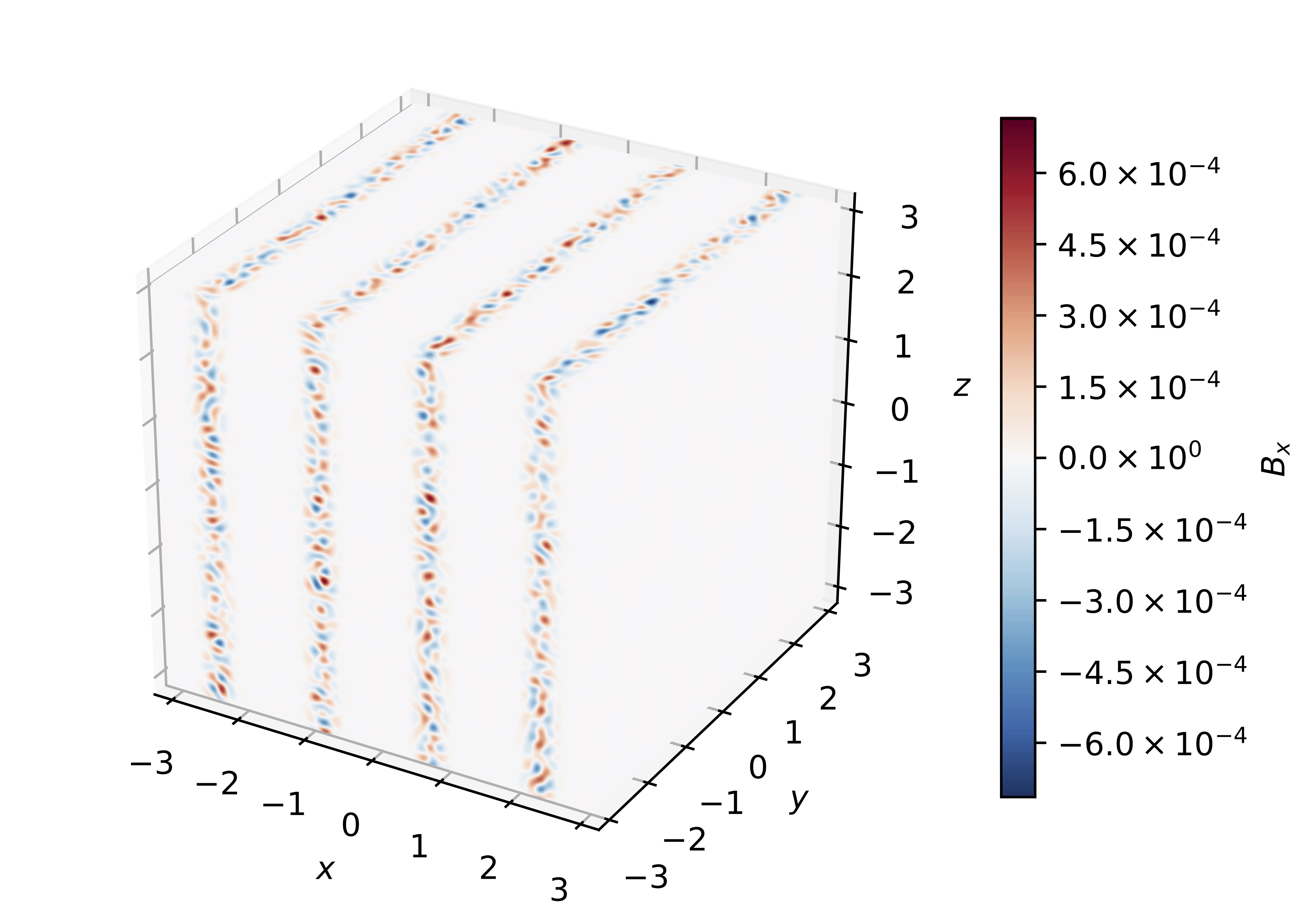}
  \includegraphics[width=0.3\textwidth]{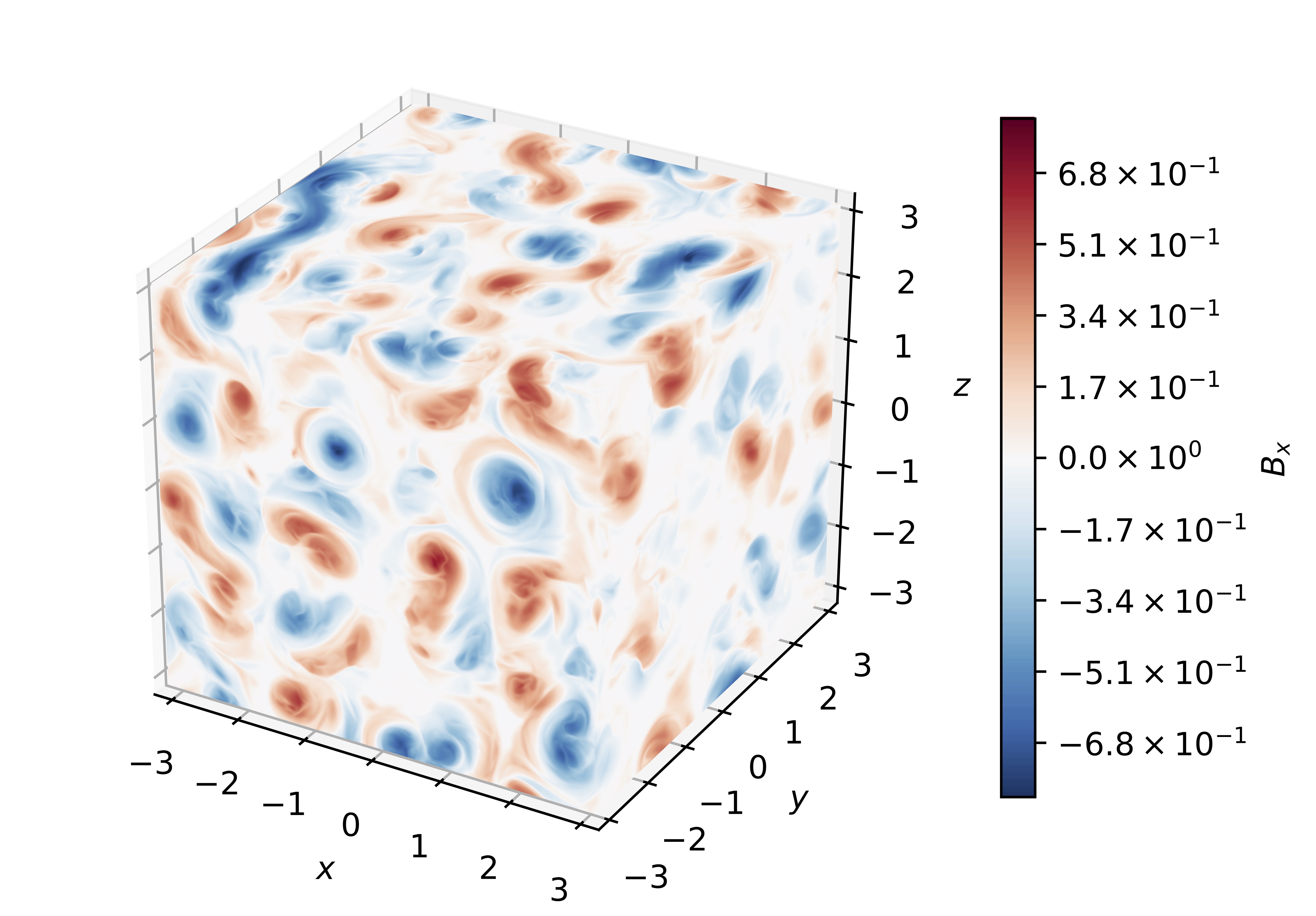}
  \includegraphics[width=0.3\textwidth]{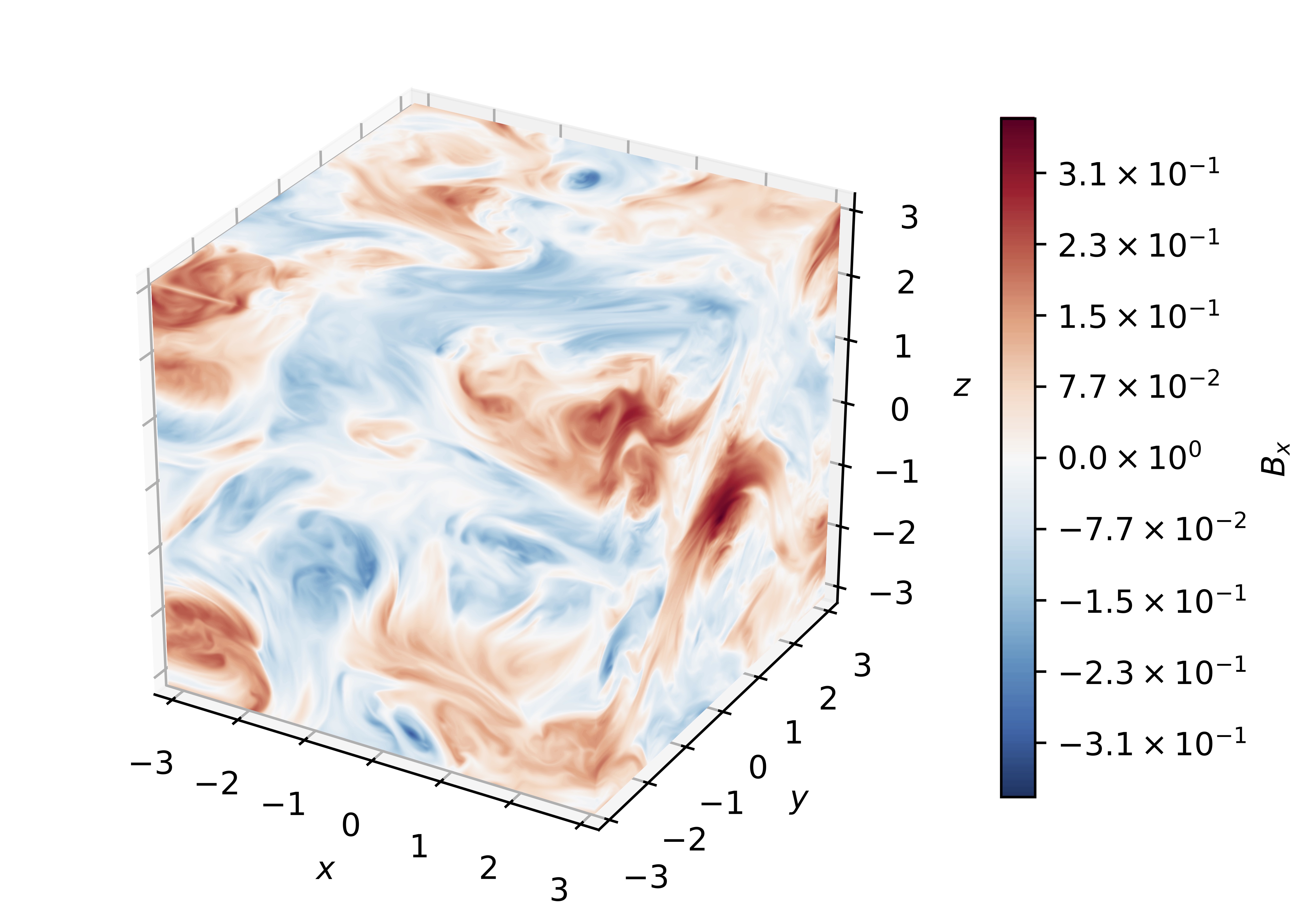} \\
\caption{Snapshots of run S23D taken at the different times: During the
kinetic dynamo phase ($t = 0.02$, left), the mean-field dynamo phase
($t = 0.04$, middle), and at the time when the inverse cascade reaches
the scale of the domain ($t = 0.06$, right).
}
\label{fig_cubestk1}
\end{figure*}

The time evolution of $B_\mathrm{rms}$ and other relevant quantities of
run S23D are presented in Fig.~\ref{fig_ts_S3D}a.
The high initial value of $\mu_{5,\mathrm{max}}$ leads to the $v_5$ dynamo
instability that amplifies $B_\mathrm{rms}$ by approximately $8$ orders
of magnitude.
Simultaneously, $U_\mathrm{rms}$ grows with twice the 
growth rate as the one of $B_\mathrm{rms}$ 
and the two fields become comparable 
at $t\approx 0.03$.
At that time, the magnetic Reynolds number has become larger than unity,
leading to the onset of turbulent effects.
In run S23D, we have initially $\langle\mu_{5}\rangle=0$,
and the mean magnetic field is not generated at the initial time.
However, $\langle\mu_{5}\rangle$ is produced at approximately twice the
rate of $B_\mathrm{rms}$; see Fig.~\ref{fig_ts_S3D}b.
Until the time $t=0.027$, the signs of both $\langle\mu_{5}\rangle$
and $\langle\AAA\cdot\BB\rangle$ are negative, but with the
onset of turbulence, the signs of $\langle\mu_{5}\rangle$ and
$\langle\AAA\cdot\BB\rangle$ are always opposite.
Eventually, $\langle\mu_{5}\rangle$ reaches a value of $+0.2$, and
hence a significant mean chiral chemical potential is produced.
Due to numerical precision, $\langle\mathcal{H}\rangle + 2 \langle\mu_5\rangle/\lambda$
grows to a value of $\approx10^{-5}$, despite the opposite signs of
$\langle\mu_{5}\rangle$ and $\langle\AAA\cdot\BB\rangle$.
We stress that $\langle\mathcal{H}\rangle + 2 \langle\mu_5\rangle/\lambda$ only 
reaches values that are below the
numerical precision and that this does not indicate a 
violation of the conservation law.

The evolution of the spatial structures of
$\mu_5$, $\AAA \cdot \BB$, and $B_x$ on the surface of the numerical domain, 
can be seen in Fig.~\ref{fig_cubestk1}.
In the $v_5$ dynamo phase (left column of Fig.~\ref{fig_cubestk1}), it
can be seen that in regions where $\mu_5<0$, a negative $\AAA \cdot \BB$
is generated, and in regions where $\mu_5>0$ also $\AAA \cdot \BB>0$.
The magnetic field is generated on small spatial scales 
($k\approx 25$) which is consistent 
with the initial amplitude of the $\mu_5$ sine function;
$\mu_{5,\mathrm{max}}(t_0)=50$.
The fastest amplification of $\AAA \cdot \BB$ and $B_x$ occurs in the regions
where the amplitude of $\mu_5$ has maxima.
The spatial correlation between the signs of $\mu_5$ and $\AAA \cdot
\BB$ can still be seen in the nonlinear phase; see the middle column of
Fig.~\ref{fig_cubestk1} which shows the snapshots at $t = 0.04$.
At this time, the characteristic scale of $B_x$ has already increased
significantly ($k\approx 5$).
The right-hand column of Fig.~\ref{fig_cubestk1} shows the simulation at the
time when the inverse cascade reaches the domain size, i.e., the first time
when $k_\mathrm{p}=k_1$.
By this time, fluctuations in $\mu_5$ have increased strongly and 
both, $\AAA \cdot \BB$ and $B_x$, exhibit a large-scale structure. 

\begin{figure}
  \includegraphics[width=0.45\textwidth]{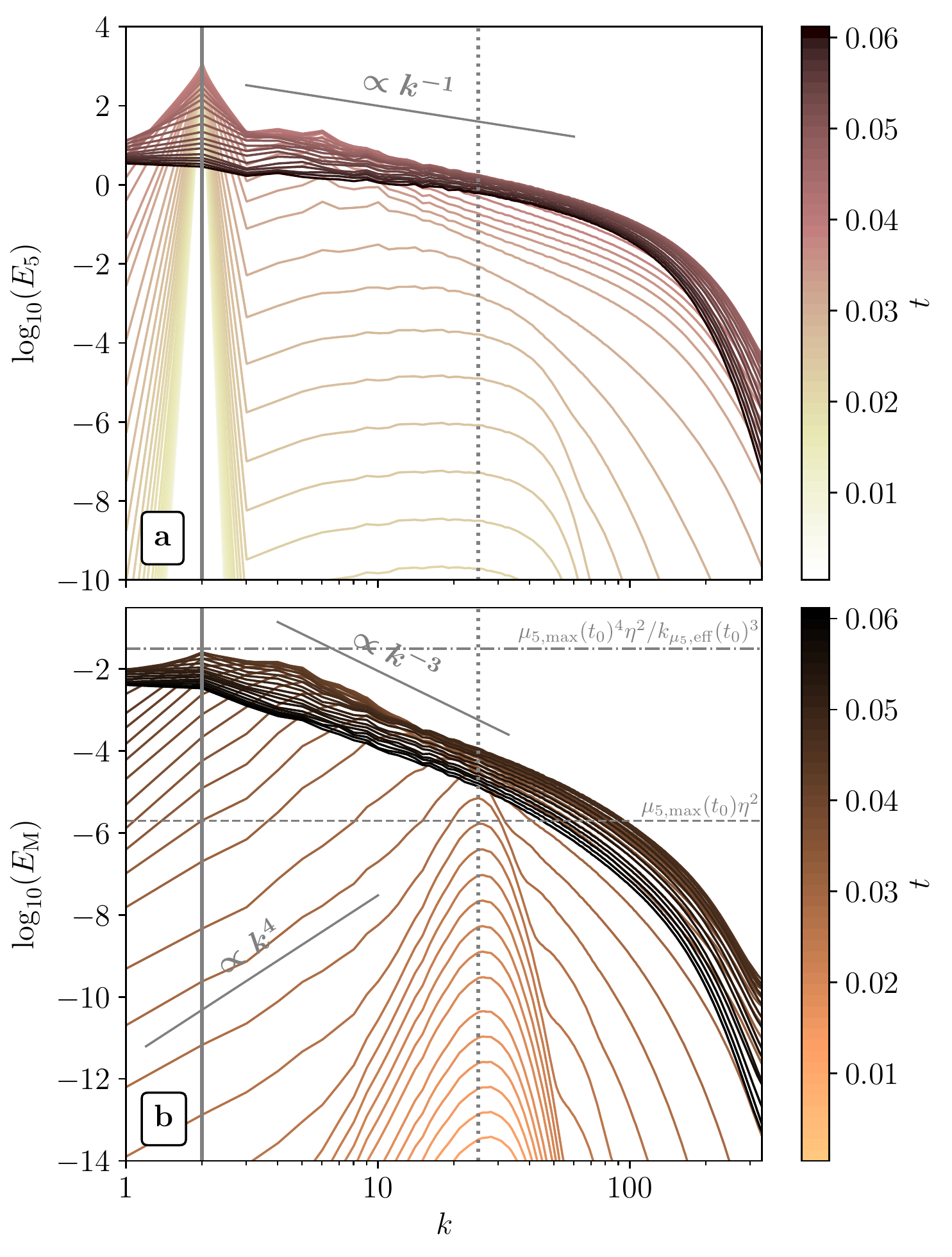}
\caption{
Evolution of power spectra in Run S23D.
\textit{(a)} Power spectra of $\mu_5$, $E_5$.
\textit{(b)} Magnetic energy spectra, $E_\mathrm{M}$.
The horizontal lines indicate the level of $E_\mathrm{M}$
at the onset of the inverse cascade (gray dashed line; as discussed in Ref.~\citep{BSRKBFRK17})
and the maximum energy (gray dashed-dotted line; see Sec.~\ref{sec_saturation}).}
\label{fig_spec_S23D}
\end{figure}

The evolution of the spatial structure
in run S23D can also be seen in the power spectra at different times. 
Figure~\ref{fig_spec_S23D}a shows the evolution of $E_5(k,t)$ and
Fig.~\ref{fig_spec_S23D}b the one of $E_\mathrm{M}(k,t)$.
The magnetic energy peaks initially at $k\approx 25$, as expected 
from the $v_5$ dynamo theory
for an amplitude of $\mu_{5,\mathrm{max}}=50$.
However, the magnetic energy grows also at smaller wave numbers and
obeys a $k^4$ spectrum.
After $t\approx 0.03$, the peak of $E_\mathrm{M}$ shifts towards
larger spatial scales, i.e., smaller $k$.
During that phase, the amplitude of $E_\mathrm{M}$ still increases
and a magnetic spectrum $E_\mathrm{M}\propto k^{-3}$ 
is established, together with
a spectrum of the chiral chemical potential $E_5\propto k^{-1}$; 
see Fig.~\ref{fig_spec_S23D}.
This is different from the 
case of a uniform $\mu_5$ field,
where the magnetic spectrum is $E_\mathrm{M}\propto k^{-2}$.
The amplitude of $E_\mathrm{M}$ decreases only after the inverse 
cascade has reached the initial scale of $\mu_5$, $k=2$. 
By the time when the inverse cascade arrives at the minimum
wave number of the numerical domain, $k=k_1=1$, the
$E_\mathrm{M}$ spectrum becomes less steep and is closer to $k^{-2}$.
We note that, already at early times $t < 0.03$, 
the spectrum of the chiral chemical potential, $E_5$,
also grows at $k\approx25$; see Fig.\ \ref{fig_spec_S23D}a. 
At late times, $E_5$ has been strongly modified by the magnetic field:
the peak at $k=2$ has vanished and an almost flat spectrum towards large
$k$ has developed.
The final scaling is approximately $E_5\propto k^{-1}$.

\begin{figure}
  \includegraphics[width=0.45\textwidth]{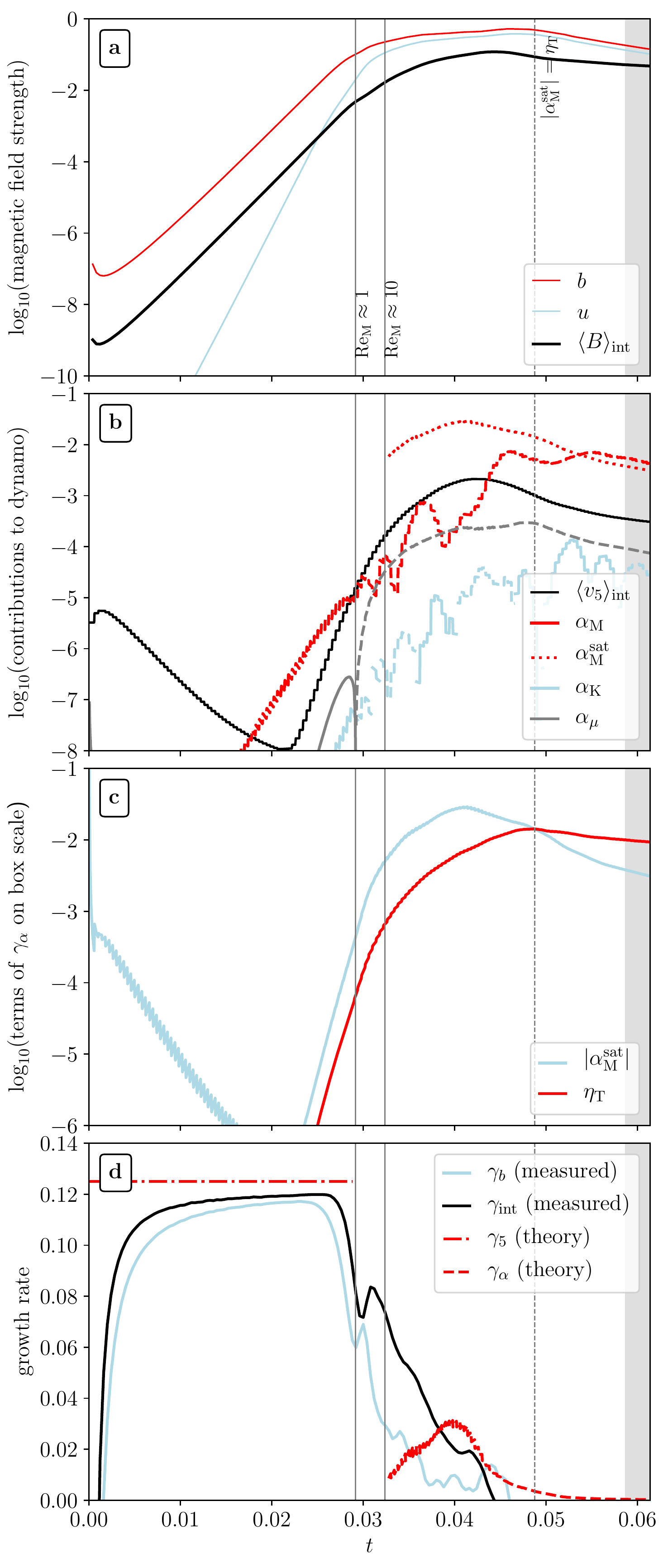}
\caption{Evolution of run S23D.
\textit{(a)} 
Time series of $\langle B\rangle_\mathrm{int}$, $b$, and $u$.
\textit{(b)} Different contributions to the mean-field dynamo:
The mean chiral velocity $\langle v_\mathrm{5}\rangle_\mathrm{int}$ 
and different estimates of the $\alpha$ effect.  
A positive sign is indicated by solid line style and a negative sign 
by dashed line style.
\textit{(c)} The two terms of the mean-field dynamo growth rate. 
\textit{(d)} The measured growth rate of 
$\langle B_\mathrm{int}\rangle$ and $b$ 
and comparison with the theoretical predictions for the 
nonturbulent dynamo phase at $t \lesssim 0.028$ and the 
turbulent mean-field dynamo phase, $t \gtrsim 0.032$. 
Note, that $\gamma_\alpha$ is based on the largest 
contribution to the mean-field dynamo, $\alpha_\mathrm{M}$.
}
\label{fig_gamma_t_S3D}
\end{figure}

We now analyze the amplification of the magnetic field on different 
scales in more detail. 
In particular, we compare the evolution of the magnetic field 
strength associated with 
the energy of magnetic fluctuations ${\bm b}$ at the wave number of 
the maximum growth rate of the $v_5$ dynamo instability
\footnote{This corresponds to the scale of the $v_5$ dynamo 
instability, which for S23D is $\mu_{5,\mathrm{max}}(t_0)/2 = 25$.}
with the one at the time-dependent integral scale of turbulence, 
$k_\mathrm{int}(t)$:
\begin{eqnarray}
   \langle B\rangle_\mathrm{int} \equiv  \left[\frac{\int_0^{k_\mathrm{max}} E_\mathrm{M}(k)^{2}~\mathrm{d}k}{\int_0^{k_\mathrm{max}} E_\mathrm{M}(k)~\mathrm{d}k}\right]^{1/2}.
\end{eqnarray}
The time evolution of $b \equiv 
\left(2 \int_{k_5}^{k_\mathrm{max}} E_\mathrm{M}(k)~\mathrm{d}k\right)^{1/2}$
and $B_\mathrm{int}$ is presented in Fig.~\ref{fig_gamma_t_S3D}a. 
With the integral scale being $k=25$ during the $v_5$ dynamo phase, 
$b$ and $B_\mathrm{int}$ are identical for $t\lesssim 0.03$.
At $t\gtrsim 0.03$, $b$ saturates while $B_\mathrm{int}$ continues
to grow at a lower rate until $t\approx 0.045$.
To understand the measured growth rates, we calculate the
different contributions to the mean-field dynamo; see Fig.~\ref{fig_gamma_t_S3D}b.
Further, positive and negative contributions to the mean-field dynamo growth rate
are presented in Fig.~\ref{fig_gamma_t_S3D}c.

The measured growth rate of the magnetic field strength on different
scales is presented in Fig.~\ref{fig_gamma_t_S3D}d.
Note that the amplification at $k=25$ stops at $t\approx 0.03$
but before that it is well described by $\gamma_{5}$ as given by
Eq.~(\ref{eq_gammalam_max}) with $\mu_5 = \mu_{5,\mathrm{max}}$.
When the maximum field strength of the $v_5$ dynamo is reached
on $k=25$, the amplification on larger scales becomes more prominent. 
However, it cannot clearly be ascribed to a mean-field dynamo
since there the chiral chemical potential is decreasing, which leads to
a decrease of the characteristic instability wave number,
$k_5(t) = \mu_{5,\mathrm{max}}(t)/2$.
To investigate the role of the mean-field dynamo in the amplification of
energy on large spatial scales, we plot the different contributions in
Fig.~\ref{fig_gamma_t_S3D}b: the mean $v_5$ based on the integral scale
of turbulence, $\alpha_\mathrm{M}$ based on the correlation time of
fluctuations on $k_\mathrm{int}$ as well as the steady state
value of the magnetic $\alpha$ effect,
$\alpha_\mathrm{M}^\mathrm{sat}$.
We also show that $\alpha_\mu$, 
based on Eq.~(\ref{eq_alpha5}), changes sign at $t\approx 0.03$. 
The dominant contribution to the mean-field dynamo 
is the magnetic $\alpha$ effect.
We also compare the measured growth rate after $t\approx 0.03$.
The theoretical curve, $\gamma_\alpha$, describes roughly the measured growth rate 
based on averaging over the integral scale,
$\gamma_\mathrm{int}$, for $0.035 \lesssim t \lesssim 0.045$.

Note, that the magnetic Reynolds number increases throughout 
both the $v_5$ dynamo phase and also the mean-field dynamo phase, because
\textit{(i)} the velocity field continues to grow and \textit{(ii)}
the wave number based on the integral scale of turbulence decreases. 
Therefore, the turbulent diffusion $\eta_\mathrm{T}= \Rm \eta/3$
increases continuously and eventually the decay term
$(\eta+\eta_\mathrm{T}) k^2$ dominates over the source term
$(\overline{v}_5 + \alpha) k$ in Eq.~(\ref{eq_gammaalpha}).
When this equilibrium is reached at the minimum wave number of the
domain, $k_1=1$, the mean-field dynamo would operate only on scales beyond the
numerical domain and the amplification of $\langle B\rangle_\mathrm{int}$
comes to an end.
Indeed, turbulent dissipation for the minimum
wave number, $\eta_\mathrm{T}$, becomes larger than
$|\alpha_\mathrm{M}^\mathrm{sat}|$ at $t\approx 0.048$ 
[see Fig.~\ref{fig_gamma_t_S3D}c], at which time the measured
$\gamma_\mathrm{int}$ has dropped below zero.

\begin{figure}
  \includegraphics[width=0.45\textwidth]{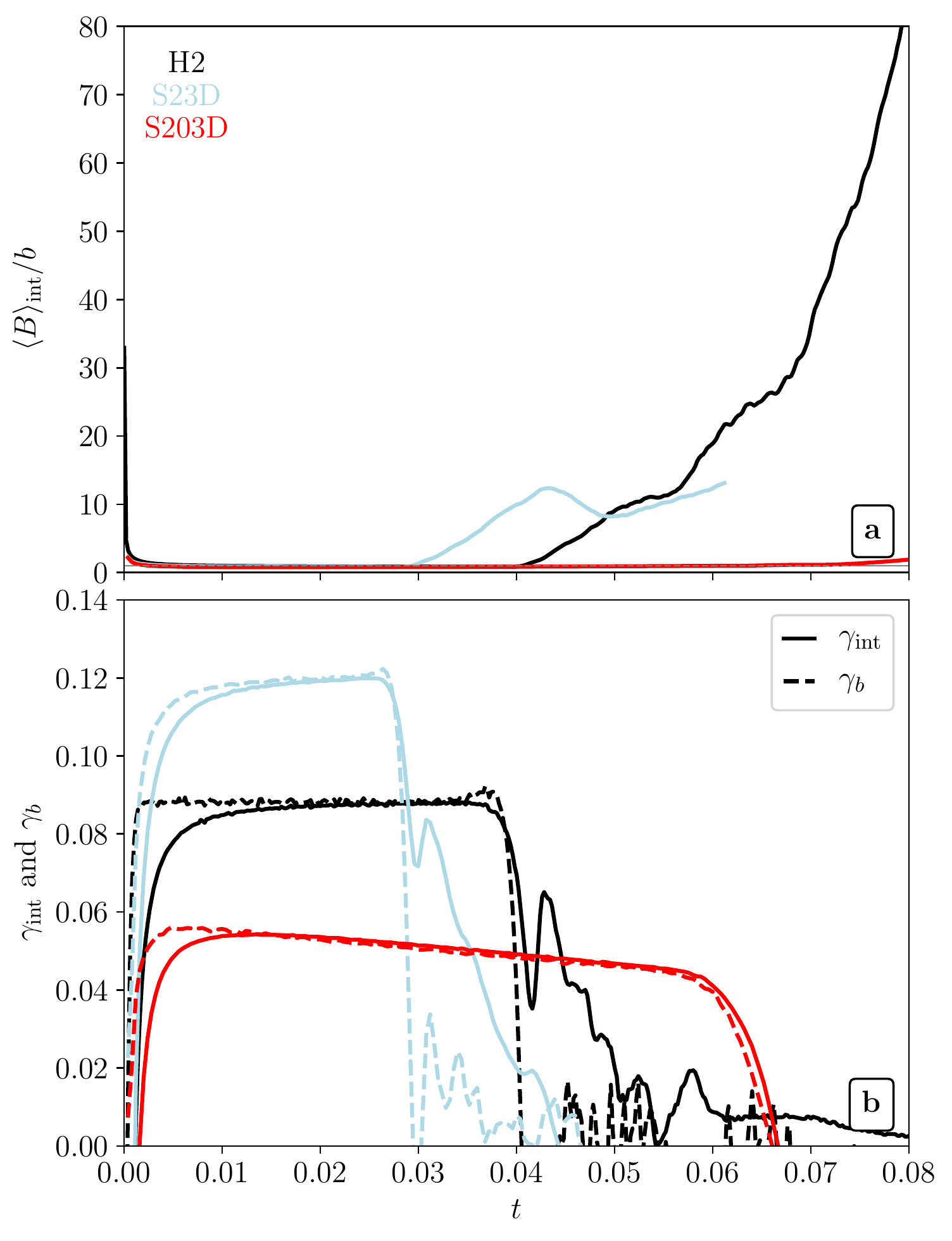}
\caption{Time evolution the magnetic field in runs H2, S23D, and S203D. 
\textit{(a)} Ratio of the magnetic field strength on the integral scale of turbulence, $
\langle B\rangle_\mathrm{int}$, and the strength of fluctuations, ${\bm b}$. 
\textit{(b)} The measured growth rates of $\langle B\rangle_\mathrm{int}$ (solid lines) 
and ${\bm b}$ (dashed lines). 
}
\label{fig_gamma_t_allS3D}
\end{figure}

In Fig.~\ref{fig_gamma_t_allS3D}, a direct comparison between the
high-resolution run with constant initial $\mu_5$ (run H2) and
inhomogeneous $\mu_5$ (run S23D) is presented.
In both cases, the ratio of $\langle B \rangle_\mathrm{int}$  
and the magnetic field $B_5$ related to the $v_5$ dynamo,
starts increasing at the onset of the mean-field dynamo; see Fig.~\ref{fig_gamma_t_allS3D}a.
For H2, the mean-field dynamo starts at $t\approx 0.04$ and for S23D at $t\approx 0.03$. 
The mean-field dynamo phase in run S23D begins earlier,
since its initial maximum value of $\mu_5$ is larger than the one in H2
($\mu_{5,\mathrm{max}}(t_0)=42$ for H2 and $\mu_{5,\mathrm{max}}(t_0)=50$
for S23D; see Table~\ref{tab_DNSoverview}).
This leads to a higher growth rate of the magnetic field, as can 
be seen in Fig.~\ref{fig_gamma_t_allS3D}b, and therefore to a faster 
generation of turbulence in the system. 
Qualitatively, the growth rates of magnetic energy for
different wave numbers evolve in a similar way in H2 and S23D.
The growth rate on the characteristic instability scale of the $v_5$ dynamo, 
$\gamma_5$, and the
one on the integral scale of turbulence, $\gamma_\mathrm{int}$,
are comparable during the $v_5$ dynamo phase. 
With the onset of turbulence, $\gamma_5$ drops to zero while 
$\gamma_\mathrm{int}$ decreases but remains positive for an extended time.

In S203D, the magnetic field on the integral scale never becomes larger than
the rms value; see the red line in Fig.~\ref{fig_gamma_t_allS3D}a.
Here, the growth rate in the $v_5$ dynamo phase is less than in S23D by a 
factor of more than $2$. 
This is consistent with the findings in
Sec.~\ref{subsec_S2D_correlationlength}, where the $v_5$ dynamo could
not develop well in setups with effective correlation wave numbers
$k_{\mu_5,\mathrm{eff}}$ that were close to the dynamo instability
scale $k_5$; see also the power spectra for run S203D in 
Figs.~\ref{fig_appendix_spec}c and \ref{fig_appendix_spec}d shown in Appendix~\ref{appendix C}.
The growth rate in S203D even decreases during the $v_5$ dynamo phase
due to diffusion at the high wave number $\mu_5$.
At $t\approx 0.06$, both $\gamma_5$ and $\gamma_\mathrm{int}$ drop to zero in 
run S203D, therefore indicating no sign of a mean-field dynamo. 
In fact, turbulence never develops in S203D and the maximum $\Rm$ over the 
entire simulation time is only $2.06$; see Table~\ref{tab_DNSoverview}.

\subsection{DNS with initial fluctuations of $\mu_5$}
\label{DNSFluctuations}

\begin{figure}
\includegraphics[width=0.45\textwidth]{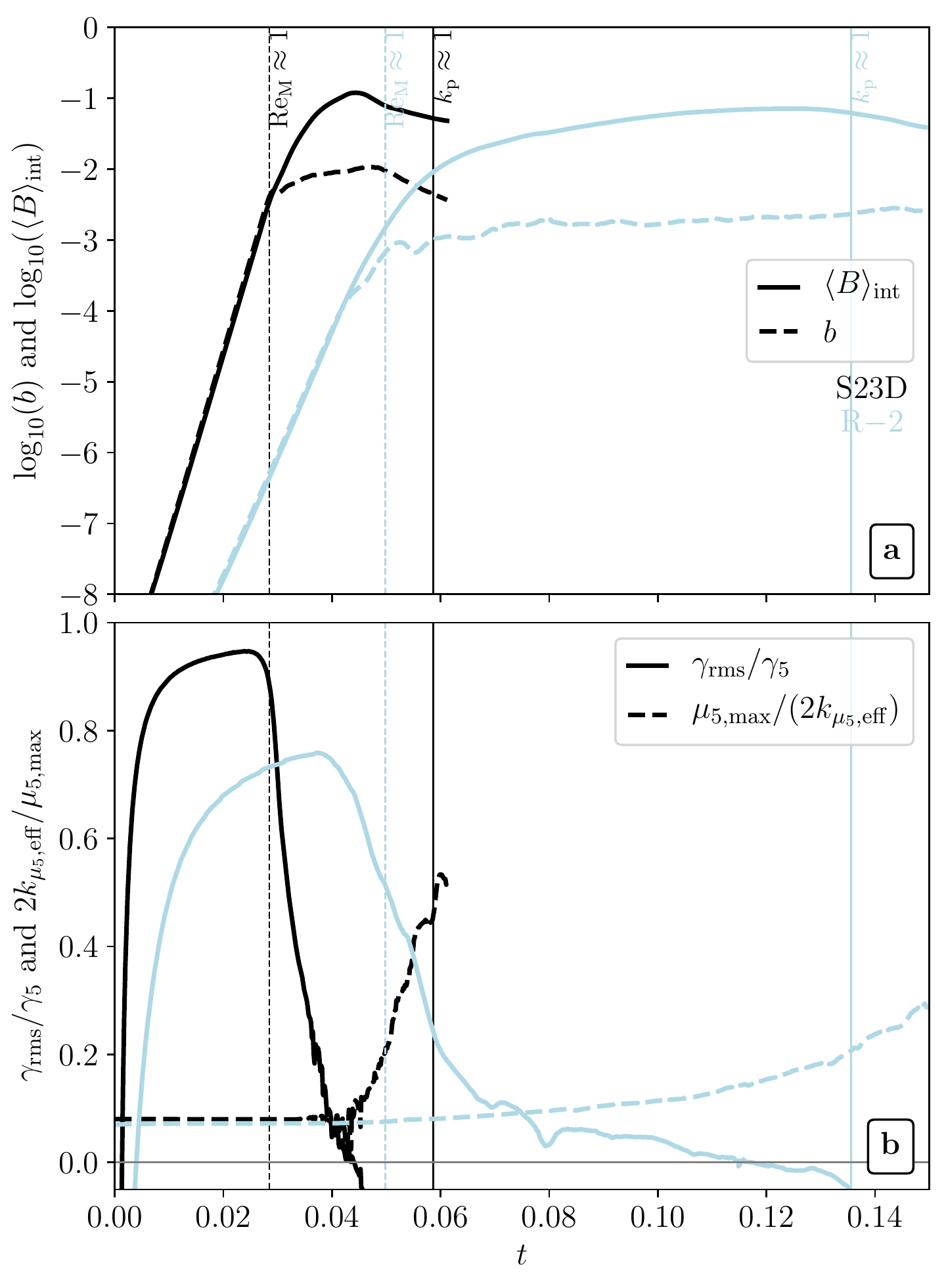}
\caption{Direct comparison between the reference run with an initial $\mu_5$ 
in the form of a sine spatial profile in the $x$ direction, S23D, and the reference run with 
random fluctuations of $\mu_5$, R$-$2.
\textit{(a)} Time evolution of $\langle B\rangle_\mathrm{int}$ and $b$.
\textit{(b)} Growth rate of $B_\mathrm{rms}$ normalized by $\gamma_5$ 
and the ratio of $\mu_{5,\mathrm{max}}/2$ and $k_{\mu_5,\mathrm{eff}}$.    
}
\label{fig_ts_R$-$2_S23D}
\end{figure}

\begin{figure}
\includegraphics[width=0.45\textwidth]{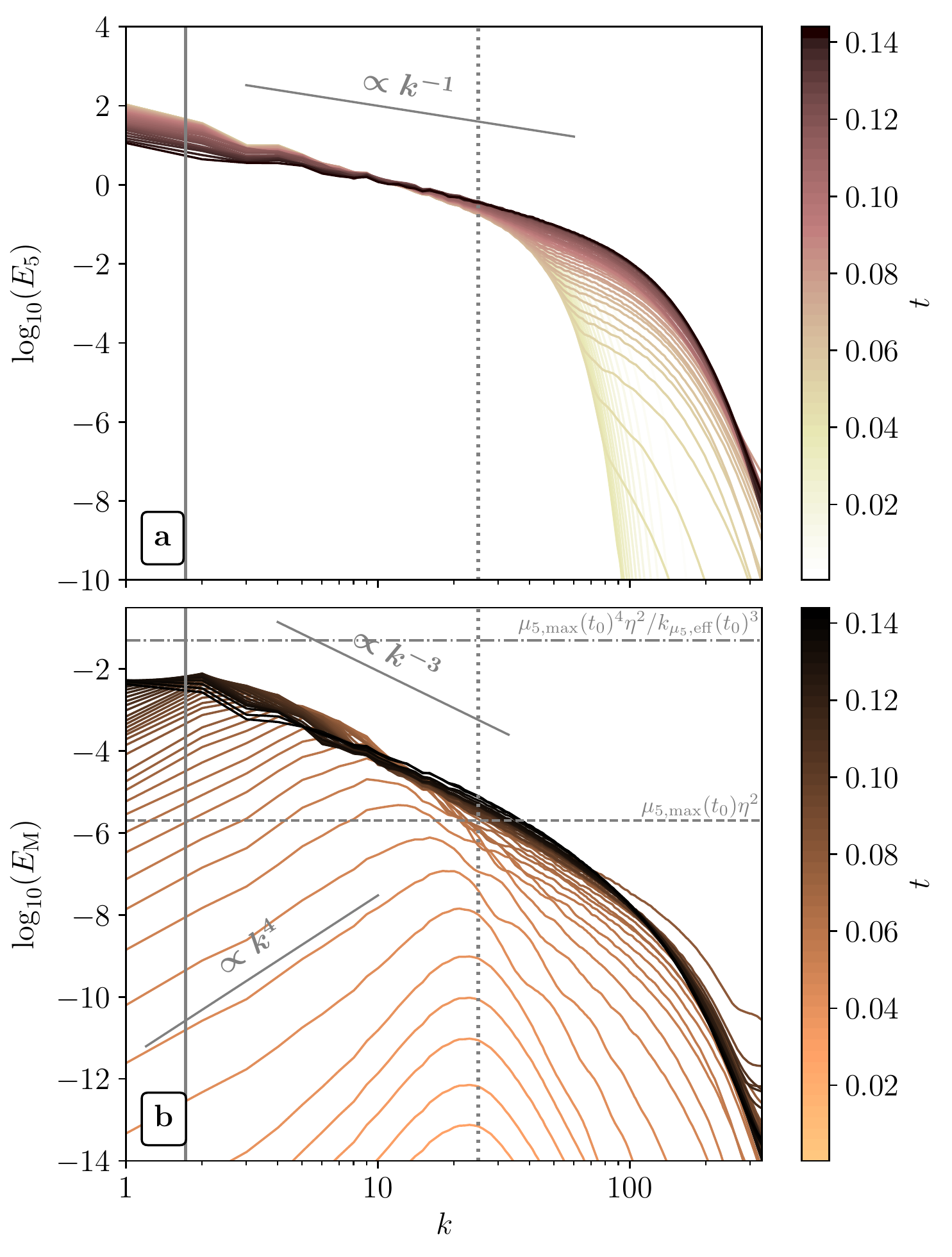}\\
\caption{
Similar to Fig.~\ref{fig_spec_S23D}, but for Run R$-$2.
}
\label{fig_spec_R$-$2}
\end{figure}

This section complements Ref.~\citep{SRB21a}, in which
we have analyzed DNS with an initially random distribution of $\mu_5$.
The existence of a mean-field dynamo phase in these scenarios has been
reported in Ref.~\citep{SRB21a} as the first demonstration of the generation
of large-scale magnetic fields from an $\mu_5$ with initially vanishing
mean value.
In this section, we analyze the properties of this instability
in greater and more technical details.

As a reference run for a DNS with initial random distributions of 
$\mu_5$ we use run R$-$2 and begin with a direct comparison to our previous 
example of a sine function initial spatial profile of $\mu_5$, run S23D.
Snapshots of $\mu_5$, $\AAA\cdot\BB$, and $B_x$ of run R$-$2 at different times 
are presented in the Appendix; see Fig.~\ref{fig_appendix_cubes_R-2}.
As shown in Fig.~\ref{fig_ts_R$-$2_S23D}, the magnetic field growth in the $v_5$ 
dynamo phase in R$-$2 is slower than in S23D despite the 
initially comparable values of $\mu_{5,\mathrm{max}}$. 
The difference in growth rates in the two runs cannot be explained by 
different separation of scales. 
The ratio of the scale of the $v_5$ dynamo instability, 
$\mu_{5,\mathrm{max}}/2$, and the effective correlation 
length of $\mu_{5}$, $k_{\mu_5,\mathrm{eff}}$, is $\approx 0.1$ in both runs. 
Therefore, the differences must come from the shape of 
the $\mu_5$ spectra; see the spectra of R$-$2 in Fig.~\ref{fig_spec_R$-$2} 
and the one of S23D in Fig.~\ref{fig_spec_S23D}.
Note, that the measured growth rate in R$-$2 increases more slowly than in S23D, so 
for a lower value of the initial magnetic seed field, the maximum ratio of 
$\gamma_\mathrm{rms}/\gamma_5$ could get closer to $1$.
Another interesting difference between S23D and R$-$2 is the fact that 
the mean-field dynamo phase starts earlier in the latter run and also lasts longer. 
In S23D, $\Rm>10$ at $t\approx 0.032$ and the maximum magnetic field is reached at $t\approx 0.04$.
In R$-$2, the turbulent dynamo operates between  
$t\approx 0.068$ and $t\approx 0.12$ (see below).

\begin{figure}
  \includegraphics[width=0.45\textwidth]{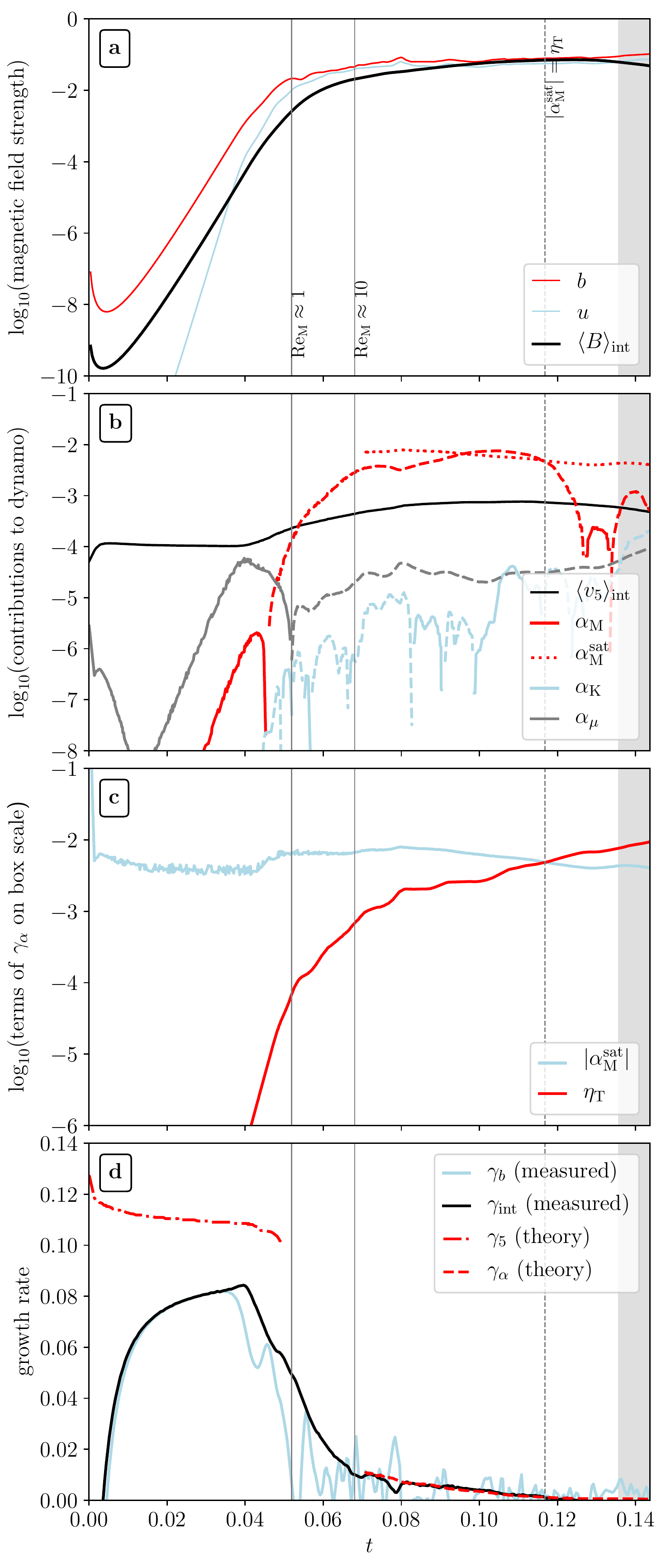}
\caption{Same as Fig.~\ref{fig_gamma_t_S3D} but for run R$-$2. 
}
\label{fig_gamma_t_R$-$2}
\end{figure}

The detailed mean-field dynamo analysis for R$-$2 is presented in Fig.~\ref{fig_gamma_t_R$-$2}.
At $t\approx 0.052$, the magnetic Reynolds number becomes larger than
unity, which coincides with the time when the magnetic energy at $k=25$
saturates; see Fig.~\ref{fig_gamma_t_R$-$2}a.
The magnetic field on the integral scale of turbulence,
$B_\mathrm{int}$, continues to grow with the predominantly
positive contribution to the growth rate being
the magnetic $\alpha$ effect; 
see Fig.~\ref{fig_gamma_t_R$-$2}b.
As in run S23D, the maximum field strength of the mean-field dynamo occurs 
once $\eta_\mathrm{T}k^2$ becomes larger than
$|\alpha_\mathrm{M}^\mathrm{sat}|k$ at $k=1$,
based on the size of the numerical domain.
This time is indicated by the vertical dashed lines in Fig.~\ref{fig_gamma_t_R$-$2}.
We stress again, that the mean-field dynamo limitation is here primarily an effect of
the finite size of the numerical domain: with increasing $\Rm$, the value
of $\eta_\mathrm{T}$ and therefore, the characteristic wave number of the
mean-field dynamo eventually become less than the minimum wave number
of the domain.
The growth rate during the mean-field dynamo phase, $\gamma_\alpha$, matches the 
measured growth rate of the magnetic field on the integral scale well between 
the time when $\Rm>10$ (vertical solid line at $t\approx 0.068$) and the time 
when $\eta_\mathrm{T} = |\alpha_\mathrm{M}^\mathrm{sat}|$ (vertical dashed 
line at $t\approx 0.117$). 

\begin{figure*}
\begin{minipage}[t]{\textwidth}
  \includegraphics[width=0.32\textwidth]{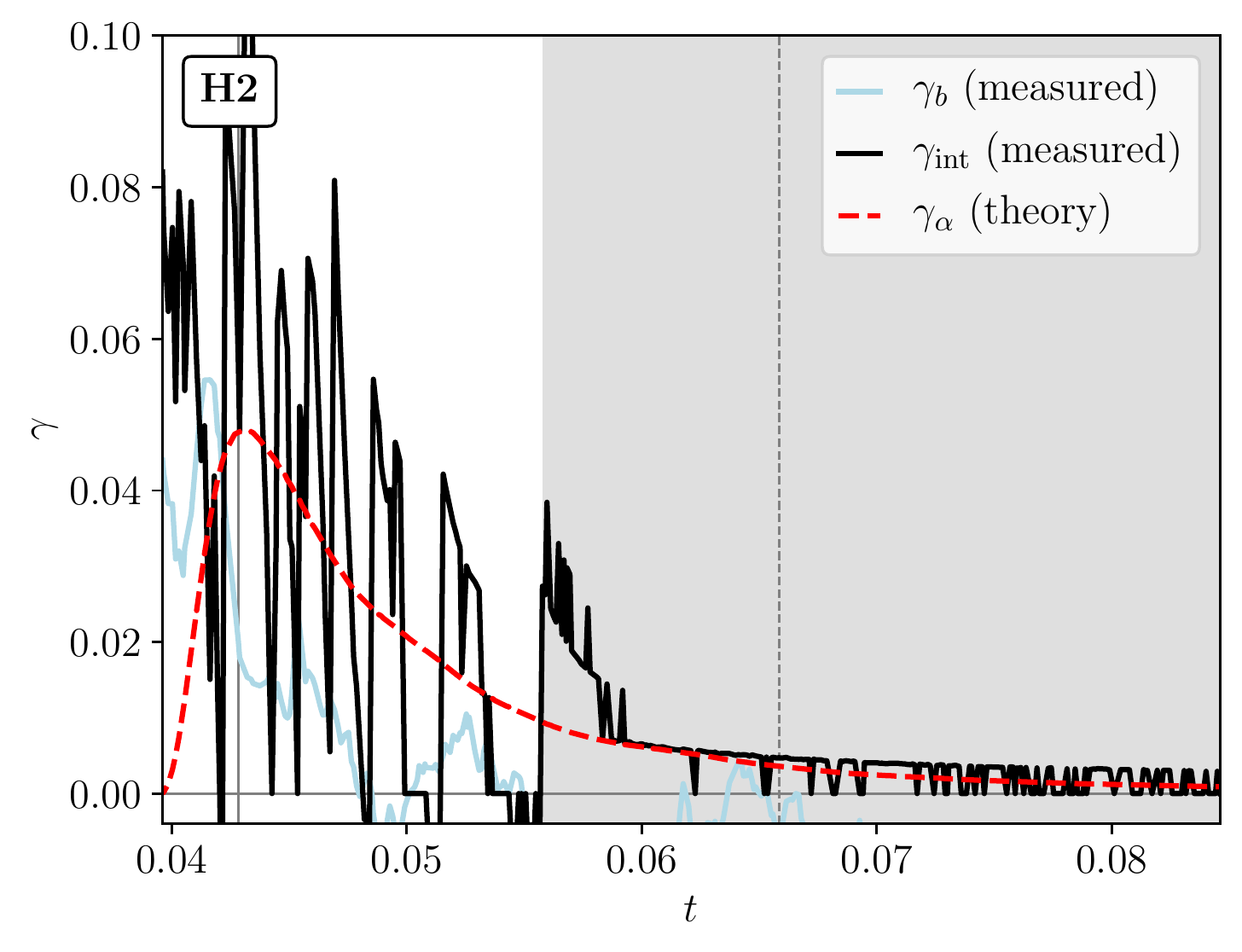}
  \includegraphics[width=0.32\textwidth]{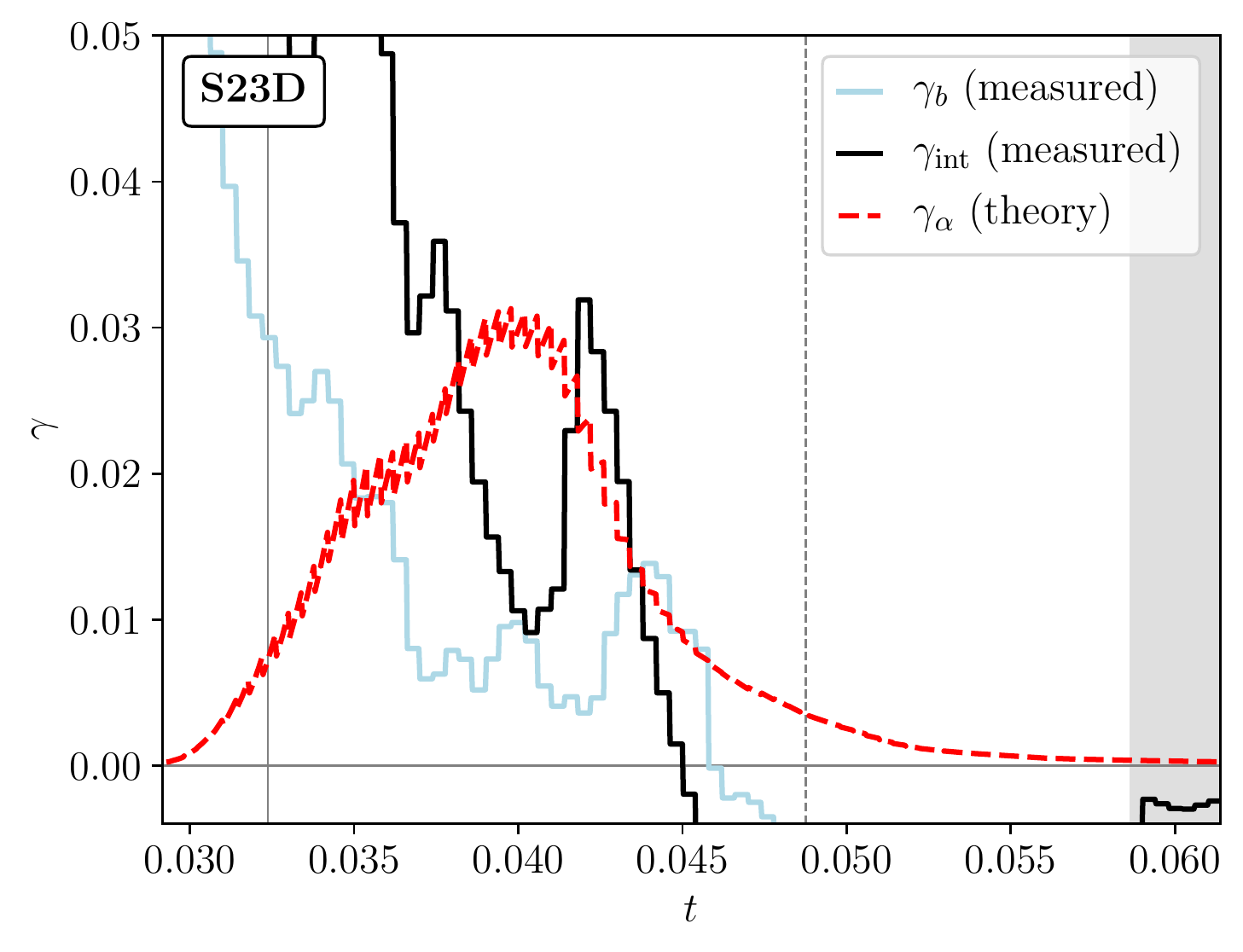} 
  \includegraphics[width=0.32\textwidth]{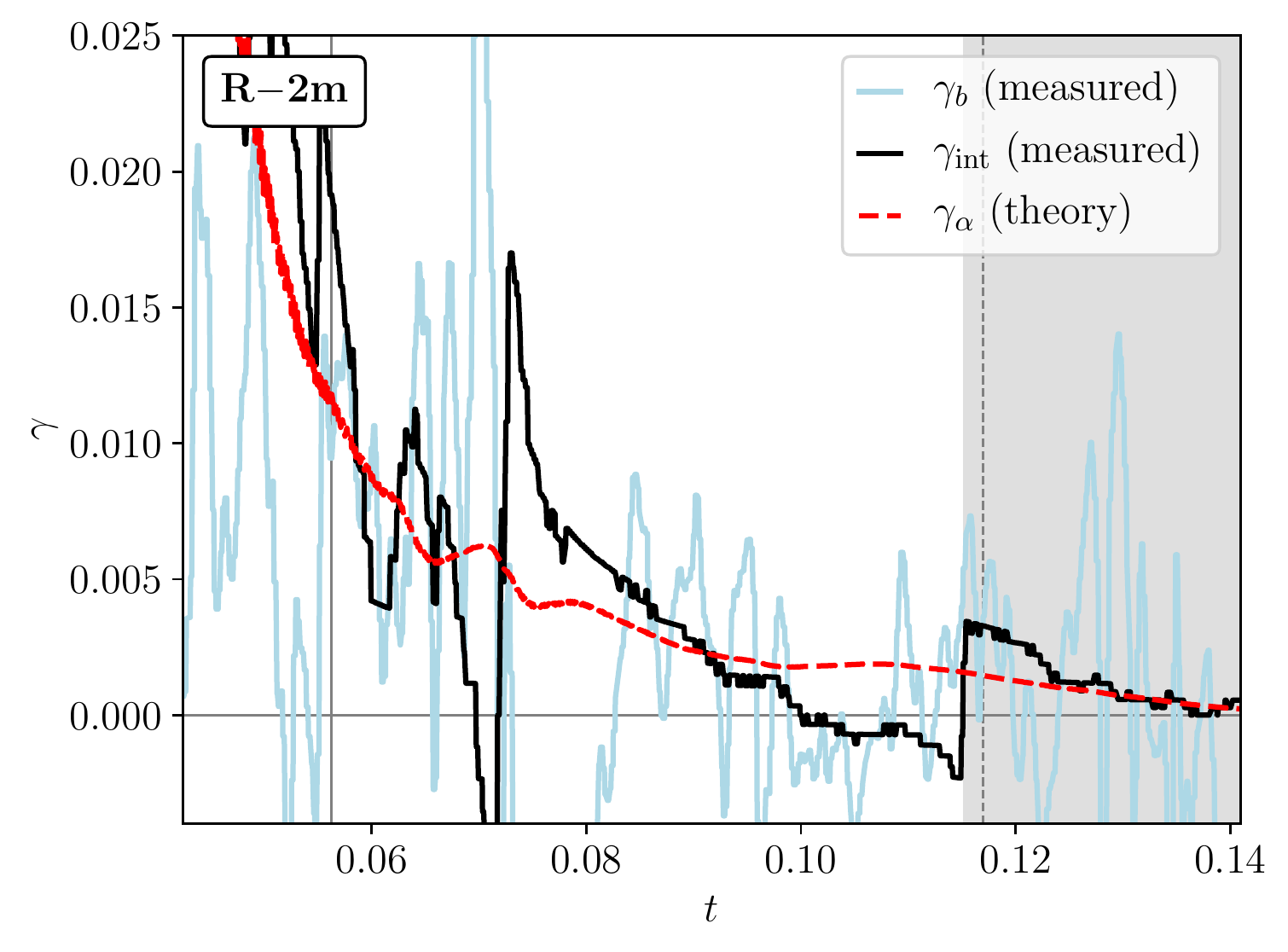} \\
  \includegraphics[width=0.32\textwidth]{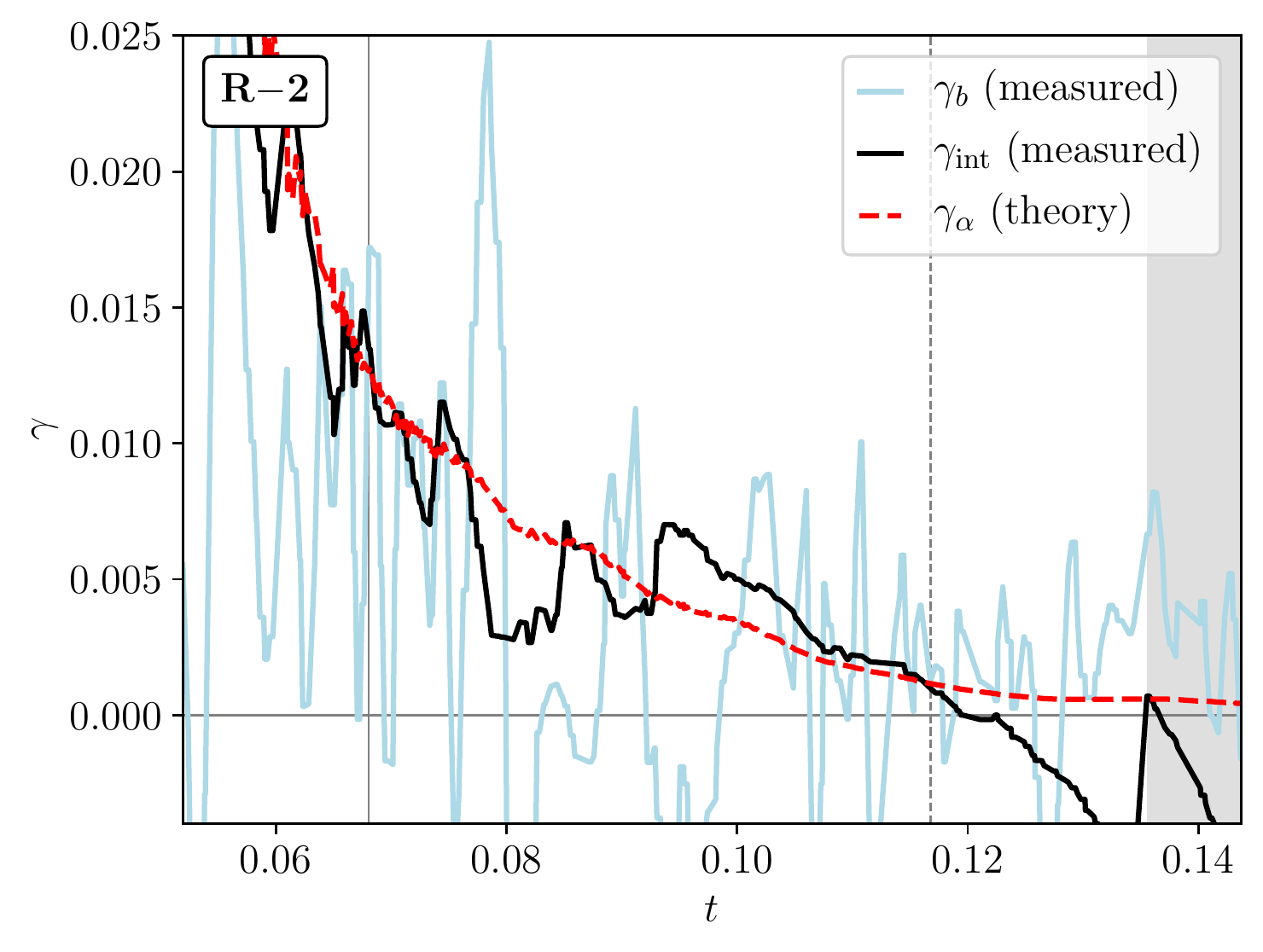} 
  \includegraphics[width=0.32\textwidth]{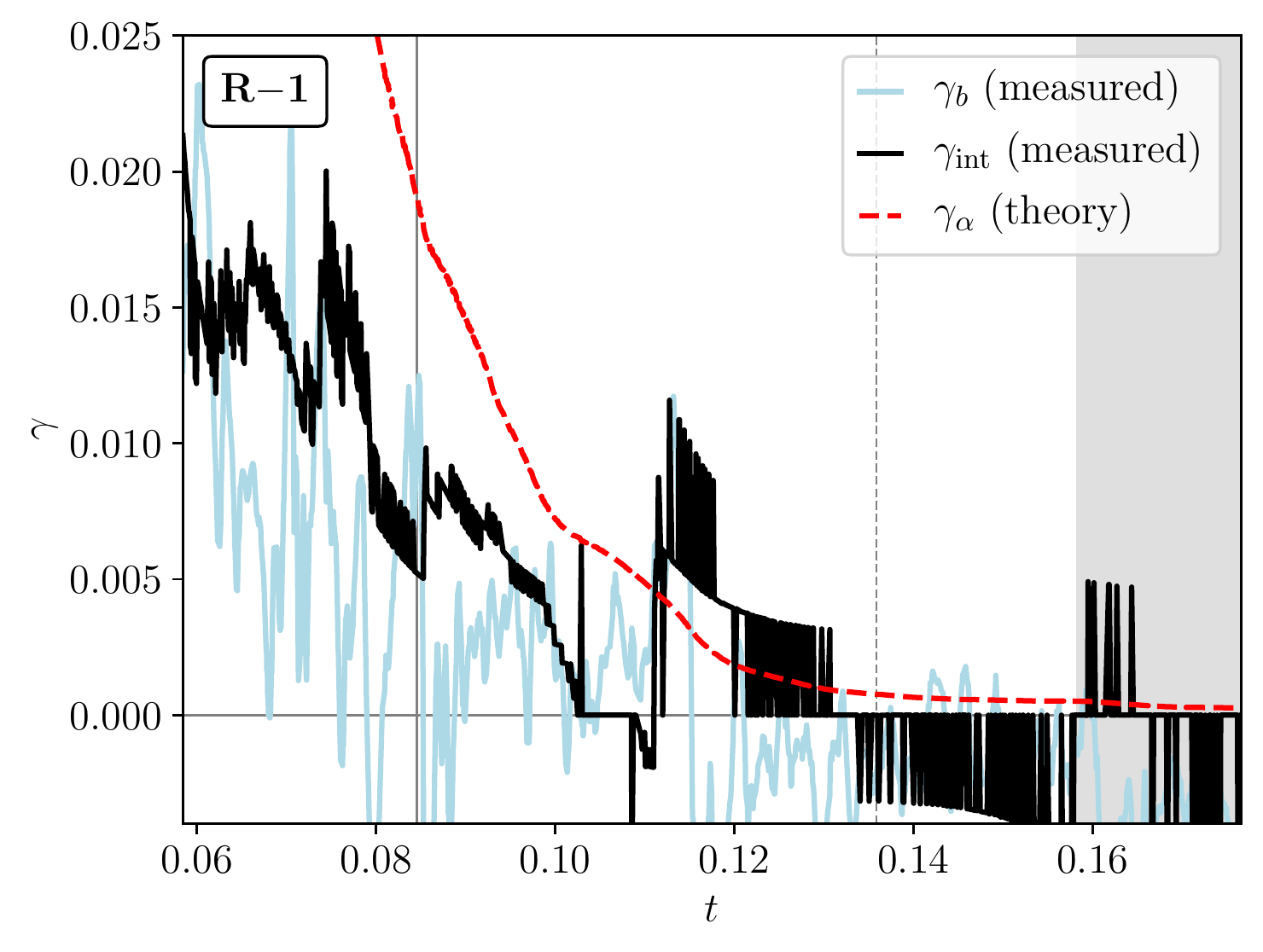}
  \includegraphics[width=0.32\textwidth]{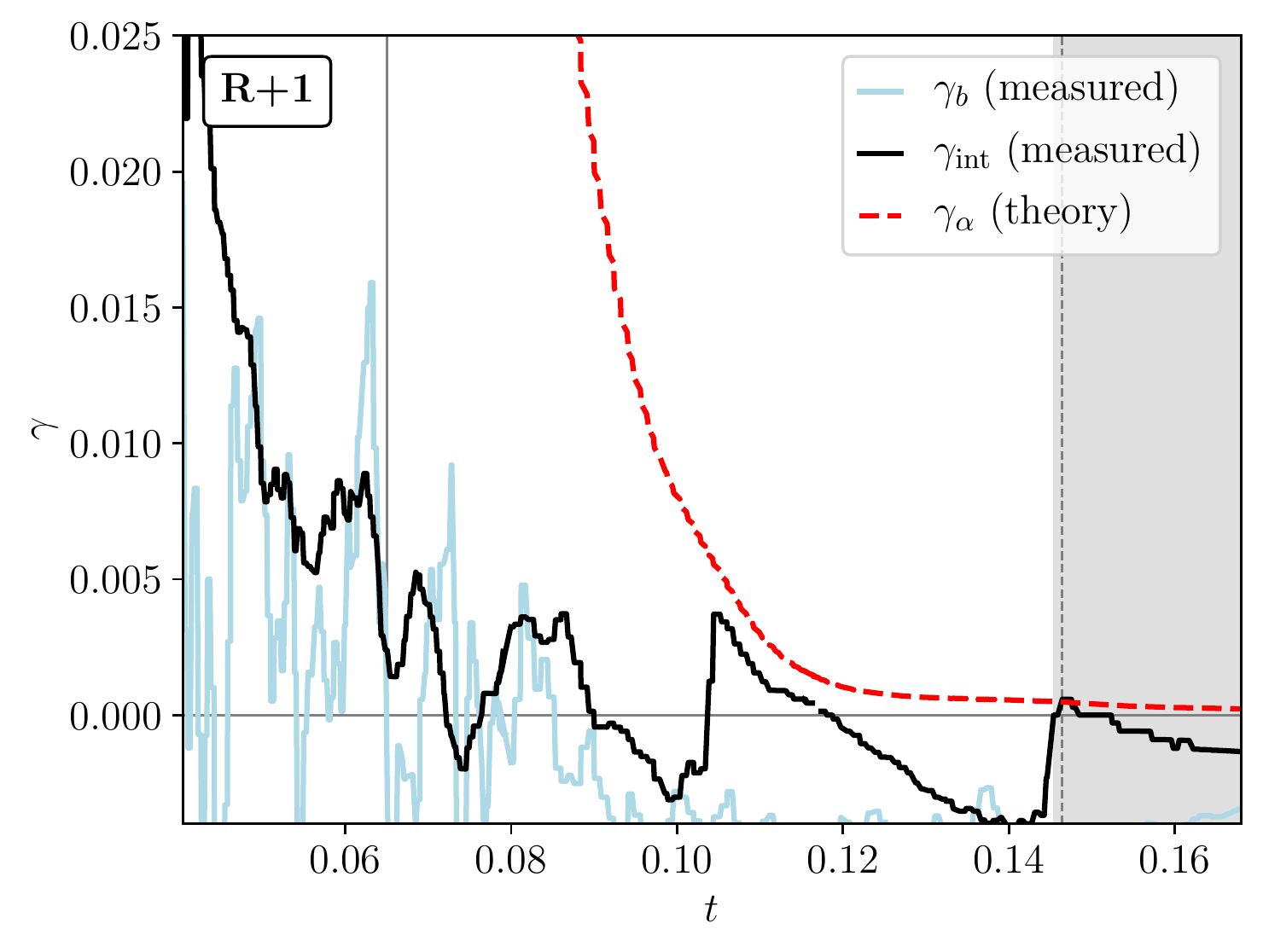}\\
\end{minipage}
\caption{Comparison between the measured growth rate and the theoretical 
prediction for all runs with turbulence. 
The blue lines show the growth rate of the magnetic field on the
characteristic scale of the $v_5$ dynamo and the black lines the rate on
the integral scale of turbulence, $k_\mathrm{int}$.
The red dashed lines show the theoretically predicted growth rate of
the mean-field dynamo, $\gamma_\alpha$.
In the case of H2, $\gamma_\alpha = \alpha_\mu^2/(4\eta_\mathrm{T})$,
while for all other runs, $\gamma_\alpha =
(\alpha_\mathrm{M}^\mathrm{sat})^2/(4\eta_\mathrm{T})$.
The time axes start at the moment when $\Rm = 1$ and the solid vertical 
lines indicate the time when $\Rm = 10$. 
The dashed vertical lines show the time when
$|\alpha_\mathrm{M}^\mathrm{sat}| <\eta_\mathrm{T}$ (and $|\alpha_\mu| <\eta_\mathrm{T}$
for run H2), i.e., when the characteristic scale of the mean-field dynamo
has increased to a length that is larger than the size of the numerical
domain and therefore growth in DNS comes to an end.
For times larger than $t_{k_1}$, i.e., when the peak of the
magnetic energy spectrum has reached the wave number $k_1$, the plots are
shaded in gray.
}
\label{fig_gamma_t_compact}
\end{figure*}

\subsection{Comparison of mean-field dynamos in DNS with different initial $\mu_5$}

Evidence for mean-field dynamos after the onset of turbulence exists
for all DNS presented in this study that reach sufficiently high Reynolds numbers.
A summary of the measured growth rates in all DNS after the onset of 
turbulence, is presented in Fig.~\ref{fig_gamma_t_compact}. 
There, blue lines show the growth rate of the characteristic 
magnetic field strength on the instability scale of the $v_5$ dynamo,
i.e., the growth rate of magnetic fluctuations $\gamma_b$.
Since the time axes start at the moment when $\Rm$ has become larger than unity, 
$\gamma_b$ quickly drops to zero in all cases, but it keeps
fluctuating in time.
The black lines show the measured growth rates on the integral scale,
$\gamma_\mathrm{int}$, which decreases more slowly than $\gamma_b$ in all runs.
The theoretically expected growth rate of the mean-field dynamo,
$\gamma_\alpha$, is shown as dashed red lines.

In the theoretical curves of $\gamma_\alpha$,
we use the maximum contributions to the dynamo growth rate.
In the case of run 
H2, the maximum contribution comes from
the $\alpha_\mu$ effect for which we use
Eq.~(\ref{eq_alpha5}) with $\langle v_5 \rangle = \eta \langle \mu_5 \rangle$.
Note that here the volume average $\langle \mu_5 \rangle$ is larger
than the average based on the integral scale of turbulence $\langle
\mu_5 \rangle_\mathrm{int}$.
In agreement with previous findings reported in Ref.~\citep{Schober2017},
the $\alpha_\mu$ effect describes the growth rate of the mean-field dynamo
in a system with a constant (homogeneous) initial $\mu_5$ well. 

The mean-field dynamo in all runs with an inhomogeneous initial $\mu_5$
is best described by the magnetic $\alpha$ effect, as given by Eq.~(\ref{eq_alpha_mag_sat}).
This has been discussed in detail for runs S23D and R$-$2
before, and is shown in Fig.~\ref{fig_gamma_t_compact}
for all other runs with high $\Rm$.

In all runs, except for run H2, we have used
$\langle \mu_5 \rangle_\mathrm{int}$ in the
analysis of the mean-field dynamo.
Like for runs S23D [Fig.~\ref{fig_gamma_t_S3D}b] and R$-$2
[Fig.~\ref{fig_gamma_t_R$-$2}b], $\alpha_\mathrm{M}$ is the dominant
contribution in all runs with an inhomogeneous initial $\mu_5$.
In the postprocessing of those runs, we have used
$\alpha_\mathrm{M}^\mathrm{sat}$, taking averages of $\mu_5$ and
$\BB$ on the integral scale of turbulence \footnote{Indeed,
even for run R$-$2m which has an initial nonvanishing component of 
\unexpanded{$\langle \mu_5\rangle$}, 
the average on $k_\mathrm{int}$, 
\unexpanded{$\langle\mu_5\rangle_\mathrm{int}$} 
dominates over the volume average, 
once
turbulence sets in.}, to calculate $\gamma_\alpha$.
For runs S23D, R$-$2m, R$-$2, and R$-$1, the theoretically expected
$\gamma_\alpha$ match the observed growth rate on the integral scale of turbulence, $\gamma_\mathrm{int}$.
For run R$+$1, $\gamma_\alpha$ is much larger than $\gamma_\mathrm{int}$,
yet they seem to vanish at the same time $t\approx0.12$.
This mismatch in R$+$1 is probably due to the low value of the
magnetic Reynolds number which only reaches $\Rm \approx 16$ at its maximum.

\subsection{Coevolution of power spectra}
\label{sec_spectra}

\begin{figure*}
\begin{minipage}[t]{\textwidth}
  \includegraphics[width=0.32\textwidth]{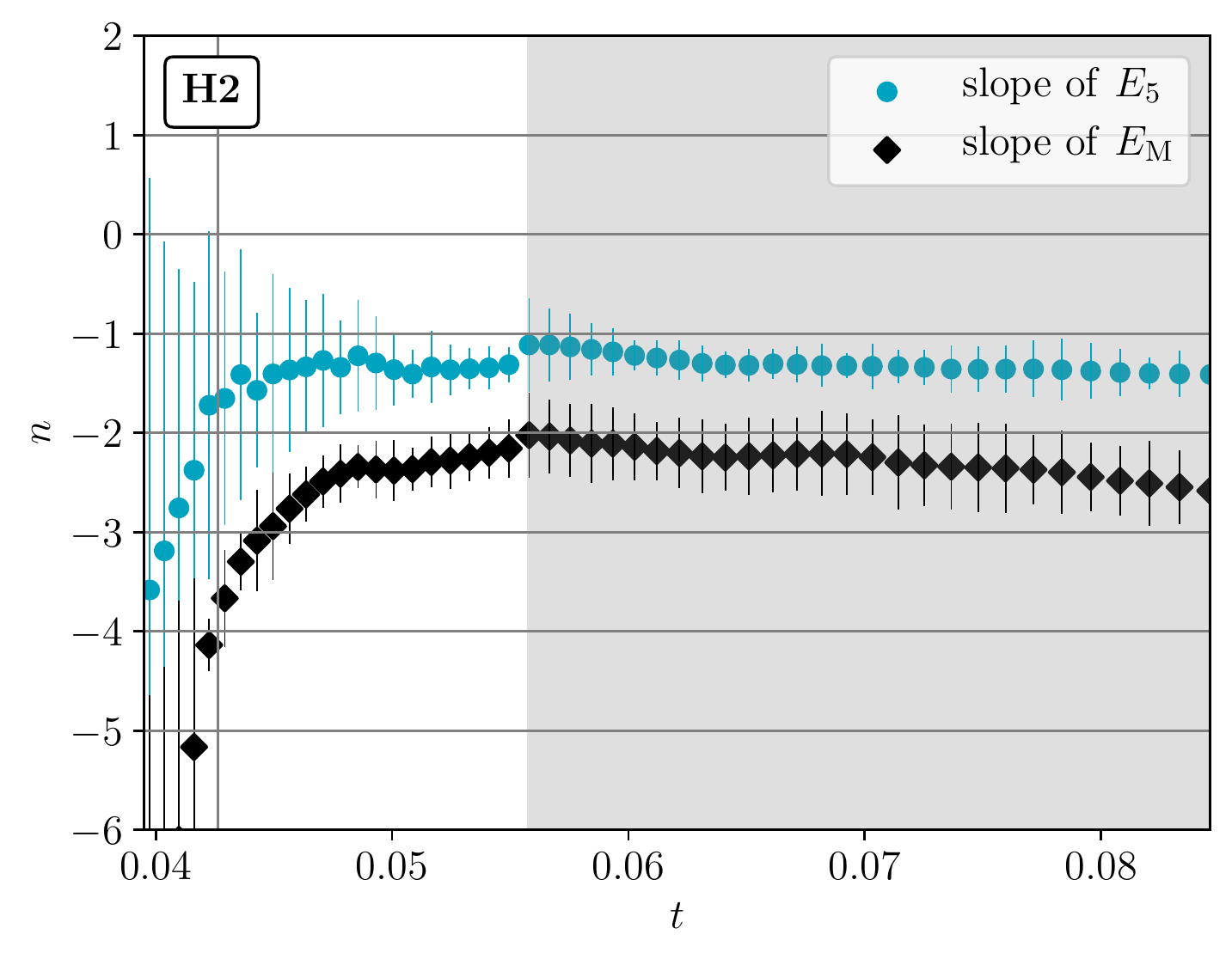}
  \includegraphics[width=0.32\textwidth]{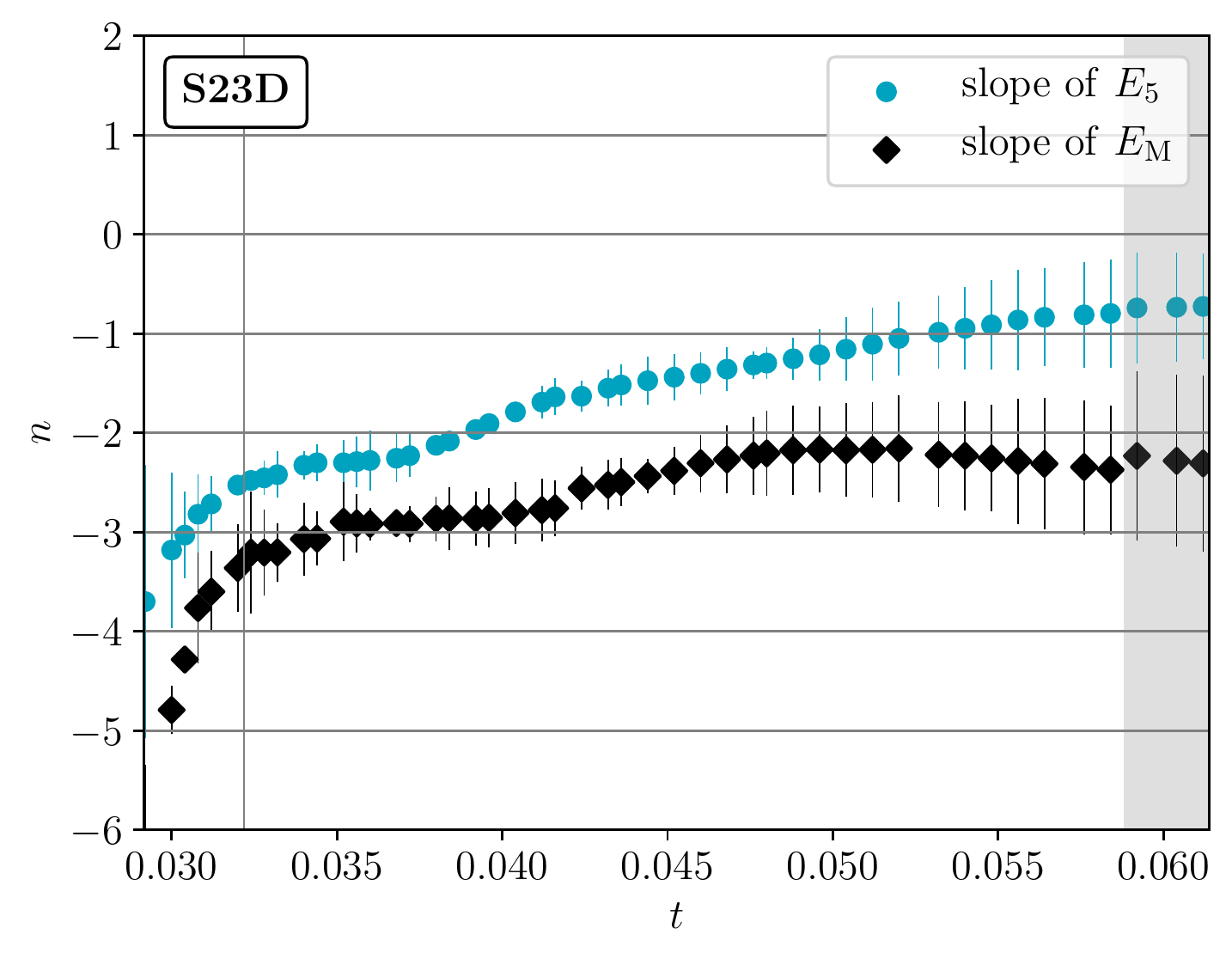} 
  \includegraphics[width=0.32\textwidth]{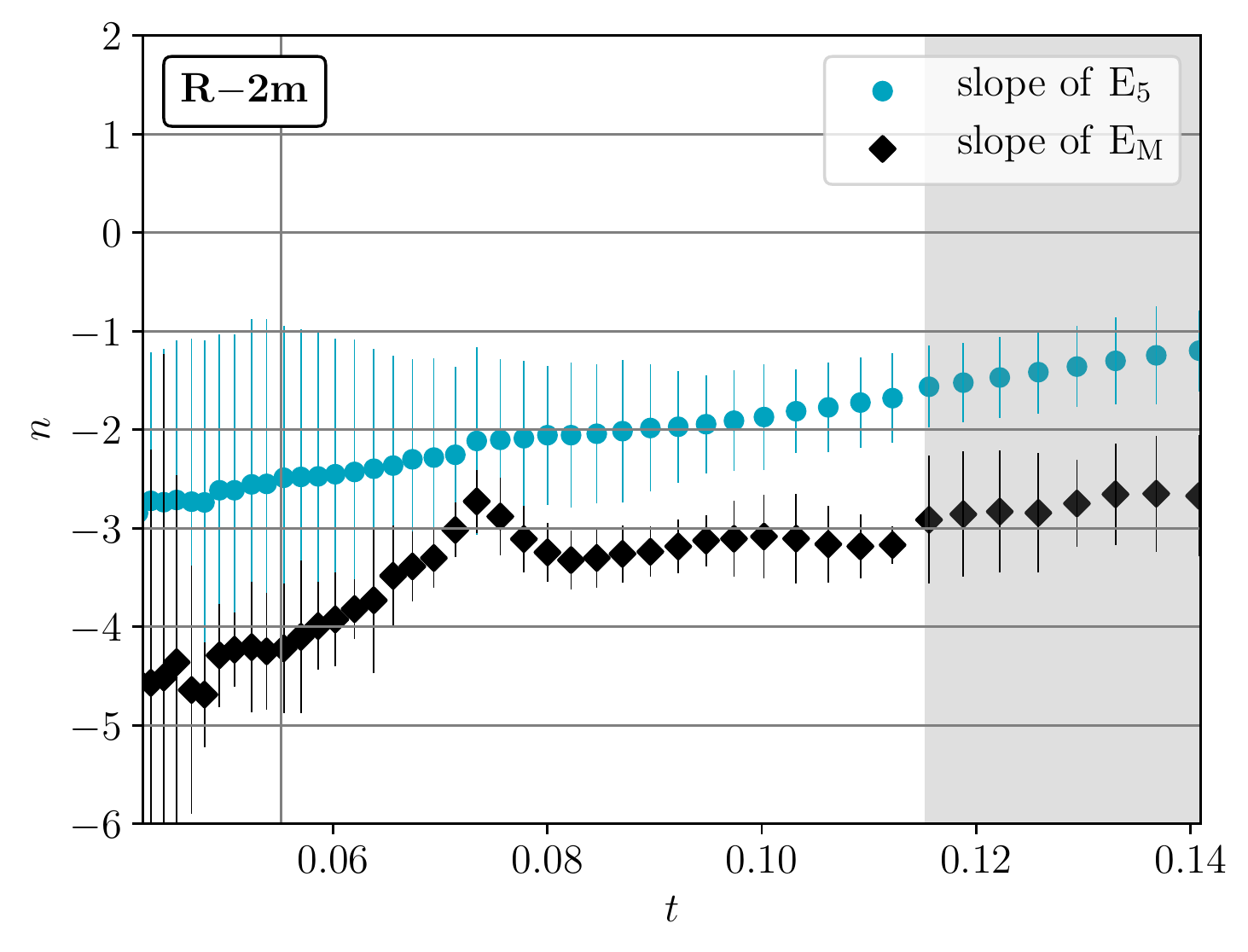} \\
  \includegraphics[width=0.32\textwidth]{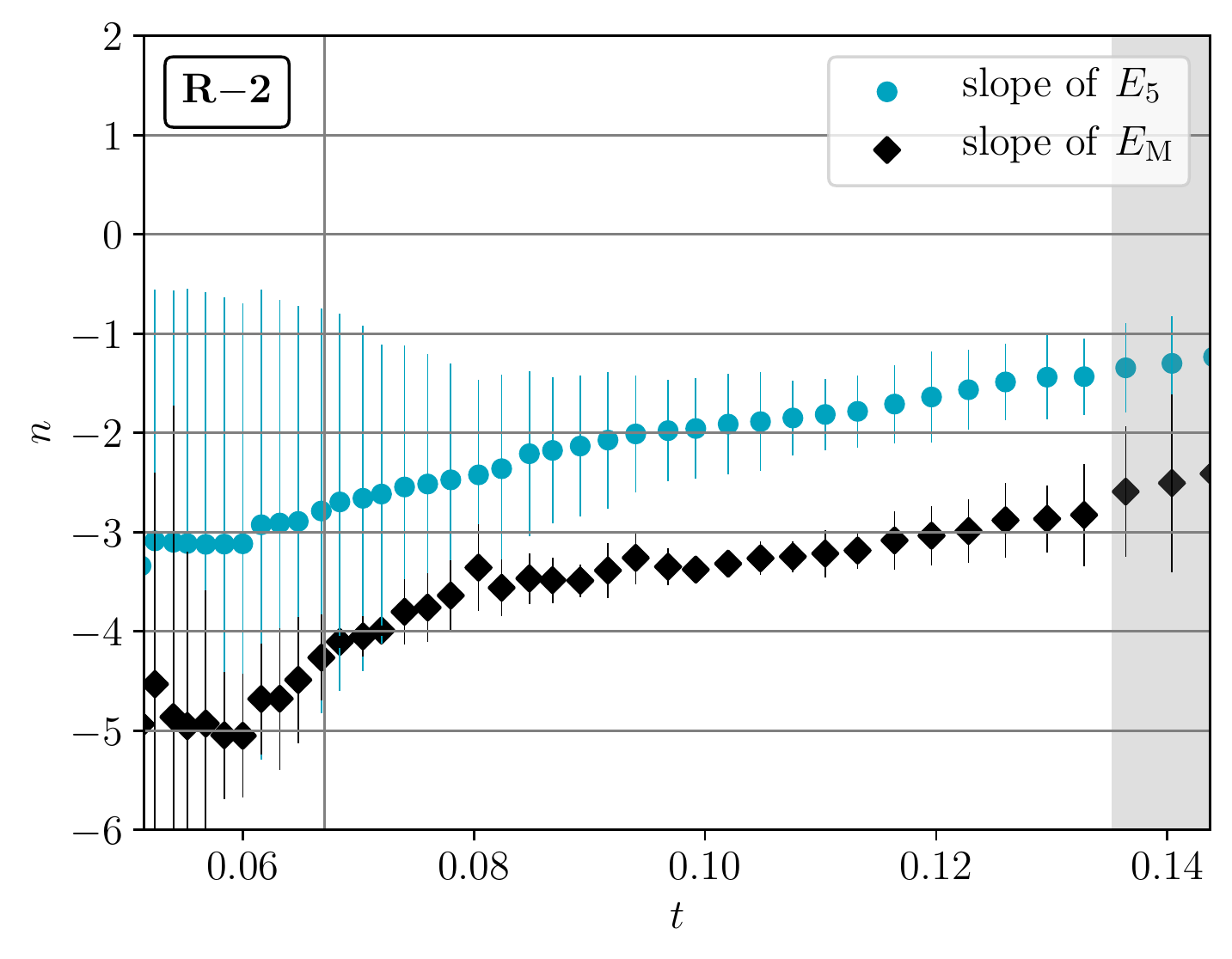} 
  \includegraphics[width=0.32\textwidth]{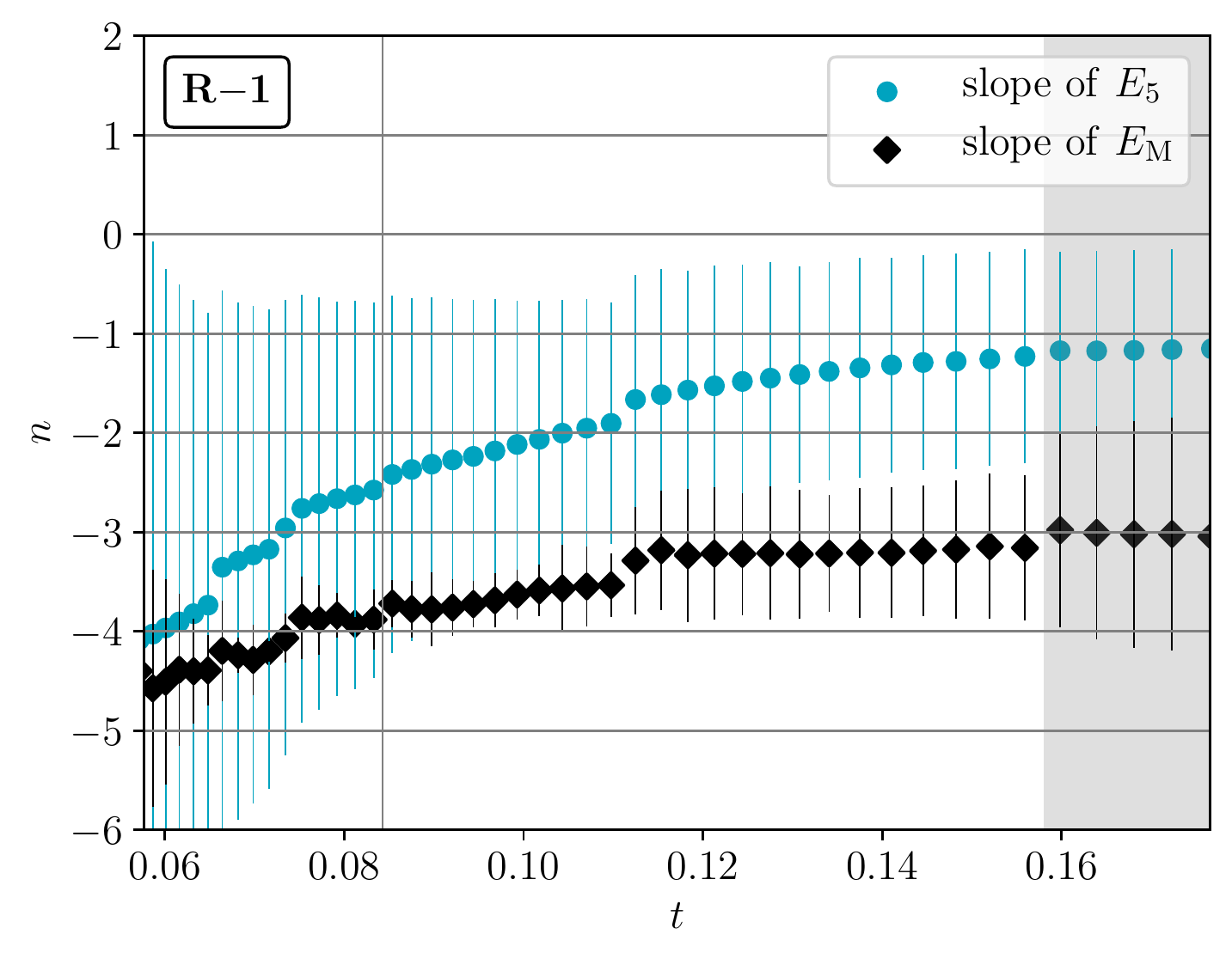}
  \includegraphics[width=0.32\textwidth]{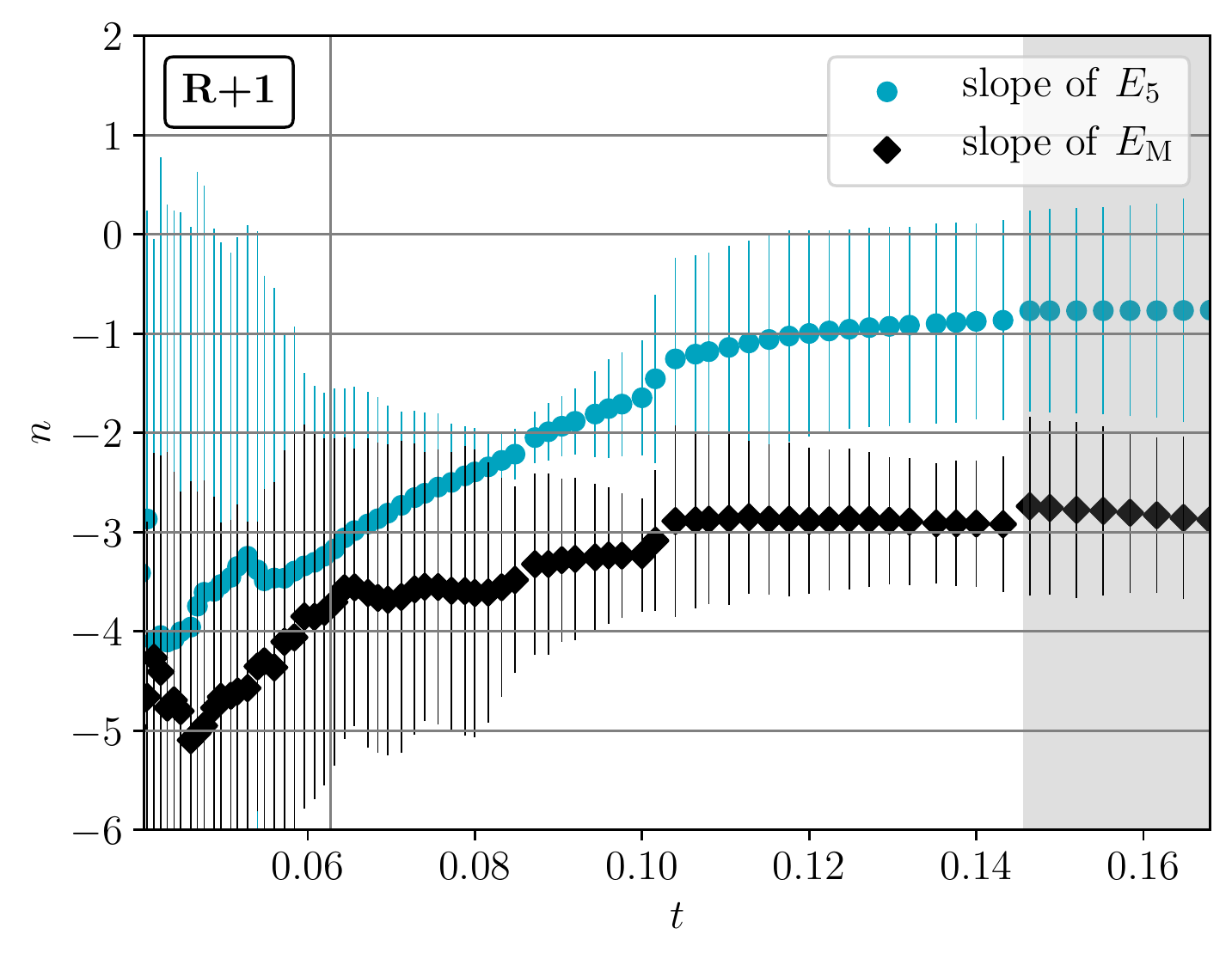}\\
\end{minipage}
\caption{Fitted slopes of the $E_\mathrm{M}$ and $E_5$ spectra as a function of 
time  for all runs with turbulence.  
The time axes start at the moment when $\Rm = 1$ and the solid vertical lines indicate the time when $\Rm = 10$. 
For times larger than $t_{k_1}$, i.e., when the peak of the
magnetic energy spectrum has reached the wave number $k_1$, the plots are greyed out.
}
\label{fig_slopes_t}
\end{figure*}

During the chiral dynamo phase, the power spectra of magnetic energy and 
the chiral chemical potential evolve in an interdependent way.
When the magnetic energy grows for the wave number $k_5$, $E_5$
is also amplified around that wave number.
This can be seen clearly in Fig.~\ref{fig_spec_S23D}, where
$E_\mathrm{M}$ is initially only concentrated at one wave number $k=2$ that coincides with the wave number of the initial sine profile of $\mu_5$.
The amplitude of the sine function, $\mu_{5,\mathrm{max}}=50$, is large
enough to cause an instability in the magnetic energy spectrum at $k=25$.
Figure~\ref{fig_spec_S23D}a shows that also $E_5$ grows at $k=25$
but with a broader peak.
With the onset of the inverse cascade a power-law scaling in
$E_\mathrm{M}$ develops and likewise a power-law slope in $E_5$ is established 
first for $k>2$ and later also for the lowest wave numbers in the system.
In the example of run S23D, we observe a coevolution of the slopes of the 
power spectra $E_\mathrm{M}$ and $E_5$.
Such a simultaneous change of slopes can also be seen for
run R$-$2, the spectra of which are presented in Fig.~\ref{fig_spec_R$-$2}.

To quantify the evolution of the $E_\mathrm{M}$ and $E_5$ spectra,
we determine their slope $n$ by fitting to a power-law $\propto k^n$.
The fits are performed for all spectra after the onset of the inverse
cascade at time $t_\mathrm{IC}$, i.e., once the peak of $E_\mathrm{M}$, $k_\mathrm{p}$, has
become less than $k_5$.
Since the power-law typically extends to wave numbers 
larger than $k_5$,
we set the fitting range at time $t$ 
to $k_\mathrm{p}(t) < k <\ 2 k_\mathrm{p}(t_\mathrm{IC})$,
where $k_\mathrm{p}(t)$ is the wave number on which $E_\mathrm{M}$ has its current maximum 
and $k_\mathrm{p}(t_\mathrm{IC})$ is the wave number at which $E_\mathrm{M}$ had its maximum 
at the onset of the inverse cascade. 
Note, that $k_\mathrm{p}(t_\mathrm{IC})\approx k_5$.
To obtain a typical error, we divide the fitting range into three 
equidistant parts, fit these parts 
separately to obtain three fitting results $n_1$, $n_2$, and $n_3$. 
As the error we use 
$\pm \mathrm{max}(|n_1-\overline{n}|,|n_2-\overline{n}|, |n_3-\overline{n}|)$
with $\overline{n}=(n_1+n_2+n_3)/3$.

The time evolution of the power-law slopes in the $E_\mathrm{M}$
and $E_5$ spectra after the onset of the chiral inverse cascade 
are presented in Fig.~\ref{fig_slopes_t}.
Here, the results for all DNS with sufficiently high Reynolds numbers are shown.
We find that the slopes evolve in an interdependent way for all cases, expect for run R$+$1. 
In this case, the slopes evolve self-similarly  
only after the maximum magnetic field strength has been reached (i.e.~after $t\approx0.1$). 
The reason for this is probably related to the original positive slope
of the $E_5$ spectrum, which requires a longer time for rearrangement
to develop a negative slope 
and to follow the $E_\mathrm{M}$ spectrum. 
The time evolution of the power spectra of run R$+$1 is presented 
in the middle panels of Fig.~\ref{fig_appendix_spec}, along with 
the spectra of run H2 (left panels) and run S203D (right panels),
see Appendix~\ref{appendix C}.

The setup with a random inhomogeneous $\mu_5$ distribution with zero mean 
results in a $\propto k^{-3}$ magnetic energy scaling, which is different 
from the case with a homogeneous $\mu_5$ distribution, where the scaling is $\propto k^{-2}$.
The two setups are rather different and have very different underlying physics.
The principal difference is the following.
In the linear stage of the chiral dynamo instability,
an initially homogeneous $\mu_5$ excites a magnetic field with a wave number whose value is around the average of $\mu_5/2$
[see Fig.~\ref{fig_appendix_spec}b],
while a random $\mu_5$ with zero mean excites a random magnetic field over a 
broad range of scales [see e.g.~Fig.~\ref{fig_spec_R$-$2}b].
In the nonlinear stage, there is an inverse cascade of the magnetic field and a magnetic driving of turbulence in both setups.
However, the properties of turbulence in both systems are distinct from each other, as discussed next.

One of the indications of the difference in these systems 
is that there are two different mechanisms of generation of
a mean-field dynamo in the resulting turbulent flows:
(i) in the case of an initial homogeneous $\mu_5$,
it is the $\alpha_\mu$ effect related to the interactions of fluctuations of $\mu_5$ and tangling magnetic fluctuations;
(ii) in the case of a random $\mu_5$ with zero mean,
it is the magnetic $\alpha$ effect, which is caused by
the current helicity of small-scale magnetic fluctuations.
Both types of $\alpha$ effect are caused by the produced turbulence with different properties in both systems, resulting in different magnetic spectra in the final stage of the magnetic field evolution in these systems.

\subsection{Maximum field strength}
\label{sec_saturation}

The observed $k^{-3}$ scaling of the magnetic energy spectra allows to estimate the maximum magnetic field strength. 
Assuming that it is controlled by $\eta$ and $\mu_{5,\mathrm{max}}(t_0)$,
dimensional arguments imply that the magnetic energy spectrum is given by
\begin{eqnarray}
   E_\mathrm{M}(k) = C \meanrho \eta^2 \mu_{5,\mathrm{max}}(t_0)^4 k^{-3},
\end{eqnarray}
where $C$ is a constant, and $\meanrho=1$ in our DNS.
We use $C=1$ for an order-of-magnitude estimate. 
The DNS indicate that the maximum value of $E_\mathrm{M}(k)$ is typically 
reached at the wave number $k\approx k_{\mu_5,\mathrm{eff}}(t_0)$ and 
therefore the maximum possible magnetic field is given by
\begin{eqnarray}
  B_\mathrm{sat,eff} \approx  \sqrt{2} \, \eta \frac{\mu_{5,\mathrm{max}}(t_0)^2}{k_{\mu_5,\mathrm{eff}}(t_0)}.
\label{eq_Bsateff}
\end{eqnarray}
For Eq.~(\ref{eq_Bsateff}) it is assumed that $k_{\mu_5,\mathrm{eff}}$ does not 
change significantly during the dynamo instability.
This is only a valid assumption if $\lambda$, i.e., the coupling between $\mu_5$ and 
$\BB$, is small. 
For large values of $\lambda$, there is a strong backreaction on the $\mu_5$ field
and the dynamo limitation occurs through the same mechanism as observed in the DNS of \citep{BSRKBFRK17},
i.e., by means of the conservation law:
\begin{eqnarray}
  B_{\mathrm{sat}, \lambda} \approx \frac{\mu_{5,\mathrm{max}}(t_0)}{\sqrt{\lambda}}.
\label{eq_Bsatlambda}
\end{eqnarray}

\begin{figure}
  \includegraphics[width=0.45\textwidth]{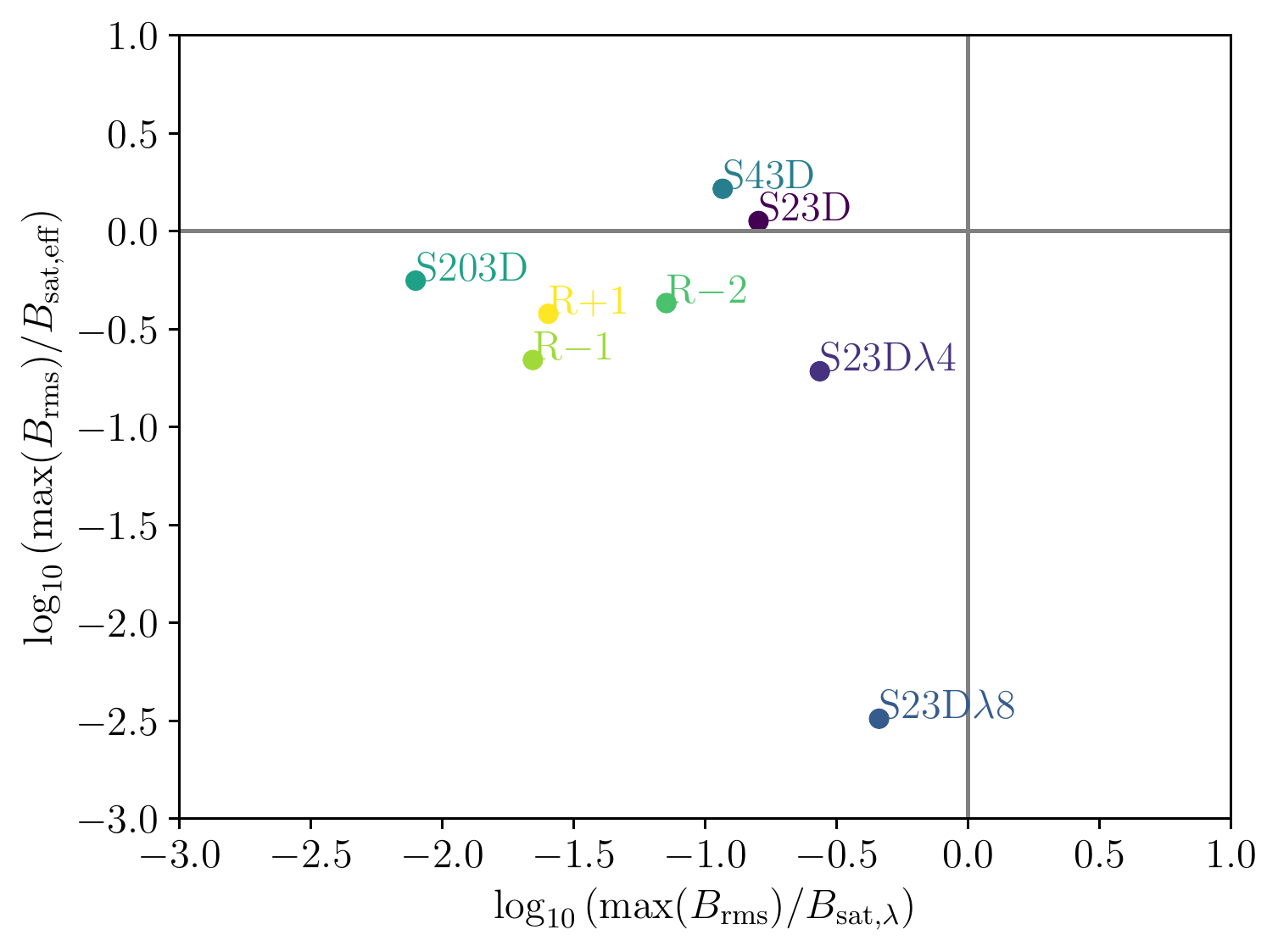}
\caption{Maximum magnetic field strength obtained in DNS over the $B_\mathrm{sat,eff}$ vs.\
maximum magnetic field strength obtained in DNS over the $B_{\mathrm{sat}, \lambda}$.}
\label{fig_BmaxBtheomax}
\end{figure}

In Fig.~\ref{fig_BmaxBtheomax} we compare the maximum value of the 
rms magnetic field strength in the two phenomenological estimates given 
in Eqs.~(\ref{eq_Bsateff}) and (\ref{eq_Bsatlambda}).
The limitation mechanism via the conservation law plays role for runs S23D$\lambda4$ and S23D$\lambda8$, while in the remaining runs, dynamo limitation
is controlled by the initial correlation length of $\mu_5$.

\subsection{Effects of chiral magnetic waves}
\label{sec_mu}

In this study, we have focused on scenarios where the dynamics is driven by the CME. 
However, there is also the chiral separation effect that describes the 
coupling between $\mu_5$ and the chemical potential $\mu = \mu_\mathrm{R} + \mu_\mathrm{L}$. 
In the presence 
of an equilibrium mean magnetic field ${\bm B}_0$,
a nonzero $\mu$ permits chiral magnetic waves (CMWs) \cite{KY11}
with the frequency
\begin{eqnarray}
   \omega_{\rm CMW} = (C_5 \, C_\mu)^{1/2} \left|{\bm k} {\bm 
\cdot} {\bm B}_0 \right| ,
\label{eq_CMW}
\end{eqnarray}
where $C_5$ and $C_\mu$ are coupling constants.
The behavior of CMWs for an initial nonuniform random $\mu_5$
has not yet been studied.
To test the effects of CMWs on the scenario of chiral plasma instabilities driven by a nonuniform $\mu_5$,
we perform two additional simulations that take the coupling to $\mu$ into account.
Therefore, Eq.~(\ref{mu-DNS}) is replaced by
\begin{eqnarray}
  \frac{{\mathrm D} \mu_5}{{\mathrm D} t} &=& \mathscr{D}_5(\mu_5)
  + \lambda \, \eta \, \left[{\BB} {\bm \cdot} (\nab   \times   {\BB})
  - \mu_5 {\BB}^2\right]
  - C_5 ({\BB} {\bm \cdot} \nab) \mu,  \nonumber \\
\label{mu5-DNS-2} 
\end{eqnarray}
which we solve together with Eqs.~(\ref{ind-DNS})--(\ref{rho-DNS})
and the evolution equation for the chemical potential
\begin{eqnarray}
  \frac{{\mathrm D} \mu}{{\mathrm D} t}  &=& D_\mu \, \Delta \mu - C_\mu (\BB {\bm \cdot} \nab)  \mu_5 .
\label{mu-DNS}
\end{eqnarray} 
We repeat run R$-$2 with the additional $\mu$ dynamics. 
As an initial condition for $\mu$ we use a uniform value of $\mu=51$, which corresponds
roughly to the initial maximum value of $\mu_5$. 
This initial condition implies that in grid cells where
$\mu_5=51$, 
all fermions have the same handedness.
For the coupling constants we use $C_5=C_\mu=0.1$ in run R-2\_CMW1,
which implies that the velocity of the CMW is roughly ten percent of
the Alfv\'en velocity.
For run~R-2\_CMW2, we use $C_5=C_\mu=1$, so the velocity of the CMW is
approximately equal to the Alfv\'en velocity.
We note that for run~R-2\_CMW2 we have used shock viscosity 
during the nonlinear phase for numerical stability.
This means that we add a bulk viscosity
$\zeta=C_{\rm shock}\delta x^2\bra{\max(0,-\nab {\bm \cdot} \UU)}$
to the stress tensor so that
$\tau_{ij}=2\nu \rho {\sf S}_{ij}+\rho\zeta\delta_{ij}\nab {\bm \cdot} \UU$.
Here, angled brackets denote a five-point running average.
The technique of shock viscosity was developed by von Neumann and Richtmyer 
\cite{vNR50}; see Ref.~\cite{QianEtAl2020} for an application to simulations of detonations with the {\sc Pencil Code}.

\begin{figure}
  \includegraphics[width=0.45\textwidth]{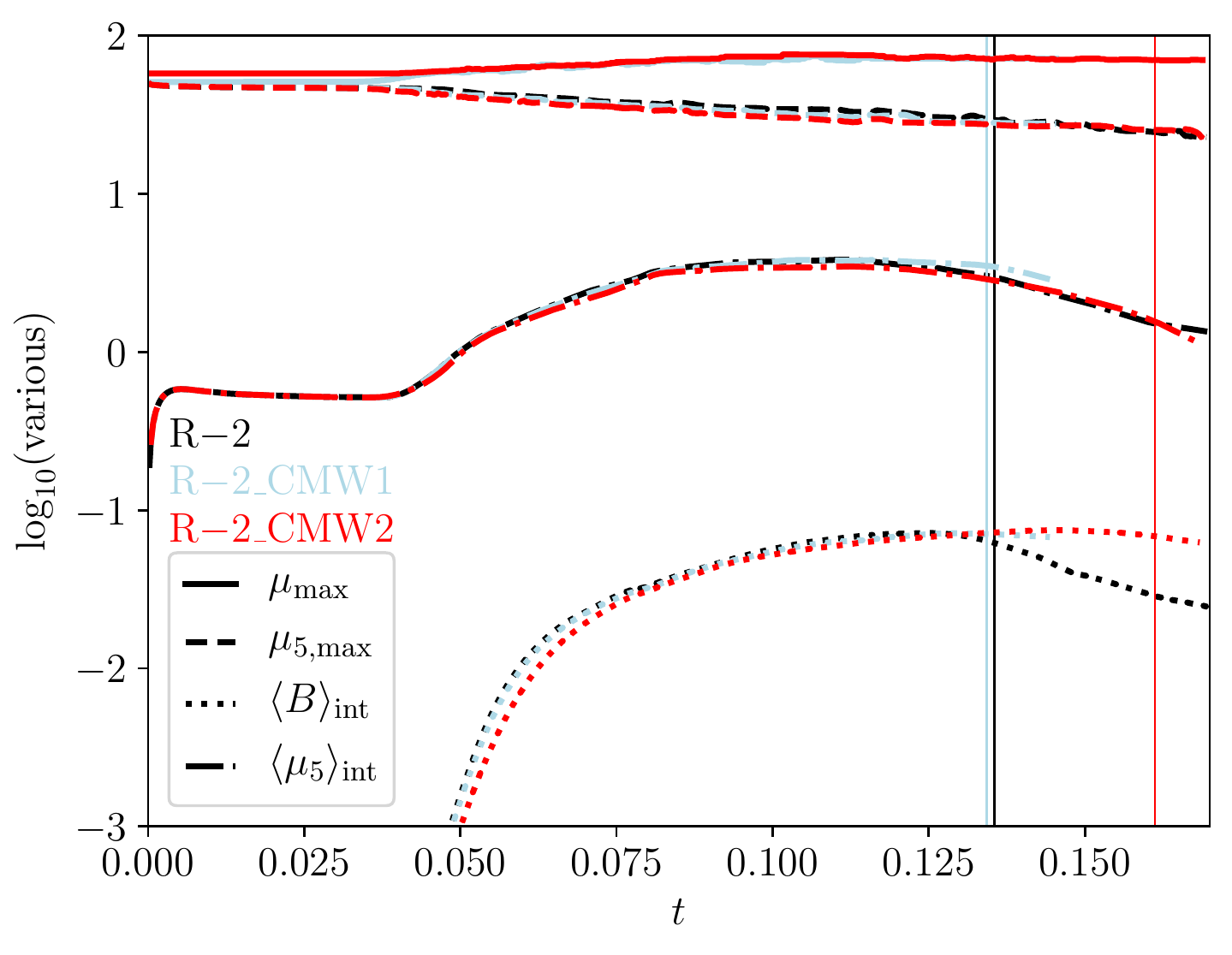}
\caption{Comparison of the time evolution of runs without (run R$-$2)
and with (runs R$-$2\_CMW1 and R$-$2\_CMW2)
the evolution of the chemical potential. 
The dynamics in all three runs are very comparable. 
Small differences can only be seen at late times, 
when a large enough $\langle B \rangle_\mathrm{int}$
has been produced that leads to chiral magnetic waves.
The thin vertical lines with colors referring to the different runs
indicate the time $t_{k_1}$.
}
\label{fig_ts_CMW}
\end{figure}

\begin{figure}
  \includegraphics[width=0.45\textwidth]{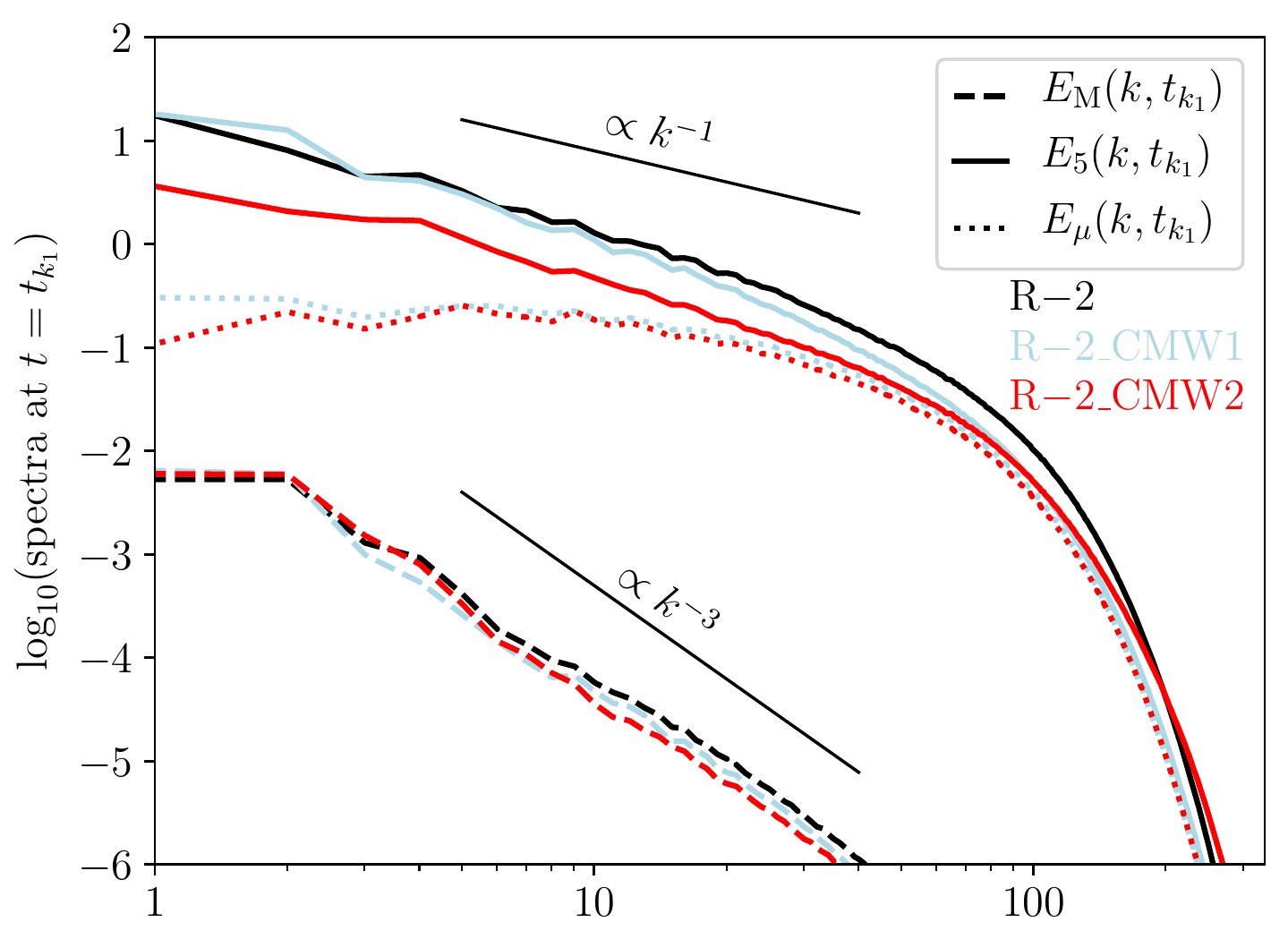}
\caption{
Comparison of different energy spectra at time $t=t_{k_1}$ in runs R$-$2,
R$-$2\_CMW1, and R$-$2\_CMW2.
Scaling relations of $E_\mathrm{M}$ and $E_5$ 
seem not to be affected for the produced 
chiral magnetic waves. 
The power spectrum of the chemical potential, $E_\mu$
which is defined in the same way as $E_5$,
approaches $E_5$ at high $k$.
But $ E_\mu\ll E_5$ at small $k$ at the time $t_{k_1}$. 
}
\label{fig_spec_t1_CMW}
\end{figure}

Our two exemplary simulations with chiral magnetic waves show that they
do not alter the dynamics of the systems presented in this work (see Figs.~\ref{fig_ts_CMW}-\ref{fig_spec_t1_CMW}).
The main reason is that these systems do not have an external magnetic
field. 
Therefore, waves can only develop at late phases of the simulations.
However, both runs, R-2\_CMW1 and R-2\_CMW2, do not show significant 
differences to run R-2 without $\mu$.
As can be seen in Fig.~\ref{fig_ts_CMW}, the maximum value of $\mu_5$
decreases a bit faster when CMWs occur.
Yet this does not affect the production of the mean magnetic field
$\langle B \rangle_\mathrm{int}$ significantly. 
In all three cases, $\langle B \rangle_\mathrm{int}$ grows up to
approximately $0.1$ by the time the inverse cascade 
reaches the minimum wave number of the numerical domain.
Throughout the simulations, the maximum value of $\mu$, 
$\mu_\mathrm{max}$ continuously grows in time. 
In Fig.~\ref{fig_spec_t1_CMW}, we demonstrate that also the magnetic
energy spectra and the $\mu_5$ spectra,
$E_\mathrm{M}$ and $E_5$, at the time $t_{k_1}$ are not 
significantly affected by the presence of CMWs.
For the runs with $\mu$ evolution, the $\mu$ spectra, $E_\mu$,
are comparable with $E_5$ at high wave numbers, while 
they are significantly lower at low wave numbers.
For R-2\_CMW2, the $E_5$ spectrum at $t_{k_1}$ has the same 
scaling of $\propto k^{-1}$ but its amplitude is almost an 
order-of-magnitude less than the ones in runs R-2
and R-2\_CMW1.
This may be related to the additional shock viscosity in
R-2\_CMW2.

\bigskip

\section{Conclusion}
\label{sec_conclusion}

In this paper we have analyzed various dynamo instabilities
that are sourced by an initial inhomogeneous distribution of the chiral chemical potential. 
To this end, we performed DNS of chiral MHD with the \textsc{Pencil Code}.
While the existence of chiral dynamo instabilities has been confirmed
with DNS before, most previous studies have assumed  
a uniform distribution of $\mu_5$.
In this paper we performed a detailed study of dynamo instabilities
caused by an inhomogeneous $\mu_5$, which clarifies and supports the
findings presented in Ref.~\citep{SRB21a}, in particular the buildup
of a mean $\mu_5$ and the occurrence
of a mean-field dynamo.
To test the necessary conditions for a small-scale chiral dynamo, we
have used a 2D toy model in which $\mu_5$ was initialized with a sine
function along one direction.
Its wave number was varied to explore the effect
of the effective correlation wave number $k_{\mu_5,\mathrm{eff}}$.
We have demonstrated that the small-scale chiral dynamo can operate if
$k_{\mu_5,\mathrm{eff}} < k_5$; see Fig.\ \ref{fig_singlesin_comp},
where $k_5$ is the wave number based on
the scale of the maximum growth rate of the small-scale chiral dynamo
instability based on the maximum value of $\mu_5$.
With larger scale separation, the measured growth rate of the rms magnetic
field approaches the maximum possible value
$\gamma_5 = \eta \mu_{5,\mathrm{max}}^2/4$ in the system.
Saturation of the dynamo occurs
once the fluctuations of the chiral chemical potential, $\mu_5'$, 
experience a backreaction from $B_\mathrm{rms}$, leading to a change of
the characteristic scale $k_{\mu_5,\mathrm{eff}}$.
When $k_{\mu_5,\mathrm{eff}}$ becomes comparable to $k_5$, the growth of
the magnetic field stops; see Fig.\ \ref{fig_singlesin_saturation_time}.

Another main focus of this work was a detailed analysis of the DNS with initial fluctuations of $\mu_5$ 
with zero mean described shortly in Ref.~\cite{SRB21a}.
In all of our DNS that develop turbulence, i.e., which reach sufficiently
large $\Rm$, we could confirm the presence of a mean-field dynamo;
see Fig.~\ref{fig_gamma_t_compact}.
Contrary to the previously studied case of homogeneous $\mu_5$ where
the mean-field dynamo is dominated by the $\alpha_\mu$ effect that is
related to fluctuations of $\mu_5$ itself, for inhomogeneous $\mu_5$
the magnetic $\alpha$ effect, $\alpha_\mathrm{M}$, 
related to the current helicity plays 
the central role in the mean-field dynamo phase
(see e.g., Fig.~\ref{fig_gamma_t_R$-$2}). 
The main reason for this effect is the additional source of
current helicity, $2 \meanv_5 \overline{{\bm b}^2}$, caused by magnetic 
fluctuations produced by inhomogeneities of $\mu_5$.
Note that in this study we had to use 
the average based on the integral scale of turbulence
which increases during the nonlinear evolution of the system.

Finally, we reported a tight connection between the evolution of the
power spectra of magnetic energy, $E_\mathrm{M}(k)$, and that of the chiral chemical potential,
$E_5(k)$; see Fig.\ \ref{fig_slopes_t}.
With the onset of turbulence, independently of their initial shape,
both power spectra develop a power-law scaling with a negative index.
Specifically, the $E_5(k)$ spectra approach a universal scaling proportional to $k^{-1}$.
For our reference run with homogeneous $\mu_5$, $E_\mathrm{M}$
approaches a $k^{-2}$ scaling, which is consistent with the results of Ref.~\citep{BSRKBFRK17}.
In the runs with an inhomogeneous 
initial $\mu_5$, a slightly steeper scaling of $k^{-3}$ develops, except
for the run with an initial $\mu_5$ in the form of a sine wave (S23D), 
where the spectrum is closer to $E_5 \propto k^{-2}$.

Our results can be employed in models of primordial plasmas.
Several models of the early Universe, e.g., specific scenarios of inflation
or cosmological phase transitions, predict the production
of primordial magnetic fields which should evolve according to the 
laws of chiral MHD as long as the temperature is $>~10~\mathrm{MeV}$.
Detailed models of the evolution of the primordial magnetic fields
are needed, if it is to be used to constrain fundamental physics at 
the time before recombination.

\begin{acknowledgements}
We have benefited from stimulating discussions with
Nathan Kleeorin and Abhijit B.\ Bendre.
J.S.~acknowledges the support by the Swiss National Science Foundation under Grant No.\ 185863.
A.B.~was supported in part through a grant from the Swedish Research Council
(Vetenskapsr{\aa}det, 2019-04234).
\end{acknowledgements}

\appendix


\section{Comparison between $-\nabla^4$ and $\nabla^2$ diffusion of $\mu_5$}
\label{sec_diffusion}
In direct numerical simulations of chiral MHD, large discretization 
errors cause phase errors in the 
advection of the high wave number contributions to $\mu_5$.
Therefore, dissipation of $\mu_5$ on small spatial scales is required.
In our previous work (e.g., Refs.~\citep{BSRKBFRK17,Schober2017}), 
where we considered an initially uniform $\mu_5$, 
the $\nabla^2$ diffusion 
never affected the evolution of $\mu_5$ significantly.
In this study, however, we consider cases where $\mu_5$ is concentrated at large 
wave numbers and therefore is affected by $\nabla^2$ diffusion. 
A $\nabla^2$ diffusion ($-k^2$ in Fourier space) constantly reduces the value of $\mu_5$ at moderately high $k$ and thereby the effects of a
fluctuating chiral chemical potential on the magnetic field.
To prevent this loss of $\mu_5$ before it can be converted into magnetic helicity, we have implemented a $-\nabla^4$ diffusion
($-k^4$ in Fourier space) that mostly acts on the highest wave numbers of the numerical domain where it is needed for numerical stability.
This allows us to study the effects of a $\mu_5$ at moderately high $k$.

\begin{figure*}
  \includegraphics[width=0.8\textwidth]{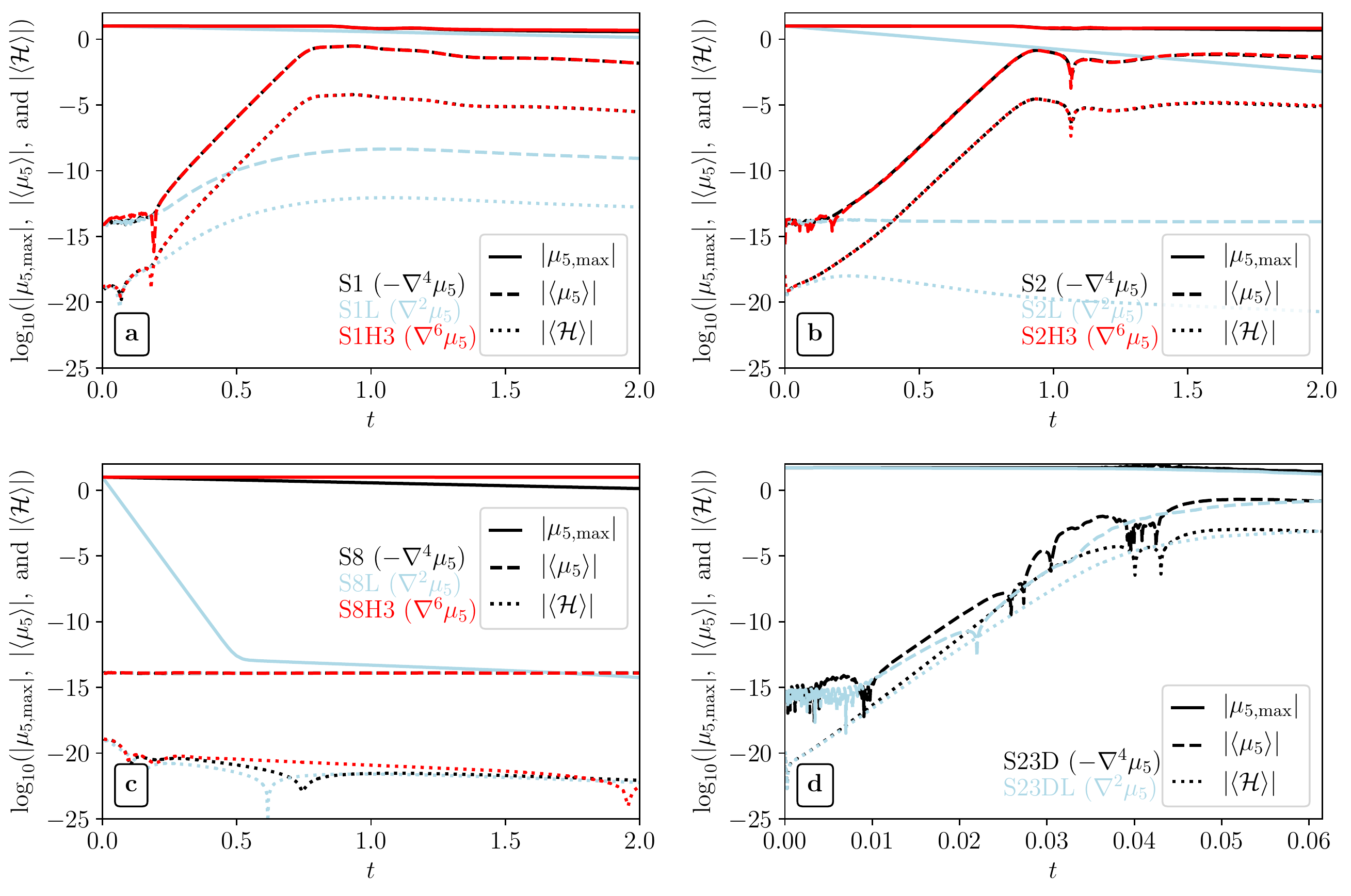}
\caption{Comparison between runs with the default
hyperdiffusion ($\propto -\nabla^4 \mu_5$, black line color) and 
Laplacian diffusion ($\propto \nabla^2 \mu_5$, blue line color) and 
third-order hyperdiffusion ($\propto \nabla^6 \mu_5$, red line color).
For each run, the time evolution of $|\mu_\mathrm{5,max}|$ (solid lines), 
$|\langle\mu_\mathrm{5}\rangle|$ (dashed lines), and 
$|\langle\mathcal{H}\rangle|$ (dotted lines) is shown.
\textit{(a)} Runs S1, S1L, and S1H3.
\textit{(b)} Runs S2, S2L, and S2H3.
\textit{(c)} Runs S8, S8L, and S8H3.
\textit{(d)} Runs S23D and S23DL.
}
\label{fig_appendix_diffusion}
\end{figure*}

In Fig.~\ref{fig_appendix_diffusion}, we present the difference between
the default second-order hyperdiffusion, Laplacian diffusion, and third 
order hyperdiffusion for selected runs.
Runs S1, S2, S8, and S23D have been repeated with Laplacian diffusion
(runs S1L, S2L, S8L, and S23DL) and we have additionally tested third
order hyperdiffusion for runs S1, S2, S8 (runs S1H3, S2H3, and S8H3).
Laplacian diffusion strongly affects an inhomogeneous
$\mu_5$, especially if its initial inverse correlation length is large in comparison to the Nyquist wave number $k_\mathrm{Ny}$.
In particular, $|\mu_\mathrm{5,max}|$ decreases faster the closer the wave number is to $k_\mathrm{Ny}$; compare the solid lines in
Figs.~\ref{fig_appendix_diffusion}a--c. 
With faster decreasing $|\mu_\mathrm{5,max}|$, the chiral dynamo
instability phase is shorter and less efficient or in extreme cases not even present when hyperdiffusion is replaced by Laplacian diffusion; see Fig.~\ref{fig_appendix_diffusion}b. 
Third-order hyperdiffusion results in very similar dynamics for the runs
with an initial sine wave for $k=1$ and $k=2$ (S1 vs.~S1H3 and S2 vs.~S2H3)
and there is only a small difference between S8 and S8H3.
For the high-resolution 3D run, S23D, the initial characteristic wave number of the dynamo instability ($k\approx25$) is much smaller than the Nyquist wave number ($k\approx336$). 
Therefore, the difference between second-order hyperdiffusion (run S23D) and Laplacian diffusion (S23DL) is noticeable but not very significant; see Fig.~\ref{fig_appendix_diffusion}d.

\section{Snapshots of run R$-$2}

\begin{figure*}
  \includegraphics[width=0.32\textwidth]{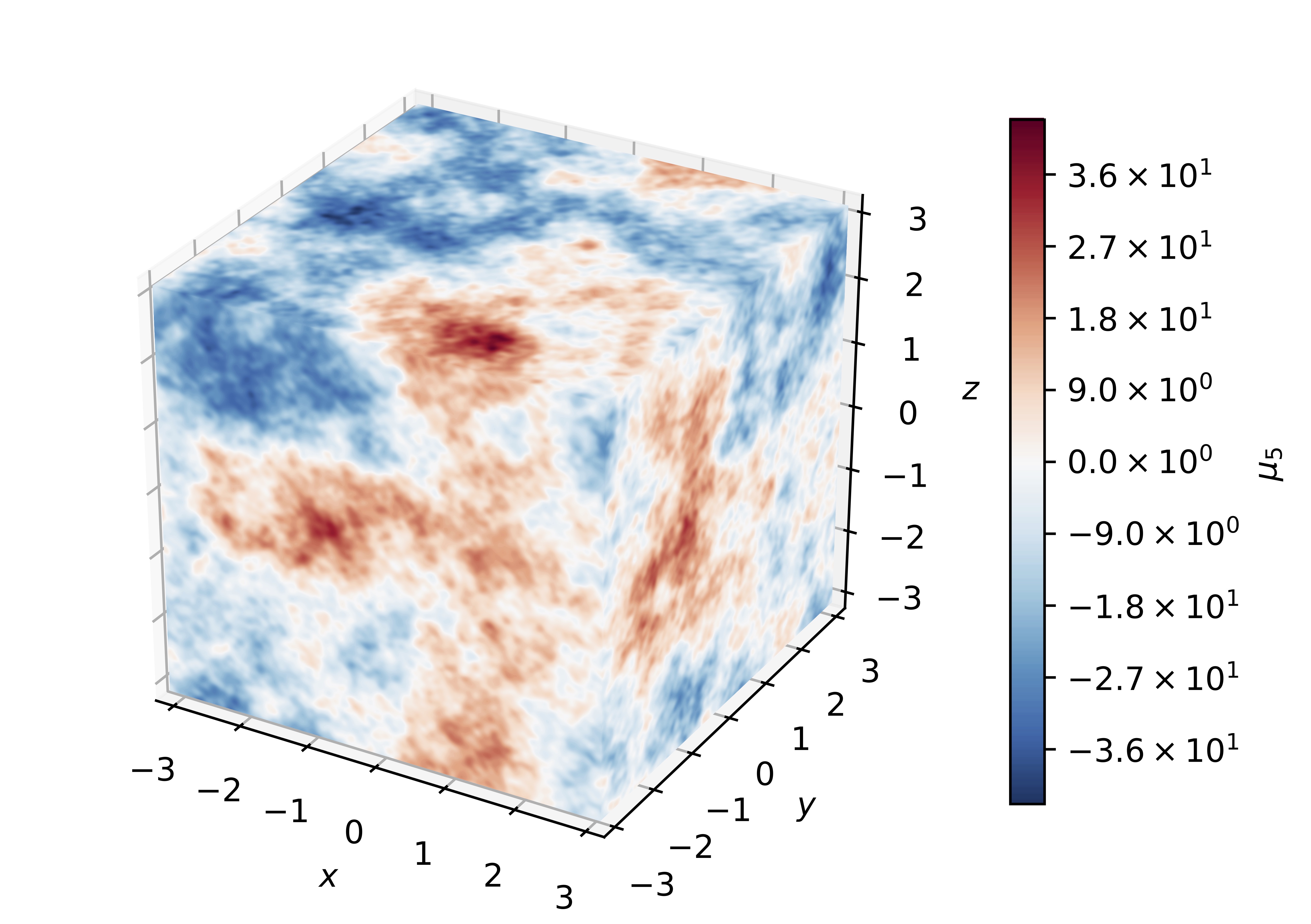}
  \includegraphics[width=0.32\textwidth]{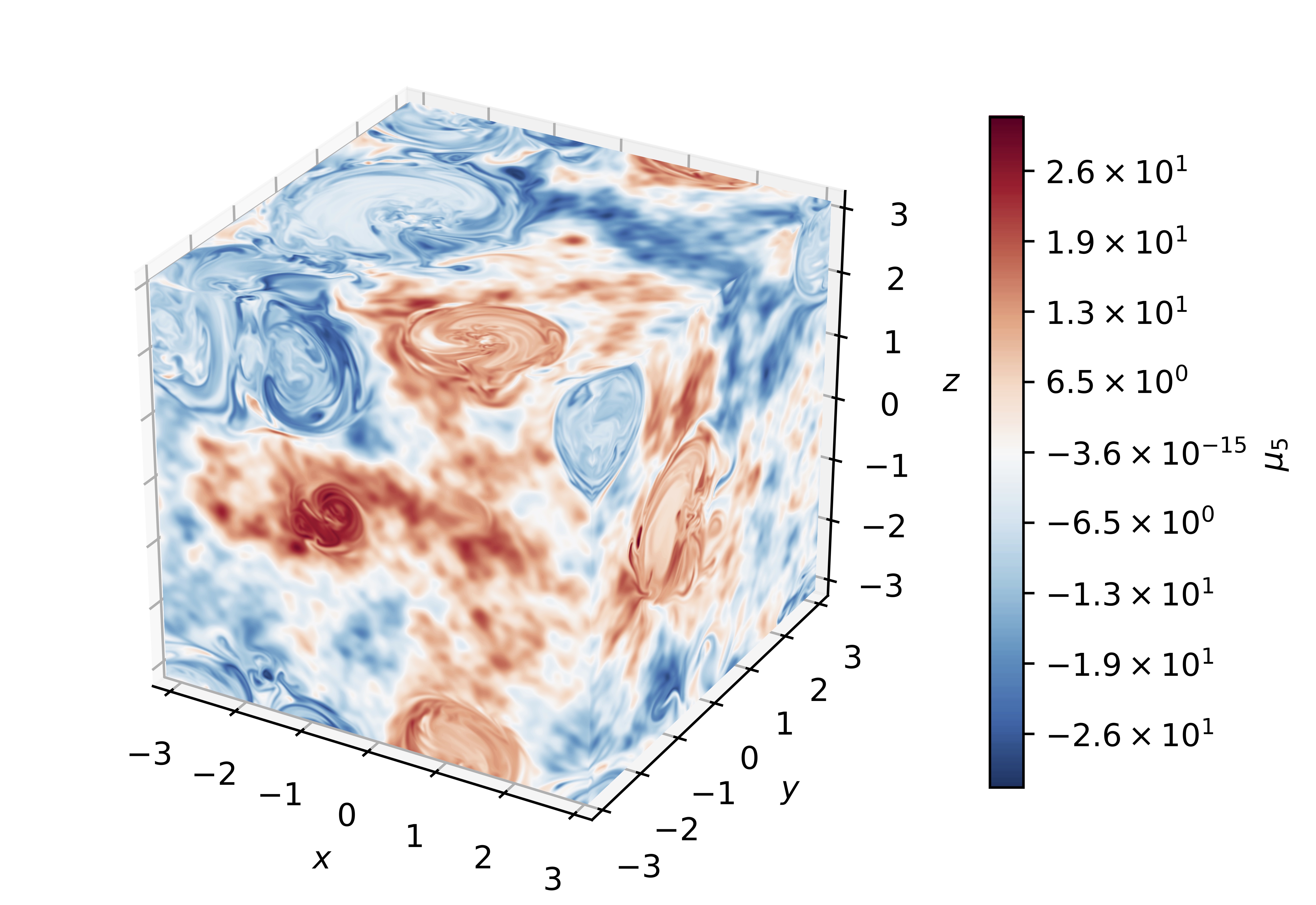}
  \includegraphics[width=0.32\textwidth]{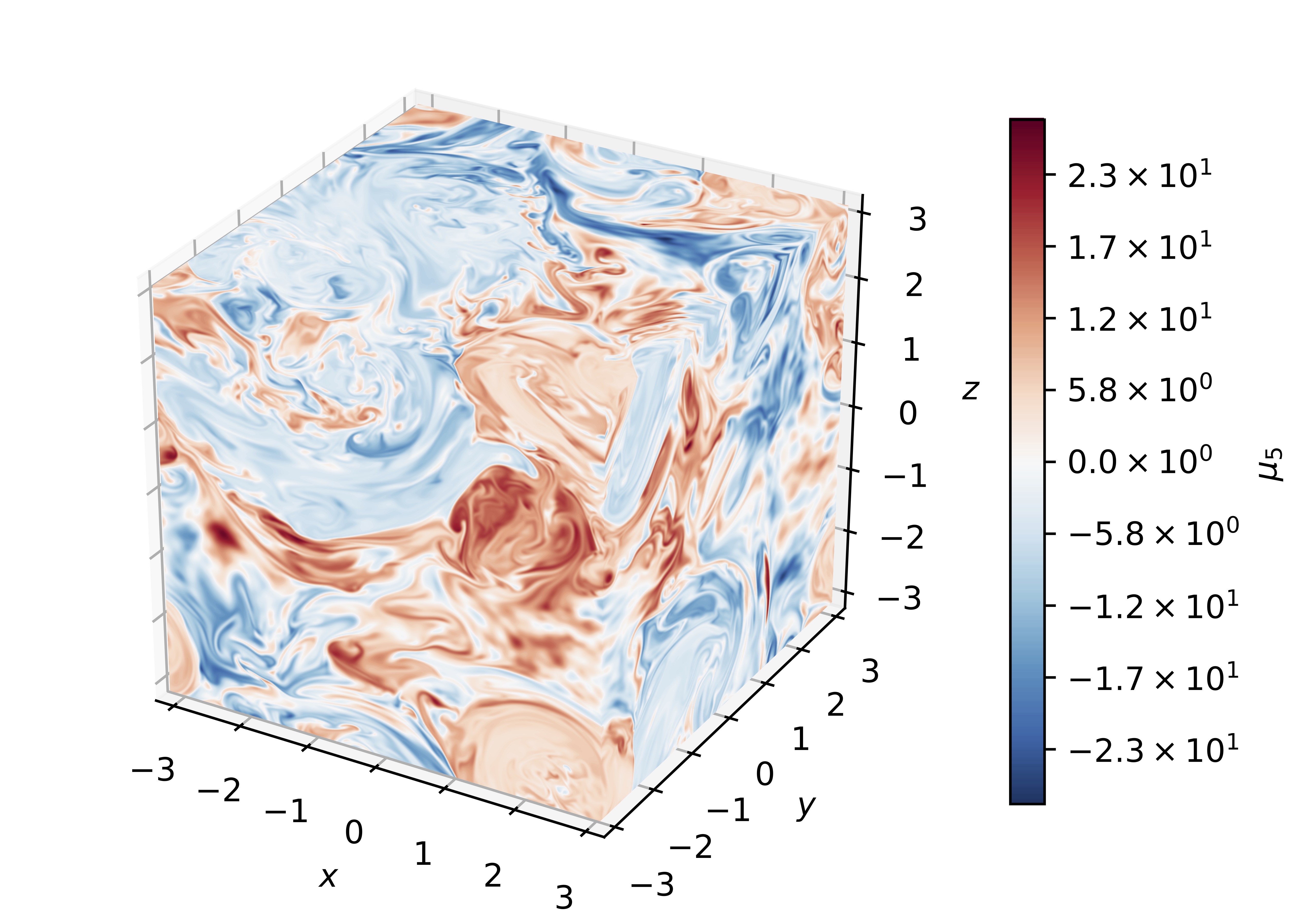} \\
  \includegraphics[width=0.32\textwidth]{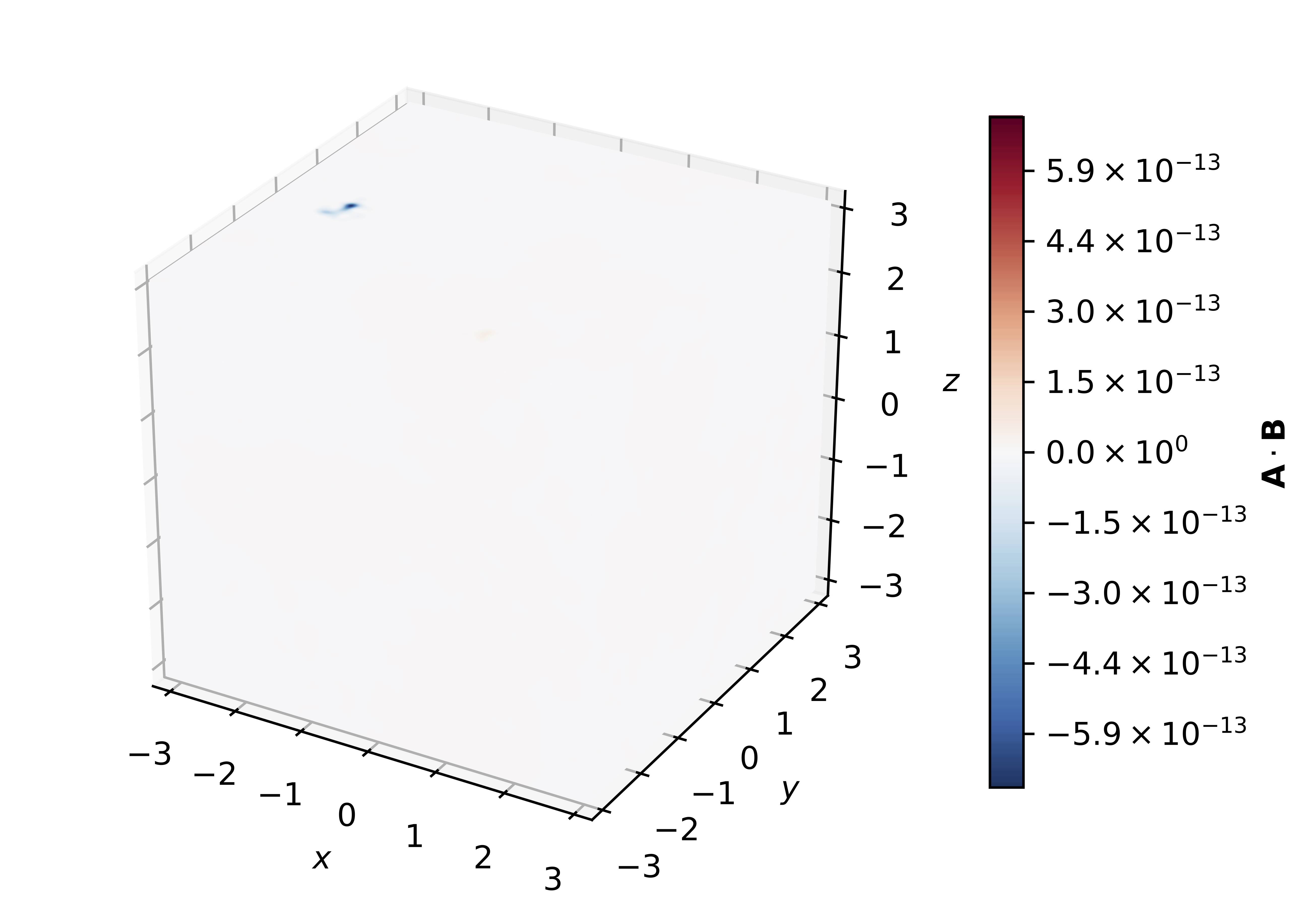}
  \includegraphics[width=0.32\textwidth]{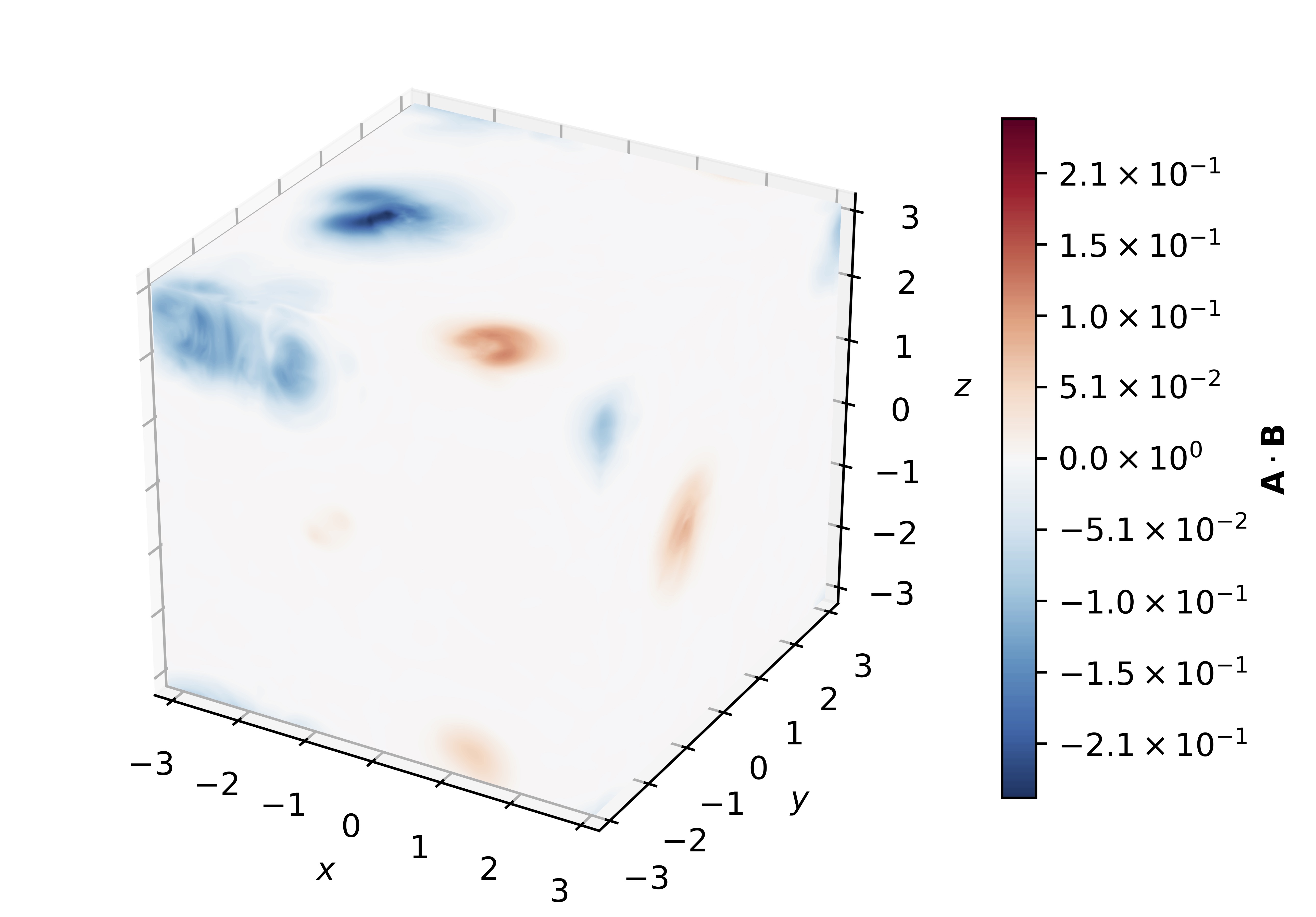}
  \includegraphics[width=0.32\textwidth]{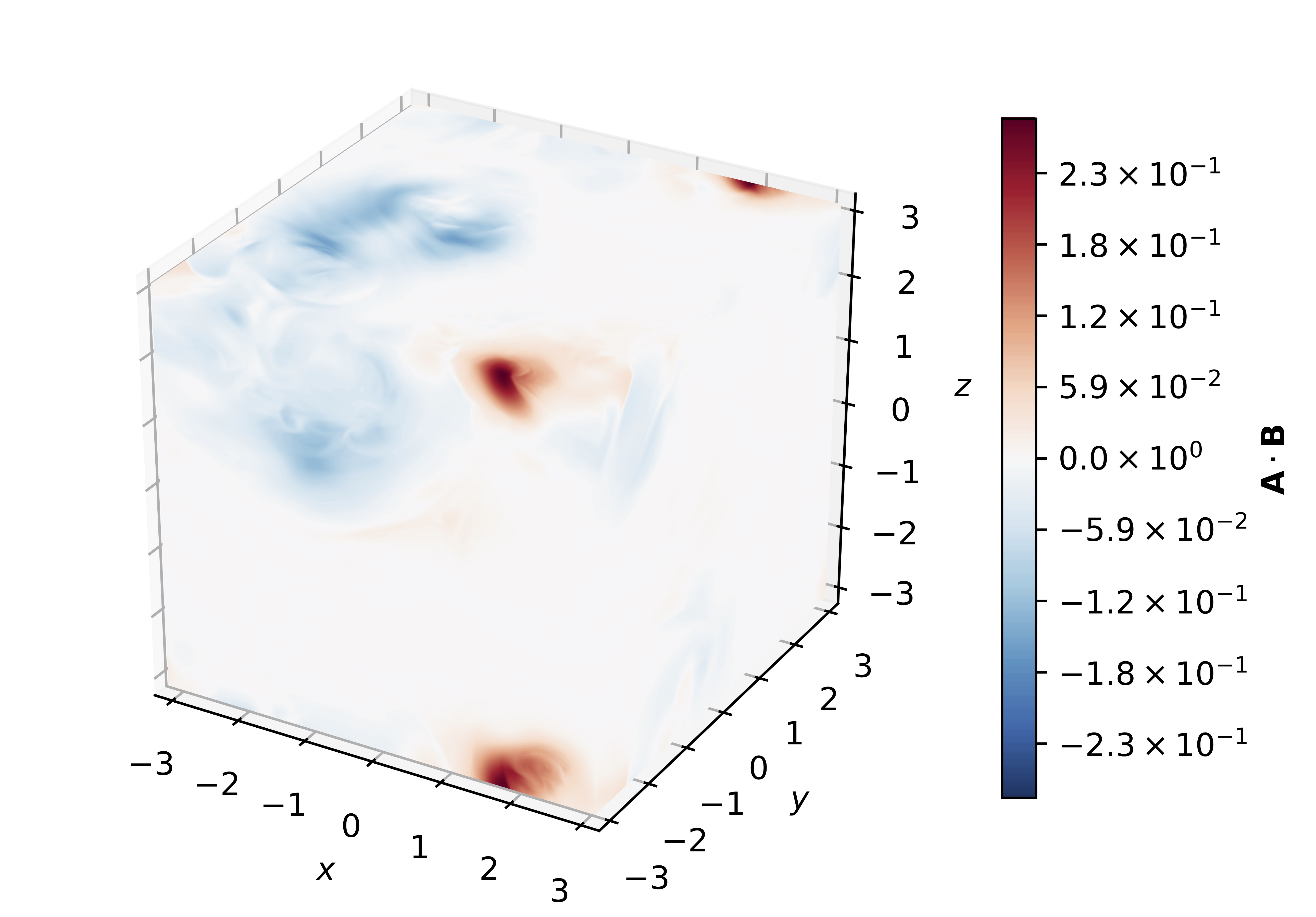} \\
  \includegraphics[width=0.32\textwidth]{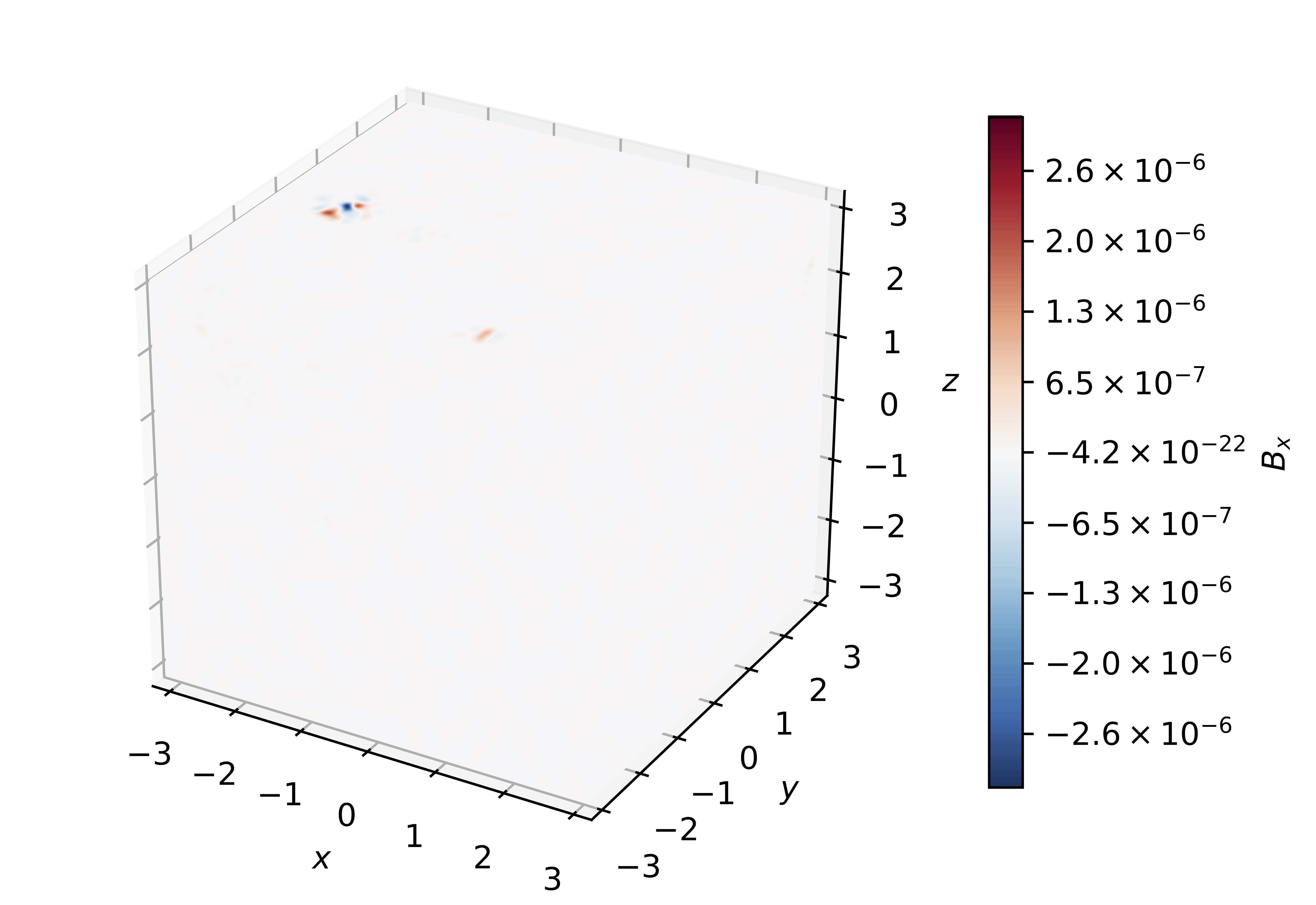}
  \includegraphics[width=0.32\textwidth]{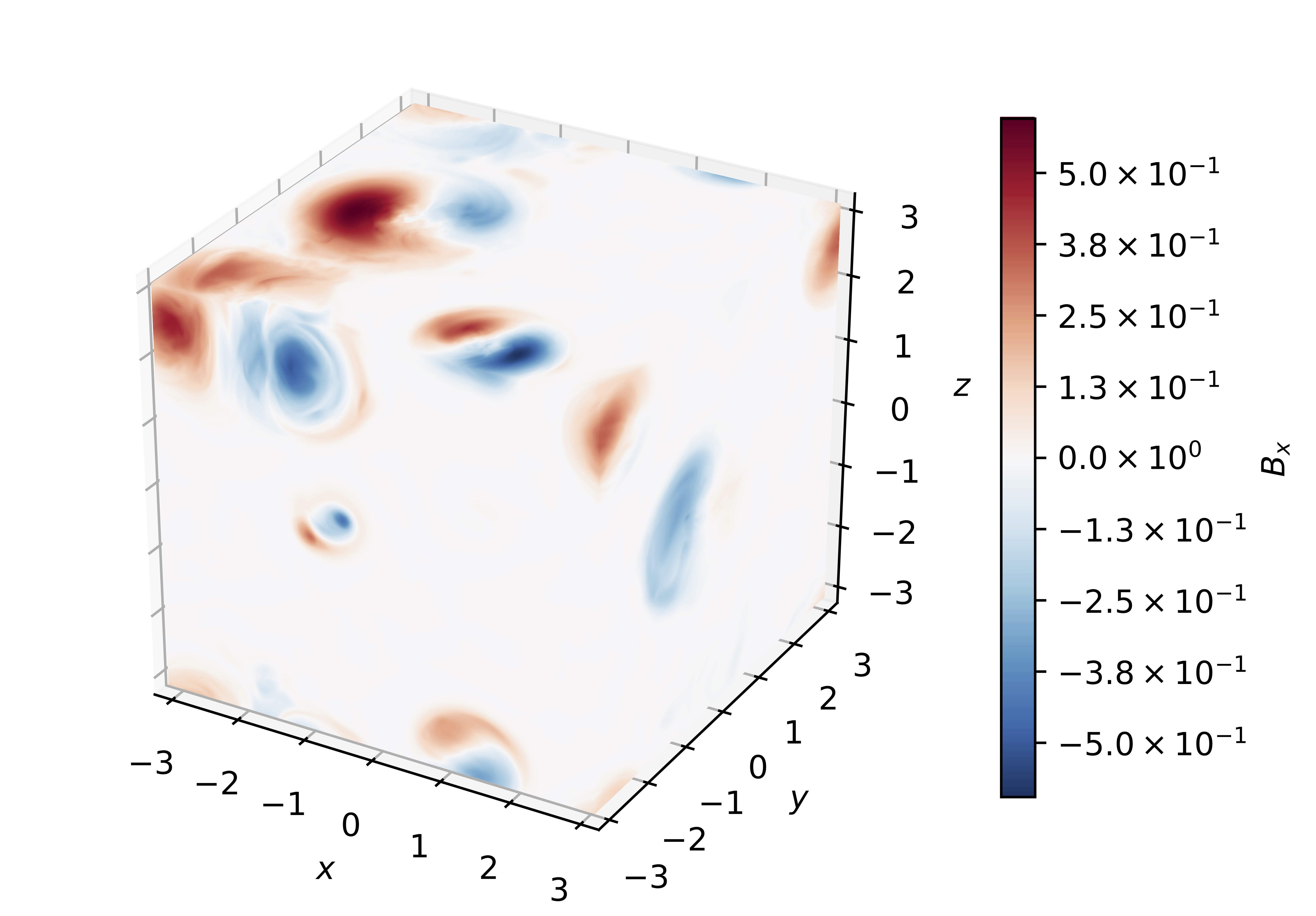}
  \includegraphics[width=0.32\textwidth]{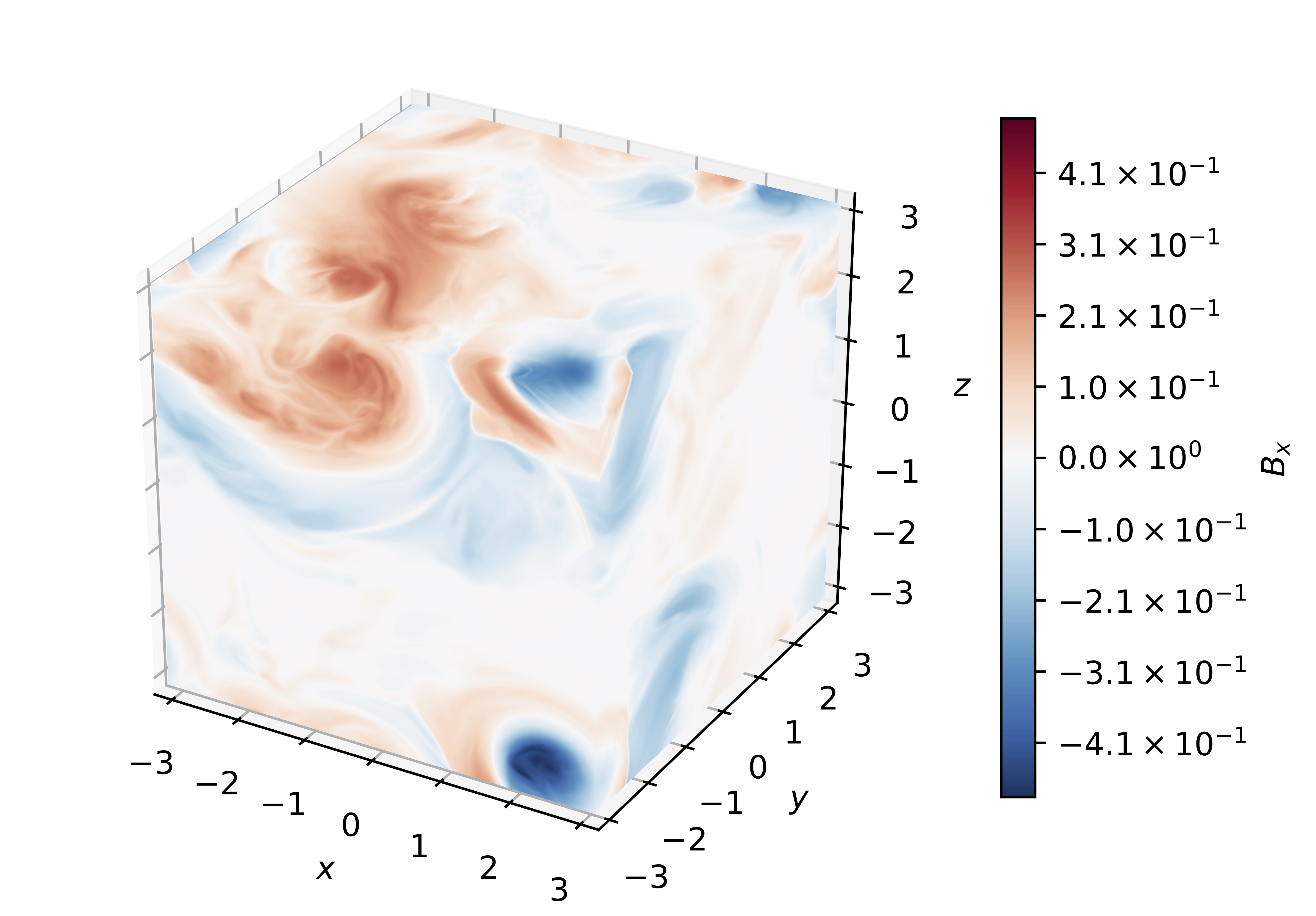} \\
\caption{Snapshots of run R$-$2 taken at the different times: during the
$v_5$ dynamo phase ($t = 0.02$, left), the mean-field dynamo phase
($t = 0.1$, middle), and at the time when the inverse cascade reaches
the scale of the domain ($t = 0.135$, right).
}
\label{fig_appendix_cubes_R-2}
\end{figure*}
In Sec.~\ref{DNSFluctuations} we discuss the time evolution
of run R$-$2, starting from the initial small-scale chiral instability to the
amplification of the magnetic field on large scales at late times.
In addition to the quantitative analysis there, we present 
in Fig.~\ref{fig_appendix_cubes_R-2} the snapshots of run R$-$2.
The values of $\mu_5$, $\AAA\cdot\BB$, and $B_x$ on the 
surfaces of the domain are shown at different times. 

\section{Time evolution of power spectra in runs H2, R$+$1, and S203D}
\label{appendix C}

\begin{figure*}
  \includegraphics[width=0.32\textwidth]{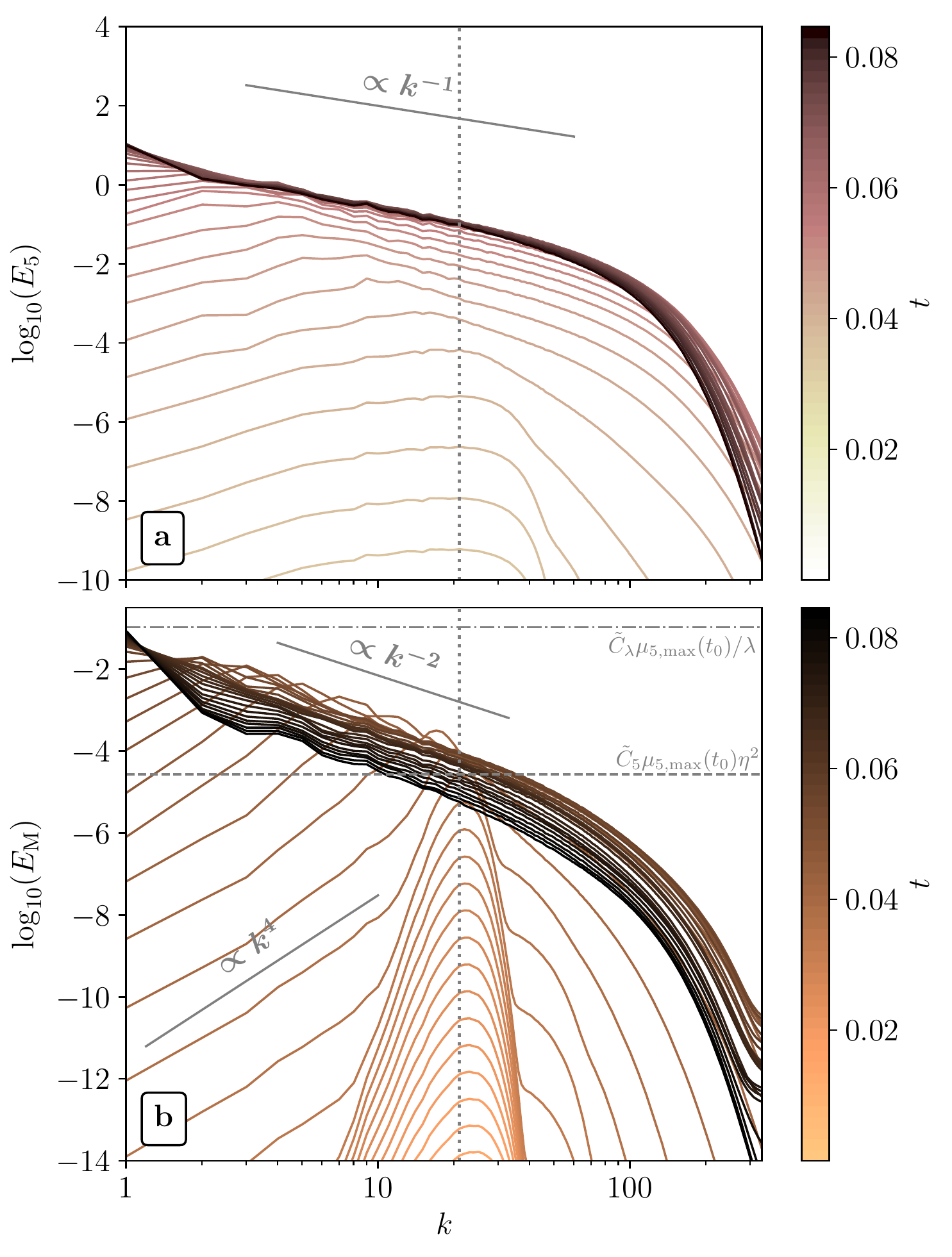}
  \includegraphics[width=0.32\textwidth]{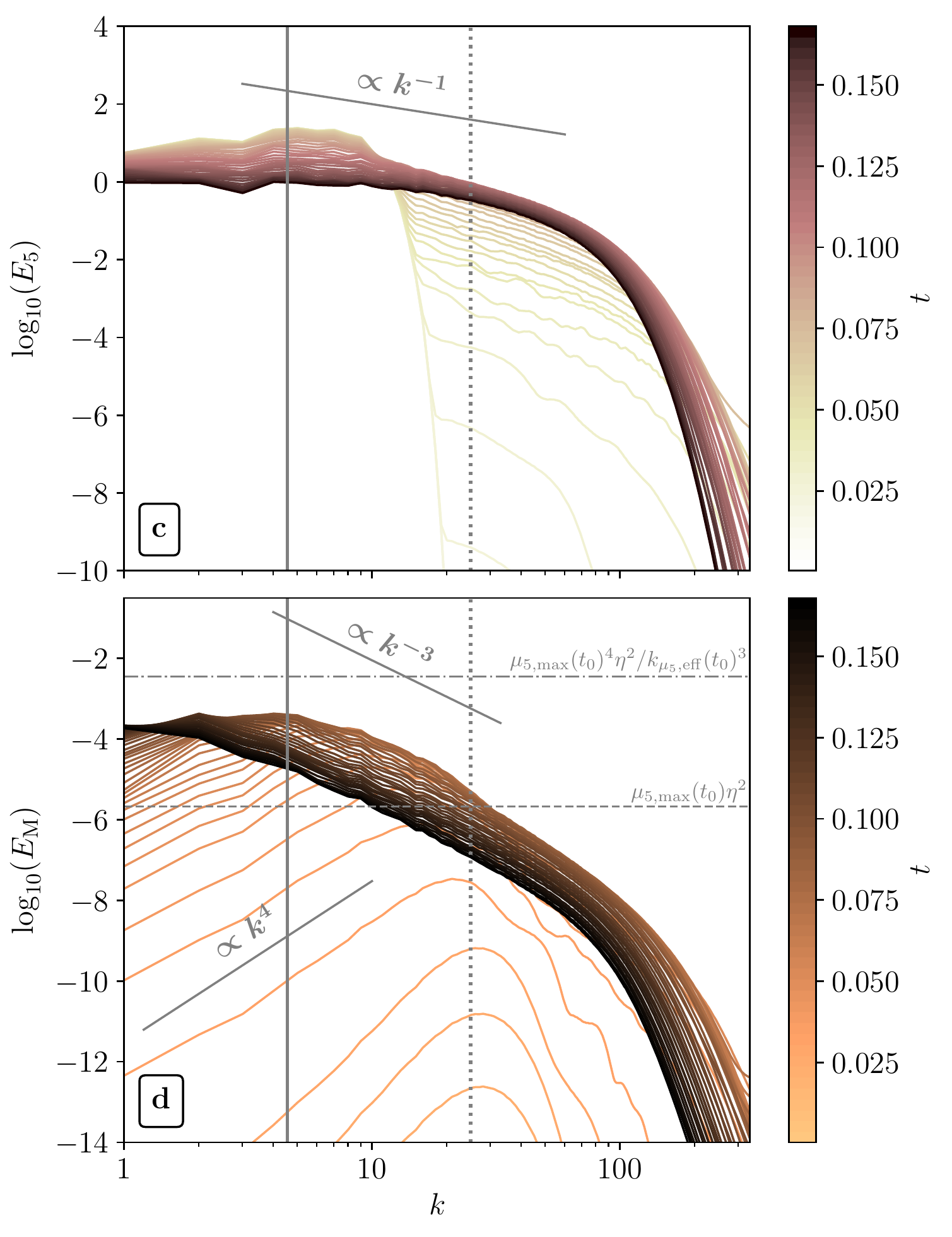} 
  \includegraphics[width=0.32\textwidth]{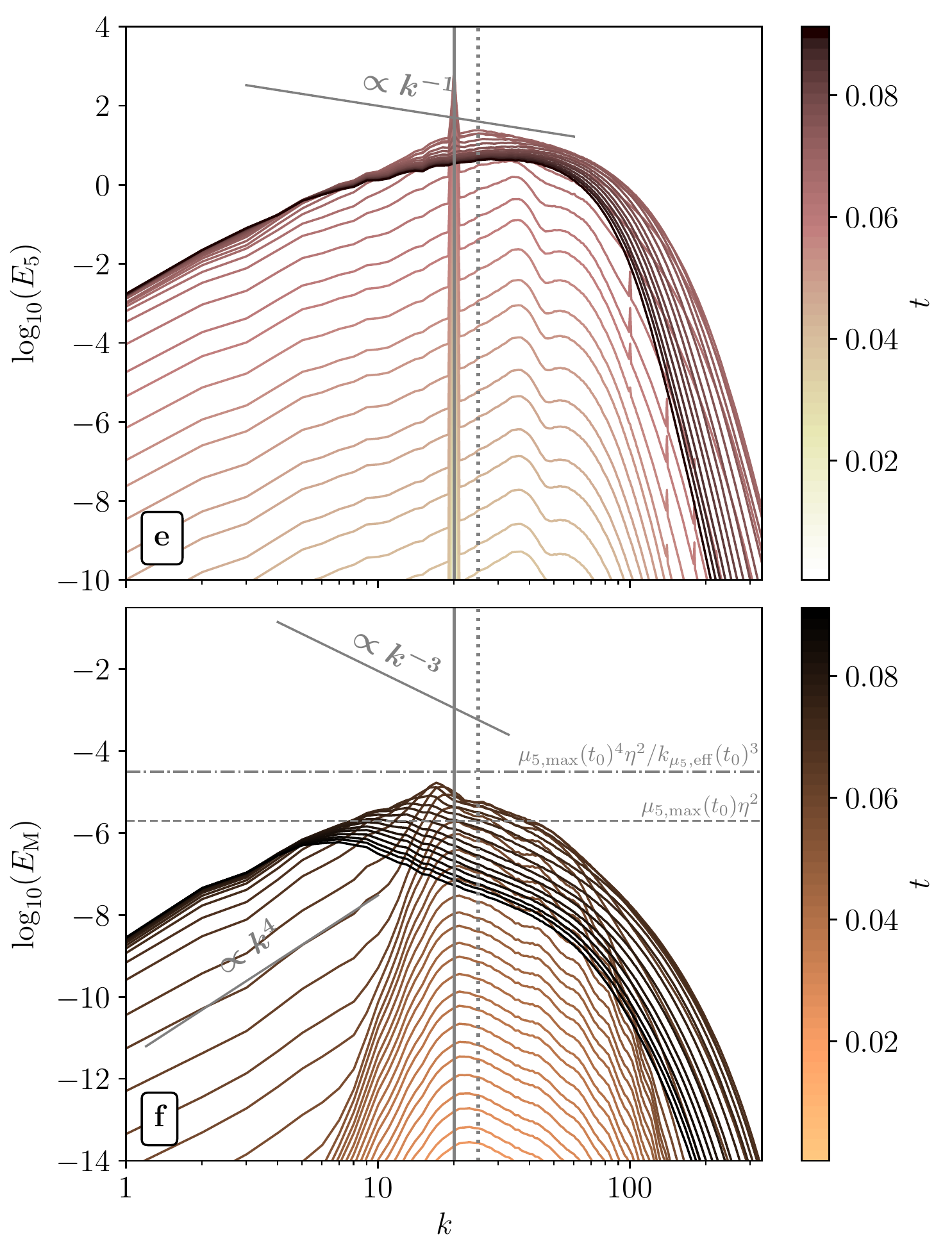}
\caption{Evolution of the spectra in runs H2 (with $\tilde{C}_5=16$ and $\tilde{C}_\lambda=1$, as reported in Ref.~\citep{BSRKBFRK17}), R$+$1, and S203D.
}
\label{fig_appendix_spec}
\end{figure*}

We have mentioned the mean-field chiral dynamo for the case of
a uniform $\mu_5$ in different places of the main text. 
For such systems the magnetic energy spectra developed a $k^{-2}$
scaling.
In Figs.~\ref{fig_appendix_spec}a and \ref{fig_appendix_spec}b, we present the energy spectra for our
comparison run H2 with initially constant $\mu_5$ and confirm the
$E_\mathrm{M}\propto k^{-2}$ scaling, which is different from the steeper
magnetic energy spectra for runs with initially 
inhomogeneous $\mu_5$ and vanishing $\langle \mu_5\rangle$.
The $\mu_5$ spectrum, on the other hand, approaches a $k^{-1}$
for all cases in which turbulence becomes sufficiently strong, hence
also for H2. 

We further present in Fig.~\ref{fig_appendix_spec} the spectra of 
run R$+$1, which are referred to 
in Sec.~\ref{sec_spectra}, and the spectra of run S203D that are 
mentioned in Sec.~\ref{sec_turb_sine}.

%

\end{document}